\documentclass[12pt]{article}
\usepackage{epsfig,amsfonts,amssymb}
\usepackage{hyperref}
\usepackage{cite}
\input epsf.sty
\usepackage{cite}

\def\ZZZ{{\hbox{ Z\kern-1.6mm Z}}}
\def\RRR{{\hbox{ R\kern-2.4mm R}}}
\def\CCC{{\hbox{ C\kern-2.0mm C}}}
\def\zzz{{\hbox{z\kern-1mm z}}}

\newcommand{\vt}{\vartheta}

\newcommand{\qeq}{{\hbox{=\kern-2.3mm ? \kern.5mm }}}
\renewcommand{\qeq}{=}

\newcommand{\eps}{\epsilon}

\newcommand{\vp}{\varphi}

\newcommand{\BB}{{\cal B}}

\newcommand{\BBB}{{\bf B}}

\newcommand{\AAA}{{\bf A}}
\newcommand{\GG}{{\cal G}}
\newcommand{\KK}{{\cal K}}

\newcommand{\FF}{{\cal F}}

\newcommand{\HH}{{\cal H}}
\newcommand{\MM}{{\cal M}}

\newcommand{\OO}{{\cal O}}

\newcommand{\PP}{{\cal P}}

\newcommand{\wt}{\widetilde}
\newcommand{\wh}{\widehat}

\newcommand{\RR}{{\cal R}}
\newcommand{\NN}{{\cal N}}

\newcommand{\be}{\begin{equation}}
\newcommand{\ee}{\end{equation}}
\newcommand{\ben}{\begin{eqnarray}\displaystyle}
\newcommand{\een}{\end{eqnarray}}

\newcommand{\refb}[1]{(\ref{#1})}
\newcommand{\p}{\partial}
\newcommand{\sectiono}[1]{\section{#1}\setcounter{equation}{0}}

\def\one{{\hbox{ 1\kern-.8mm l}}}
\def\zero{{\hbox{ 0\kern-1.5mm 0}}}

\newcommand{\bea}[1]{\begin{eqnarray}\label{#1} }
\newcommand{\eea}{\end{eqnarray}}

\newcommand{\eqref}{\refb}

\newcommand{\VVV}{{\cal X}}

\usepackage{bm}
\usepackage[table]{xcolor}

\def\figmove{

\def\JPicScale{0.4}
\ifx\JPicScale\undefined\def\JPicScale{1}\fi
\unitlength \JPicScale mm

}

\def\figrboundary{

\def\JPicScale{0.8}
\ifx\JPicScale\undefined\def\JPicScale{1}\fi
\unitlength \JPicScale mm
\begin{picture}(155,65)(0,0)
\linethickness{0.3mm}
\put(20,60){\line(1,0){120}}
\put(20,60){\vector(-1,0){0.12}}
\linethickness{0.3mm}
\put(20,40){\line(0,1){20}}
\put(20,40){\vector(0,-1){0.12}}
\linethickness{0.3mm}
\put(20,40){\line(1,0){120}}
\put(140,40){\vector(1,0){0.12}}
\linethickness{0.3mm}
\put(140,40){\line(0,1){20}}
\put(140,60){\vector(0,1){0.12}}
\put(80,65){\makebox(0,0)[cc]{$\wh S$}}

\put(80,35){\makebox(0,0)[cc]{$-S$}}

\put(28,50){\makebox(0,0)[cc]{$(\p\RR)'$}}

\put(131,50){\makebox(0,0)[cc]{$(\p\RR)'$}}

\put(80,50){\makebox(0,0)[cc]{$\RR$}}

\linethickness{0.3mm}
\put(155,40){\line(0,1){20}}
\put(155,60){\vector(0,1){0.12}}
\put(156,35){\makebox(0,0)[cc]{$V_f$}}

\end{picture}

}

\def\figanew{
\def\JPicScale{0.8}
\ifx\JPicScale\undefined\def\JPicScale{1}\fi
\unitlength \JPicScale mm

}

\def\figtorus{
\def\JPicScale{0.8}
\ifx\JPicScale\undefined\def\JPicScale{1}\fi
\unitlength \JPicScale mm
\begin{picture}(150,85)(0,0)
\linethickness{0.3mm}
\multiput(40,30)(0.12,0.12){333}{\line(1,0){0.12}}
\linethickness{0.3mm}
\multiput(100,30)(0.12,0.12){333}{\line(1,0){0.12}}
\linethickness{0.3mm}
\put(40,30){\line(1,0){60}}
\linethickness{0.3mm}
\put(80,70){\line(1,0){60}}
\linethickness{0.3mm}
\multiput(70,60)(1.97,0){31}{\line(1,0){0.98}}
\linethickness{0.3mm}
\put(75.4,42.25){\line(0,1){0.5}}
\multiput(75.31,43.74)(0.06,-0.49){1}{\line(0,-1){0.49}}
\multiput(75.09,44.71)(0.12,-0.48){1}{\line(0,-1){0.48}}
\multiput(74.76,45.64)(0.09,-0.23){2}{\line(0,-1){0.23}}
\multiput(74.3,46.52)(0.12,-0.22){2}{\line(0,-1){0.22}}
\multiput(73.75,47.35)(0.15,-0.2){2}{\line(0,-1){0.2}}
\multiput(73.09,48.09)(0.11,-0.12){3}{\line(0,-1){0.12}}
\multiput(72.35,48.75)(0.13,-0.11){3}{\line(1,0){0.13}}
\multiput(71.52,49.3)(0.21,-0.13){2}{\line(1,0){0.21}}
\multiput(70.64,49.76)(0.22,-0.11){2}{\line(1,0){0.22}}
\multiput(69.71,50.09)(0.47,-0.15){1}{\line(1,0){0.47}}
\multiput(68.74,50.31)(0.49,-0.09){1}{\line(1,0){0.49}}
\multiput(67.75,50.4)(0.5,-0.03){1}{\line(1,0){0.5}}
\put(67.25,50.4){\line(1,0){0.5}}
\multiput(66.76,50.37)(0.5,0.03){1}{\line(1,0){0.5}}
\multiput(65.78,50.22)(0.49,0.09){1}{\line(1,0){0.49}}
\multiput(64.82,49.94)(0.47,0.15){1}{\line(1,0){0.47}}
\multiput(63.91,49.54)(0.22,0.11){2}{\line(1,0){0.22}}
\multiput(63.06,49.04)(0.21,0.13){2}{\line(1,0){0.21}}
\multiput(62.27,48.43)(0.13,0.11){3}{\line(1,0){0.13}}
\multiput(61.57,47.73)(0.11,0.12){3}{\line(0,1){0.12}}
\multiput(60.96,46.94)(0.15,0.2){2}{\line(0,1){0.2}}
\multiput(60.46,46.09)(0.12,0.22){2}{\line(0,1){0.22}}
\multiput(60.06,45.18)(0.09,0.23){2}{\line(0,1){0.23}}
\multiput(59.78,44.22)(0.12,0.48){1}{\line(0,1){0.48}}
\multiput(59.63,43.24)(0.06,0.49){1}{\line(0,1){0.49}}
\put(59.6,42.25){\line(0,1){0.5}}
\multiput(59.6,42.25)(0.03,-0.5){1}{\line(0,-1){0.5}}
\multiput(59.69,41.26)(0.09,-0.49){1}{\line(0,-1){0.49}}
\multiput(59.91,40.29)(0.15,-0.47){1}{\line(0,-1){0.47}}
\multiput(60.24,39.36)(0.11,-0.22){2}{\line(0,-1){0.22}}
\multiput(60.7,38.48)(0.13,-0.21){2}{\line(0,-1){0.21}}
\multiput(61.25,37.65)(0.11,-0.13){3}{\line(0,-1){0.13}}
\multiput(61.91,36.91)(0.12,-0.11){3}{\line(1,0){0.12}}
\multiput(62.65,36.25)(0.2,-0.15){2}{\line(1,0){0.2}}
\multiput(63.48,35.7)(0.22,-0.12){2}{\line(1,0){0.22}}
\multiput(64.36,35.24)(0.23,-0.09){2}{\line(1,0){0.23}}
\multiput(65.29,34.91)(0.48,-0.12){1}{\line(1,0){0.48}}
\multiput(66.26,34.69)(0.49,-0.06){1}{\line(1,0){0.49}}
\put(67.25,34.6){\line(1,0){0.5}}
\multiput(67.75,34.6)(0.5,0.03){1}{\line(1,0){0.5}}
\multiput(68.74,34.69)(0.49,0.09){1}{\line(1,0){0.49}}
\multiput(69.71,34.91)(0.47,0.15){1}{\line(1,0){0.47}}
\multiput(70.64,35.24)(0.22,0.11){2}{\line(1,0){0.22}}
\multiput(71.52,35.7)(0.21,0.13){2}{\line(1,0){0.21}}
\multiput(72.35,36.25)(0.13,0.11){3}{\line(1,0){0.13}}
\multiput(73.09,36.91)(0.11,0.12){3}{\line(0,1){0.12}}
\multiput(73.75,37.65)(0.15,0.2){2}{\line(0,1){0.2}}
\multiput(74.3,38.48)(0.12,0.22){2}{\line(0,1){0.22}}
\multiput(74.76,39.36)(0.09,0.23){2}{\line(0,1){0.23}}
\multiput(75.09,40.29)(0.12,0.48){1}{\line(0,1){0.48}}
\multiput(75.31,41.26)(0.06,0.49){1}{\line(0,1){0.49}}

\put(67,42){\makebox(0,0)[cc]{$D_1$}}

\put(90,45){\makebox(0,0)[cc]{$D_2$}}

\put(95,65){\makebox(0,0)[cc]{$D_3$}}

\put(76,50){\makebox(0,0)[cc]{$C_1$}}

\put(106,55){\makebox(0,0)[cc]{$C_2$}}

\put(115,75){\makebox(0,0)[cc]{$C_3$}}

\put(80,25){\makebox(0,0)[cc]{$C_3$}}

\linethickness{0.3mm}
\qbezier(130,70)(137.82,77.9)(142.03,76.09)
\qbezier(142.03,76.09)(146.24,74.29)(147.5,62.5)
\put(130,70){\vector(-1,-1){0.12}}
\qbezier(147.5,62.5)(148.86,50.8)(145.25,42.38)
\qbezier(145.25,42.38)(141.64,33.95)(132.5,27.5)
\qbezier(132.5,27.5)(123.4,20.96)(116.78,19.16)
\qbezier(116.78,19.16)(110.16,17.35)(105,20)
\qbezier(105,20)(99.8,22.59)(96.19,25)
\qbezier(96.19,25)(92.58,27.41)(90,30)
\put(90,30){\vector(-1,1){0.12}}
\end{picture}
}

\def\figxa{
\def\JPicScale{0.8}
\ifx\JPicScale\undefined\def\JPicScale{1}\fi
\unitlength \JPicScale mm

}

\def\figy{
\def\JPicScale{0.8}
\ifx\JPicScale\undefined\def\JPicScale{1}\fi
\unitlength \JPicScale mm
\begin{picture}(130,60)(0,0)
\put(130,60){\makebox(0,0)[cc]{}}

\linethickness{0.3mm}
\put(31.01,39.25){\line(0,1){0.5}}
\multiput(30.99,40.25)(0.02,-0.5){1}{\line(0,-1){0.5}}
\multiput(30.94,40.75)(0.05,-0.5){1}{\line(0,-1){0.5}}
\multiput(30.87,41.24)(0.07,-0.5){1}{\line(0,-1){0.5}}
\multiput(30.77,41.73)(0.09,-0.49){1}{\line(0,-1){0.49}}
\multiput(30.65,42.22)(0.12,-0.49){1}{\line(0,-1){0.49}}
\multiput(30.51,42.7)(0.14,-0.48){1}{\line(0,-1){0.48}}
\multiput(30.35,43.17)(0.16,-0.47){1}{\line(0,-1){0.47}}
\multiput(30.16,43.64)(0.09,-0.23){2}{\line(0,-1){0.23}}
\multiput(29.96,44.09)(0.1,-0.23){2}{\line(0,-1){0.23}}
\multiput(29.73,44.54)(0.11,-0.22){2}{\line(0,-1){0.22}}
\multiput(29.48,44.97)(0.13,-0.22){2}{\line(0,-1){0.22}}
\multiput(29.21,45.39)(0.14,-0.21){2}{\line(0,-1){0.21}}
\multiput(28.92,45.8)(0.15,-0.2){2}{\line(0,-1){0.2}}
\multiput(28.61,46.19)(0.1,-0.13){3}{\line(0,-1){0.13}}
\multiput(28.28,46.57)(0.11,-0.13){3}{\line(0,-1){0.13}}
\multiput(27.93,46.93)(0.12,-0.12){3}{\line(0,-1){0.12}}
\multiput(27.57,47.28)(0.12,-0.12){3}{\line(1,0){0.12}}
\multiput(27.19,47.61)(0.13,-0.11){3}{\line(1,0){0.13}}
\multiput(26.8,47.92)(0.13,-0.1){3}{\line(1,0){0.13}}
\multiput(26.39,48.21)(0.2,-0.15){2}{\line(1,0){0.2}}
\multiput(25.97,48.48)(0.21,-0.14){2}{\line(1,0){0.21}}
\multiput(25.54,48.73)(0.22,-0.13){2}{\line(1,0){0.22}}
\multiput(25.09,48.96)(0.22,-0.11){2}{\line(1,0){0.22}}
\multiput(24.64,49.16)(0.23,-0.1){2}{\line(1,0){0.23}}
\multiput(24.17,49.35)(0.23,-0.09){2}{\line(1,0){0.23}}
\multiput(23.7,49.51)(0.47,-0.16){1}{\line(1,0){0.47}}
\multiput(23.22,49.65)(0.48,-0.14){1}{\line(1,0){0.48}}
\multiput(22.73,49.77)(0.49,-0.12){1}{\line(1,0){0.49}}
\multiput(22.24,49.87)(0.49,-0.09){1}{\line(1,0){0.49}}
\multiput(21.75,49.94)(0.5,-0.07){1}{\line(1,0){0.5}}
\multiput(21.25,49.99)(0.5,-0.05){1}{\line(1,0){0.5}}
\multiput(20.75,50.01)(0.5,-0.02){1}{\line(1,0){0.5}}
\put(20.25,50.01){\line(1,0){0.5}}
\multiput(19.75,49.99)(0.5,0.02){1}{\line(1,0){0.5}}
\multiput(19.25,49.94)(0.5,0.05){1}{\line(1,0){0.5}}
\multiput(18.76,49.87)(0.5,0.07){1}{\line(1,0){0.5}}
\multiput(18.27,49.77)(0.49,0.09){1}{\line(1,0){0.49}}
\multiput(17.78,49.65)(0.49,0.12){1}{\line(1,0){0.49}}
\multiput(17.3,49.51)(0.48,0.14){1}{\line(1,0){0.48}}
\multiput(16.83,49.35)(0.47,0.16){1}{\line(1,0){0.47}}
\multiput(16.36,49.16)(0.23,0.09){2}{\line(1,0){0.23}}
\multiput(15.91,48.96)(0.23,0.1){2}{\line(1,0){0.23}}
\multiput(15.46,48.73)(0.22,0.11){2}{\line(1,0){0.22}}
\multiput(15.03,48.48)(0.22,0.13){2}{\line(1,0){0.22}}
\multiput(14.61,48.21)(0.21,0.14){2}{\line(1,0){0.21}}
\multiput(14.2,47.92)(0.2,0.15){2}{\line(1,0){0.2}}
\multiput(13.81,47.61)(0.13,0.1){3}{\line(1,0){0.13}}
\multiput(13.43,47.28)(0.13,0.11){3}{\line(1,0){0.13}}
\multiput(13.07,46.93)(0.12,0.12){3}{\line(1,0){0.12}}
\multiput(12.72,46.57)(0.12,0.12){3}{\line(0,1){0.12}}
\multiput(12.39,46.19)(0.11,0.13){3}{\line(0,1){0.13}}
\multiput(12.08,45.8)(0.1,0.13){3}{\line(0,1){0.13}}
\multiput(11.79,45.39)(0.15,0.2){2}{\line(0,1){0.2}}
\multiput(11.52,44.97)(0.14,0.21){2}{\line(0,1){0.21}}
\multiput(11.27,44.54)(0.13,0.22){2}{\line(0,1){0.22}}
\multiput(11.04,44.09)(0.11,0.22){2}{\line(0,1){0.22}}
\multiput(10.84,43.64)(0.1,0.23){2}{\line(0,1){0.23}}
\multiput(10.65,43.17)(0.09,0.23){2}{\line(0,1){0.23}}
\multiput(10.49,42.7)(0.16,0.47){1}{\line(0,1){0.47}}
\multiput(10.35,42.22)(0.14,0.48){1}{\line(0,1){0.48}}
\multiput(10.23,41.73)(0.12,0.49){1}{\line(0,1){0.49}}
\multiput(10.13,41.24)(0.09,0.49){1}{\line(0,1){0.49}}
\multiput(10.06,40.75)(0.07,0.5){1}{\line(0,1){0.5}}
\multiput(10.01,40.25)(0.05,0.5){1}{\line(0,1){0.5}}
\multiput(9.99,39.75)(0.02,0.5){1}{\line(0,1){0.5}}
\put(9.99,39.25){\line(0,1){0.5}}
\multiput(9.99,39.25)(0.02,-0.5){1}{\line(0,-1){0.5}}
\multiput(10.01,38.75)(0.05,-0.5){1}{\line(0,-1){0.5}}
\multiput(10.06,38.25)(0.07,-0.5){1}{\line(0,-1){0.5}}
\multiput(10.13,37.76)(0.09,-0.49){1}{\line(0,-1){0.49}}
\multiput(10.23,37.27)(0.12,-0.49){1}{\line(0,-1){0.49}}
\multiput(10.35,36.78)(0.14,-0.48){1}{\line(0,-1){0.48}}
\multiput(10.49,36.3)(0.16,-0.47){1}{\line(0,-1){0.47}}
\multiput(10.65,35.83)(0.09,-0.23){2}{\line(0,-1){0.23}}
\multiput(10.84,35.36)(0.1,-0.23){2}{\line(0,-1){0.23}}
\multiput(11.04,34.91)(0.11,-0.22){2}{\line(0,-1){0.22}}
\multiput(11.27,34.46)(0.13,-0.22){2}{\line(0,-1){0.22}}
\multiput(11.52,34.03)(0.14,-0.21){2}{\line(0,-1){0.21}}
\multiput(11.79,33.61)(0.15,-0.2){2}{\line(0,-1){0.2}}
\multiput(12.08,33.2)(0.1,-0.13){3}{\line(0,-1){0.13}}
\multiput(12.39,32.81)(0.11,-0.13){3}{\line(0,-1){0.13}}
\multiput(12.72,32.43)(0.12,-0.12){3}{\line(0,-1){0.12}}
\multiput(13.07,32.07)(0.12,-0.12){3}{\line(1,0){0.12}}
\multiput(13.43,31.72)(0.13,-0.11){3}{\line(1,0){0.13}}
\multiput(13.81,31.39)(0.13,-0.1){3}{\line(1,0){0.13}}
\multiput(14.2,31.08)(0.2,-0.15){2}{\line(1,0){0.2}}
\multiput(14.61,30.79)(0.21,-0.14){2}{\line(1,0){0.21}}
\multiput(15.03,30.52)(0.22,-0.13){2}{\line(1,0){0.22}}
\multiput(15.46,30.27)(0.22,-0.11){2}{\line(1,0){0.22}}
\multiput(15.91,30.04)(0.23,-0.1){2}{\line(1,0){0.23}}
\multiput(16.36,29.84)(0.23,-0.09){2}{\line(1,0){0.23}}
\multiput(16.83,29.65)(0.47,-0.16){1}{\line(1,0){0.47}}
\multiput(17.3,29.49)(0.48,-0.14){1}{\line(1,0){0.48}}
\multiput(17.78,29.35)(0.49,-0.12){1}{\line(1,0){0.49}}
\multiput(18.27,29.23)(0.49,-0.09){1}{\line(1,0){0.49}}
\multiput(18.76,29.13)(0.5,-0.07){1}{\line(1,0){0.5}}
\multiput(19.25,29.06)(0.5,-0.05){1}{\line(1,0){0.5}}
\multiput(19.75,29.01)(0.5,-0.02){1}{\line(1,0){0.5}}
\put(20.25,28.99){\line(1,0){0.5}}
\multiput(20.75,28.99)(0.5,0.02){1}{\line(1,0){0.5}}
\multiput(21.25,29.01)(0.5,0.05){1}{\line(1,0){0.5}}
\multiput(21.75,29.06)(0.5,0.07){1}{\line(1,0){0.5}}
\multiput(22.24,29.13)(0.49,0.09){1}{\line(1,0){0.49}}
\multiput(22.73,29.23)(0.49,0.12){1}{\line(1,0){0.49}}
\multiput(23.22,29.35)(0.48,0.14){1}{\line(1,0){0.48}}
\multiput(23.7,29.49)(0.47,0.16){1}{\line(1,0){0.47}}
\multiput(24.17,29.65)(0.23,0.09){2}{\line(1,0){0.23}}
\multiput(24.64,29.84)(0.23,0.1){2}{\line(1,0){0.23}}
\multiput(25.09,30.04)(0.22,0.11){2}{\line(1,0){0.22}}
\multiput(25.54,30.27)(0.22,0.13){2}{\line(1,0){0.22}}
\multiput(25.97,30.52)(0.21,0.14){2}{\line(1,0){0.21}}
\multiput(26.39,30.79)(0.2,0.15){2}{\line(1,0){0.2}}
\multiput(26.8,31.08)(0.13,0.1){3}{\line(1,0){0.13}}
\multiput(27.19,31.39)(0.13,0.11){3}{\line(1,0){0.13}}
\multiput(27.57,31.72)(0.12,0.12){3}{\line(1,0){0.12}}
\multiput(27.93,32.07)(0.12,0.12){3}{\line(0,1){0.12}}
\multiput(28.28,32.43)(0.11,0.13){3}{\line(0,1){0.13}}
\multiput(28.61,32.81)(0.1,0.13){3}{\line(0,1){0.13}}
\multiput(28.92,33.2)(0.15,0.2){2}{\line(0,1){0.2}}
\multiput(29.21,33.61)(0.14,0.21){2}{\line(0,1){0.21}}
\multiput(29.48,34.03)(0.13,0.22){2}{\line(0,1){0.22}}
\multiput(29.73,34.46)(0.11,0.22){2}{\line(0,1){0.22}}
\multiput(29.96,34.91)(0.1,0.23){2}{\line(0,1){0.23}}
\multiput(30.16,35.36)(0.09,0.23){2}{\line(0,1){0.23}}
\multiput(30.35,35.83)(0.16,0.47){1}{\line(0,1){0.47}}
\multiput(30.51,36.3)(0.14,0.48){1}{\line(0,1){0.48}}
\multiput(30.65,36.78)(0.12,0.49){1}{\line(0,1){0.49}}
\multiput(30.77,37.27)(0.09,0.49){1}{\line(0,1){0.49}}
\multiput(30.87,37.76)(0.07,0.5){1}{\line(0,1){0.5}}
\multiput(30.94,38.25)(0.05,0.5){1}{\line(0,1){0.5}}
\multiput(30.99,38.75)(0.02,0.5){1}{\line(0,1){0.5}}

\linethickness{0.3mm}
\put(60.51,39.75){\line(0,1){0.5}}
\multiput(60.49,40.75)(0.02,-0.5){1}{\line(0,-1){0.5}}
\multiput(60.44,41.25)(0.05,-0.5){1}{\line(0,-1){0.5}}
\multiput(60.37,41.74)(0.07,-0.5){1}{\line(0,-1){0.5}}
\multiput(60.27,42.23)(0.09,-0.49){1}{\line(0,-1){0.49}}
\multiput(60.15,42.72)(0.12,-0.49){1}{\line(0,-1){0.49}}
\multiput(60.01,43.2)(0.14,-0.48){1}{\line(0,-1){0.48}}
\multiput(59.85,43.67)(0.16,-0.47){1}{\line(0,-1){0.47}}
\multiput(59.66,44.14)(0.09,-0.23){2}{\line(0,-1){0.23}}
\multiput(59.46,44.59)(0.1,-0.23){2}{\line(0,-1){0.23}}
\multiput(59.23,45.04)(0.11,-0.22){2}{\line(0,-1){0.22}}
\multiput(58.98,45.47)(0.13,-0.22){2}{\line(0,-1){0.22}}
\multiput(58.71,45.89)(0.14,-0.21){2}{\line(0,-1){0.21}}
\multiput(58.42,46.3)(0.15,-0.2){2}{\line(0,-1){0.2}}
\multiput(58.11,46.69)(0.1,-0.13){3}{\line(0,-1){0.13}}
\multiput(57.78,47.07)(0.11,-0.13){3}{\line(0,-1){0.13}}
\multiput(57.43,47.43)(0.12,-0.12){3}{\line(0,-1){0.12}}
\multiput(57.07,47.78)(0.12,-0.12){3}{\line(1,0){0.12}}
\multiput(56.69,48.11)(0.13,-0.11){3}{\line(1,0){0.13}}
\multiput(56.3,48.42)(0.13,-0.1){3}{\line(1,0){0.13}}
\multiput(55.89,48.71)(0.2,-0.15){2}{\line(1,0){0.2}}
\multiput(55.47,48.98)(0.21,-0.14){2}{\line(1,0){0.21}}
\multiput(55.04,49.23)(0.22,-0.13){2}{\line(1,0){0.22}}
\multiput(54.59,49.46)(0.22,-0.11){2}{\line(1,0){0.22}}
\multiput(54.14,49.66)(0.23,-0.1){2}{\line(1,0){0.23}}
\multiput(53.67,49.85)(0.23,-0.09){2}{\line(1,0){0.23}}
\multiput(53.2,50.01)(0.47,-0.16){1}{\line(1,0){0.47}}
\multiput(52.72,50.15)(0.48,-0.14){1}{\line(1,0){0.48}}
\multiput(52.23,50.27)(0.49,-0.12){1}{\line(1,0){0.49}}
\multiput(51.74,50.37)(0.49,-0.09){1}{\line(1,0){0.49}}
\multiput(51.25,50.44)(0.5,-0.07){1}{\line(1,0){0.5}}
\multiput(50.75,50.49)(0.5,-0.05){1}{\line(1,0){0.5}}
\multiput(50.25,50.51)(0.5,-0.02){1}{\line(1,0){0.5}}
\put(49.75,50.51){\line(1,0){0.5}}
\multiput(49.25,50.49)(0.5,0.02){1}{\line(1,0){0.5}}
\multiput(48.75,50.44)(0.5,0.05){1}{\line(1,0){0.5}}
\multiput(48.26,50.37)(0.5,0.07){1}{\line(1,0){0.5}}
\multiput(47.77,50.27)(0.49,0.09){1}{\line(1,0){0.49}}
\multiput(47.28,50.15)(0.49,0.12){1}{\line(1,0){0.49}}
\multiput(46.8,50.01)(0.48,0.14){1}{\line(1,0){0.48}}
\multiput(46.33,49.85)(0.47,0.16){1}{\line(1,0){0.47}}
\multiput(45.86,49.66)(0.23,0.09){2}{\line(1,0){0.23}}
\multiput(45.41,49.46)(0.23,0.1){2}{\line(1,0){0.23}}
\multiput(44.96,49.23)(0.22,0.11){2}{\line(1,0){0.22}}
\multiput(44.53,48.98)(0.22,0.13){2}{\line(1,0){0.22}}
\multiput(44.11,48.71)(0.21,0.14){2}{\line(1,0){0.21}}
\multiput(43.7,48.42)(0.2,0.15){2}{\line(1,0){0.2}}
\multiput(43.31,48.11)(0.13,0.1){3}{\line(1,0){0.13}}
\multiput(42.93,47.78)(0.13,0.11){3}{\line(1,0){0.13}}
\multiput(42.57,47.43)(0.12,0.12){3}{\line(1,0){0.12}}
\multiput(42.22,47.07)(0.12,0.12){3}{\line(0,1){0.12}}
\multiput(41.89,46.69)(0.11,0.13){3}{\line(0,1){0.13}}
\multiput(41.58,46.3)(0.1,0.13){3}{\line(0,1){0.13}}
\multiput(41.29,45.89)(0.15,0.2){2}{\line(0,1){0.2}}
\multiput(41.02,45.47)(0.14,0.21){2}{\line(0,1){0.21}}
\multiput(40.77,45.04)(0.13,0.22){2}{\line(0,1){0.22}}
\multiput(40.54,44.59)(0.11,0.22){2}{\line(0,1){0.22}}
\multiput(40.34,44.14)(0.1,0.23){2}{\line(0,1){0.23}}
\multiput(40.15,43.67)(0.09,0.23){2}{\line(0,1){0.23}}
\multiput(39.99,43.2)(0.16,0.47){1}{\line(0,1){0.47}}
\multiput(39.85,42.72)(0.14,0.48){1}{\line(0,1){0.48}}
\multiput(39.73,42.23)(0.12,0.49){1}{\line(0,1){0.49}}
\multiput(39.63,41.74)(0.09,0.49){1}{\line(0,1){0.49}}
\multiput(39.56,41.25)(0.07,0.5){1}{\line(0,1){0.5}}
\multiput(39.51,40.75)(0.05,0.5){1}{\line(0,1){0.5}}
\multiput(39.49,40.25)(0.02,0.5){1}{\line(0,1){0.5}}
\put(39.49,39.75){\line(0,1){0.5}}
\multiput(39.49,39.75)(0.02,-0.5){1}{\line(0,-1){0.5}}
\multiput(39.51,39.25)(0.05,-0.5){1}{\line(0,-1){0.5}}
\multiput(39.56,38.75)(0.07,-0.5){1}{\line(0,-1){0.5}}
\multiput(39.63,38.26)(0.09,-0.49){1}{\line(0,-1){0.49}}
\multiput(39.73,37.77)(0.12,-0.49){1}{\line(0,-1){0.49}}
\multiput(39.85,37.28)(0.14,-0.48){1}{\line(0,-1){0.48}}
\multiput(39.99,36.8)(0.16,-0.47){1}{\line(0,-1){0.47}}
\multiput(40.15,36.33)(0.09,-0.23){2}{\line(0,-1){0.23}}
\multiput(40.34,35.86)(0.1,-0.23){2}{\line(0,-1){0.23}}
\multiput(40.54,35.41)(0.11,-0.22){2}{\line(0,-1){0.22}}
\multiput(40.77,34.96)(0.13,-0.22){2}{\line(0,-1){0.22}}
\multiput(41.02,34.53)(0.14,-0.21){2}{\line(0,-1){0.21}}
\multiput(41.29,34.11)(0.15,-0.2){2}{\line(0,-1){0.2}}
\multiput(41.58,33.7)(0.1,-0.13){3}{\line(0,-1){0.13}}
\multiput(41.89,33.31)(0.11,-0.13){3}{\line(0,-1){0.13}}
\multiput(42.22,32.93)(0.12,-0.12){3}{\line(0,-1){0.12}}
\multiput(42.57,32.57)(0.12,-0.12){3}{\line(1,0){0.12}}
\multiput(42.93,32.22)(0.13,-0.11){3}{\line(1,0){0.13}}
\multiput(43.31,31.89)(0.13,-0.1){3}{\line(1,0){0.13}}
\multiput(43.7,31.58)(0.2,-0.15){2}{\line(1,0){0.2}}
\multiput(44.11,31.29)(0.21,-0.14){2}{\line(1,0){0.21}}
\multiput(44.53,31.02)(0.22,-0.13){2}{\line(1,0){0.22}}
\multiput(44.96,30.77)(0.22,-0.11){2}{\line(1,0){0.22}}
\multiput(45.41,30.54)(0.23,-0.1){2}{\line(1,0){0.23}}
\multiput(45.86,30.34)(0.23,-0.09){2}{\line(1,0){0.23}}
\multiput(46.33,30.15)(0.47,-0.16){1}{\line(1,0){0.47}}
\multiput(46.8,29.99)(0.48,-0.14){1}{\line(1,0){0.48}}
\multiput(47.28,29.85)(0.49,-0.12){1}{\line(1,0){0.49}}
\multiput(47.77,29.73)(0.49,-0.09){1}{\line(1,0){0.49}}
\multiput(48.26,29.63)(0.5,-0.07){1}{\line(1,0){0.5}}
\multiput(48.75,29.56)(0.5,-0.05){1}{\line(1,0){0.5}}
\multiput(49.25,29.51)(0.5,-0.02){1}{\line(1,0){0.5}}
\put(49.75,29.49){\line(1,0){0.5}}
\multiput(50.25,29.49)(0.5,0.02){1}{\line(1,0){0.5}}
\multiput(50.75,29.51)(0.5,0.05){1}{\line(1,0){0.5}}
\multiput(51.25,29.56)(0.5,0.07){1}{\line(1,0){0.5}}
\multiput(51.74,29.63)(0.49,0.09){1}{\line(1,0){0.49}}
\multiput(52.23,29.73)(0.49,0.12){1}{\line(1,0){0.49}}
\multiput(52.72,29.85)(0.48,0.14){1}{\line(1,0){0.48}}
\multiput(53.2,29.99)(0.47,0.16){1}{\line(1,0){0.47}}
\multiput(53.67,30.15)(0.23,0.09){2}{\line(1,0){0.23}}
\multiput(54.14,30.34)(0.23,0.1){2}{\line(1,0){0.23}}
\multiput(54.59,30.54)(0.22,0.11){2}{\line(1,0){0.22}}
\multiput(55.04,30.77)(0.22,0.13){2}{\line(1,0){0.22}}
\multiput(55.47,31.02)(0.21,0.14){2}{\line(1,0){0.21}}
\multiput(55.89,31.29)(0.2,0.15){2}{\line(1,0){0.2}}
\multiput(56.3,31.58)(0.13,0.1){3}{\line(1,0){0.13}}
\multiput(56.69,31.89)(0.13,0.11){3}{\line(1,0){0.13}}
\multiput(57.07,32.22)(0.12,0.12){3}{\line(1,0){0.12}}
\multiput(57.43,32.57)(0.12,0.12){3}{\line(0,1){0.12}}
\multiput(57.78,32.93)(0.11,0.13){3}{\line(0,1){0.13}}
\multiput(58.11,33.31)(0.1,0.13){3}{\line(0,1){0.13}}
\multiput(58.42,33.7)(0.15,0.2){2}{\line(0,1){0.2}}
\multiput(58.71,34.11)(0.14,0.21){2}{\line(0,1){0.21}}
\multiput(58.98,34.53)(0.13,0.22){2}{\line(0,1){0.22}}
\multiput(59.23,34.96)(0.11,0.22){2}{\line(0,1){0.22}}
\multiput(59.46,35.41)(0.1,0.23){2}{\line(0,1){0.23}}
\multiput(59.66,35.86)(0.09,0.23){2}{\line(0,1){0.23}}
\multiput(59.85,36.33)(0.16,0.47){1}{\line(0,1){0.47}}
\multiput(60.01,36.8)(0.14,0.48){1}{\line(0,1){0.48}}
\multiput(60.15,37.28)(0.12,0.49){1}{\line(0,1){0.49}}
\multiput(60.27,37.77)(0.09,0.49){1}{\line(0,1){0.49}}
\multiput(60.37,38.26)(0.07,0.5){1}{\line(0,1){0.5}}
\multiput(60.44,38.75)(0.05,0.5){1}{\line(0,1){0.5}}
\multiput(60.49,39.25)(0.02,0.5){1}{\line(0,1){0.5}}

\put(10,40){\makebox(0,0)[cc]{$\bullet$}}

\put(60,40){\makebox(0,0)[cc]{$\bullet$}}

\put(55,49.5){\makebox(0,0)[cc]{$\bullet$}}

\put(59,45){\makebox(0,0)[cc]{$\bullet$}}

\put(55,30.5){\makebox(0,0)[cc]{$\bullet$}}

\put(59,35){\makebox(0,0)[cc]{$\bullet$}}

\linethickness{0.3mm}
\put(30,40){\line(1,0){10}}
\put(20,40){\makebox(0,0)[cc]{$\delta\HH$}}

\put(50,40){\makebox(0,0)[cc]{$\Gamma$}}

\end{picture}

}

\def\figvertical{
\def\JPicScale{0.4}
\ifx\JPicScale\undefined\def\JPicScale{1}\fi
\unitlength \JPicScale mm
\begin{picture}(110,75)(0,0)
\linethickness{0.3mm}
\qbezier(60,40)(60.55,40.74)(67.2,45.45)
\qbezier(67.2,45.45)(73.85,50.17)(80,50)
\qbezier(80,50)(84.92,48.37)(87.49,42.33)
\qbezier(87.49,42.33)(90.07,36.29)(95,35)
\qbezier(95,35)(99.33,35.16)(102.48,39.47)
\qbezier(102.48,39.47)(105.62,43.78)(110,45)
\linethickness{0.3mm}
\qbezier(60,60)(59.63,59.32)(54.76,54.78)
\qbezier(54.76,54.78)(49.9,50.23)(45,50)
\qbezier(45,50)(40.06,51.22)(37.45,57.04)
\qbezier(37.45,57.04)(34.84,62.86)(30,65)
\qbezier(30,65)(24.89,66.24)(19.78,64.51)
\qbezier(19.78,64.51)(14.66,62.78)(10,60)
\linethickness{0.3mm}
\put(60,40){\line(0,1){20}}
\linethickness{0.3mm}
\put(40,20){\line(1,0){50}}
\put(90,20){\vector(1,0){0.12}}
\put(105,20){\makebox(0,0)[cc]{$\MM_{g,n}$}}

\put(70,75){\makebox(0,0)[cc]{$\wt\PP_{g,n}$}}

\end{picture}

}

\def\figvertsec{

\def\JPicScale{0.5}
\ifx\JPicScale\undefined\def\JPicScale{1}\fi
\unitlength \JPicScale mm
\begin{picture}(115,85)(0,0)
\linethickness{0.3mm}
\qbezier(40,60)(39.25,65.13)(44.31,69.22)
\qbezier(44.31,69.22)(49.37,73.31)(55,75)
\qbezier(55,75)(61.89,77.07)(70.11,76.38)
\qbezier(70.11,76.38)(78.33,75.69)(80,70)
\qbezier(80,70)(80.38,63.29)(71.8,59.55)
\qbezier(71.8,59.55)(63.21,55.81)(55,55)
\qbezier(55,55)(50.65,54.46)(45.93,55.41)
\qbezier(45.93,55.41)(41.21,56.36)(40,60)
\linethickness{0.3mm}
\qbezier(40,30)(42.42,35.4)(52.32,32.3)
\qbezier(52.32,32.3)(62.23,29.2)(70,25)
\qbezier(70,25)(73.67,23.25)(77.47,20.48)
\qbezier(77.47,20.48)(81.26,17.72)(80,15)
\qbezier(80,15)(72.73,7.92)(56.3,14.29)
\qbezier(56.3,14.29)(39.88,20.67)(40,30)
\linethickness{0.3mm}
\put(40,30){\line(0,1){30}}
\linethickness{0.3mm}
\put(80,15){\line(0,1){55}}
\put(60,83){\makebox(0,0)[cc]{$  C_2$}}

\put(46,13){\makebox(0,0)[cc]{$  C_1$}}

\put(85,40){\makebox(0,0)[cc]{$  C$}}

\put(94,55){\makebox(0,0)[cc]{$  L$}}

\linethickness{0.3mm}
\put(40,0){\line(1,0){75}}
\put(115,0){\vector(1,0){0.12}}
\linethickness{0.3mm}
\put(20,10){\line(0,1){50}}
\put(20,60){\vector(0,1){0.12}}
\put(110,5){\makebox(0,0)[cc]{$\MM_{g,n}$}}

\put(15,70){\makebox(0,0)[cc]{$z_i$}}

\put(80,5){\makebox(0,0)[cc]{}}

\linethickness{0.1mm}
\qbezier(60,20)(64.4,19.71)(65.52,27.26)
\qbezier(65.52,27.26)(66.64,34.82)(70,40)
\qbezier(70,40)(76.15,47.01)(83.63,52.14)
\qbezier(83.63,52.14)(91.11,57.27)(100,60)
\linethickness{0.3mm}
\multiput(40,0)(0.48,0.12){125}{\line(1,0){0.48}}
\put(100,15){\vector(4,1){0.12}}
\end{picture}

}

\def\figtwodsection{

\def\JPicScale{0.5}
\ifx\JPicScale\undefined\def\JPicScale{1}\fi
\unitlength \JPicScale mm
\begin{picture}(145,70)(0,0)
\linethickness{0.3mm}
\put(80,20){\line(0,1){50}}
\linethickness{0.3mm}
\put(50,30){\line(0,1){30}}
\linethickness{0.3mm}
\qbezier(80,20)(90.41,27.87)(97.62,28.47)
\qbezier(97.62,28.47)(104.84,29.07)(110,22.5)
\qbezier(110,22.5)(115.14,15.93)(123.56,16.53)
\qbezier(123.56,16.53)(131.98,17.13)(145,25)
\linethickness{0.3mm}
\qbezier(80,70)(74.84,59.53)(67.62,57.12)
\qbezier(67.62,57.12)(60.41,54.72)(50,60)
\linethickness{0.3mm}
\qbezier(50,30)(42.19,35.25)(37.38,35.25)
\qbezier(37.38,35.25)(32.56,35.25)(30,30)
\qbezier(30,30)(27.45,24.73)(21.44,25.94)
\qbezier(21.44,25.94)(15.42,27.14)(5,35)

\put(20,35){\makebox(0,0)[cc]{$\wh S$}}

\put(45,45){\makebox(0,0)[cc]{$C$}}

\put(85,45){\makebox(0,0)[cc]{$C$}}

\put(120,30){\makebox(0,0)[cc]{$\wh S$}}

\put(62,65){\makebox(0,0)[cc]{$\wt S$}}

\end{picture}
}

\def\figsquare{
\def\JPicScale{0.8}
\ifx\JPicScale\undefined\def\JPicScale{1}\fi
\unitlength \JPicScale mm
\begin{picture}(100,85)(0,0)
\linethickness{0.3mm}
\put(30,30){\line(0,1){30}}
\put(30,60){\vector(0,1){0.12}}
\linethickness{0.3mm}
\put(30,30){\line(1,0){50}}
\put(80,30){\vector(1,0){0.12}}
\linethickness{0.3mm}
\put(80,30){\line(0,1){30}}
\put(80,60){\vector(0,1){0.12}}
\linethickness{0.3mm}
\put(30,60){\line(1,0){50}}
\put(80,60){\vector(1,0){0.12}}
\linethickness{0.3mm}
\put(10,20){\line(0,1){60}}
\put(10,80){\vector(0,1){0.12}}
\linethickness{0.3mm}
\put(10,20){\line(1,0){80}}
\put(90,20){\vector(1,0){0.12}}
\put(100,20){\makebox(0,0)[cc]{$z^{}_1$}}

\put(5,85){\makebox(0,0)[cc]{$z^{}_2$}}

\put(25,25){\makebox(0,0)[cc]{$(u_1, v_1)$}}

\put(75,25){\makebox(0,0)[cc]{$(v_1, u_2)$}}

\put(25,65){\makebox(0,0)[cc]{$(u_1,v_2)$}}

\put(75,65){\makebox(0,0)[cc]{$(u_2,v_2)$}}

\linethickness{0.03mm}
\qbezier(25,45)(35.41,39.81)(42.62,35)
\qbezier(42.62,35)(49.84,30.19)(55,25)
\put(45,40){\makebox(0,0)[cc]{$L$}}

\end{picture}
}

\begin{document}

\baselineskip 24pt

\begin{center}
{\Large \bf  Off-shell Amplitudes in Superstring Theory}

\end{center}

\vskip .6cm
\medskip

\vspace*{4.0ex}

\baselineskip=18pt

\centerline{\large \rm Ashoke Sen}

\vspace*{4.0ex}

\centerline{\large \it Harish-Chandra Research Institute}
\centerline{\large \it  Chhatnag Road, Jhusi,
Allahabad 211019, India}

\vspace*{1.0ex}
\centerline{\small E-mail:  sen@mri.ernet.in}

\vspace*{5.0ex}

\centerline{\bf Abstract} \bigskip

Computing the renormalized masses and S-matrix elements in string theory,
involving states whose masses are not protected from quantum corrections, requires
defining  off-shell amplitude with certain factorization
properties. While in the bosonic string
theory one can in principle construct such an amplitude from string field theory,
there is no fully consistent field theory for  type II and heterotic
string theory.
In this paper we give a practical construction of off-shell amplitudes
satisfying the desired factorization property using the formalism of picture changing
operators. We describe a systematic procedure for dealing with the spurious singularities
of the integration measure that we encounter in superstring perturbation theory. 
This procedure is also useful for computing on-shell
amplitudes, as we demonstrate by computing the effect of Fayet-Iliopoulos D-terms
in four dimensional heterotic string theory compactifications using this
formalism.

\vfill \eject

\baselineskip=18pt

\baselineskip=18pt

\tableofcontents

\sectiono{Introduction} \label{sintro}

The usual formulation of critical string theories allows us to compute
S-matrix elements of external states which do not suffer any mass renormalization.
However generic states of string theory do undergo mass renormalization and 
for these states the usual string amplitudes do not compute S-matrix elements
beyond tree level. The main reason for this is that the conformal invariance of the
vertex operators requires us to set the momenta $k_i$ carried by the external states to
satisfy $k_i^2 = - m_i^2$ where $m_i$ is the {\it tree level} mass of the state.
On the other hand computing S-matrix elements via the LSZ prescription requires
us to impose the constraints $k_i^2 = - m_{i,p}^2$ where $m_{i,p}$ is the renormalized
mass of the state. Thus if $m_{i,p}\ne m_i$ there is an apparent conflict between the
two conditions.

If we had an underlying string field theory then one could use this
to define off-shell amplitudes and then use the standard LSZ prescription to compute
S-matrix elements. Even in the absence of a string field theory 
one can give an
ad hoc definition of off-shell amplitudes in string theory\cite{nelson} (see also
\cite{Vafa1,Vafa2,Cohen:1985sm,Cohen:1986pv,AG1,AG2,Polchinski:1988jq}). 
This has been fully developed in the context of bosonic string theory.
The main problem with these amplitudes however is that the result depends on
additional spurious data encoded in the choice of local coordinate system at the
punctures where the vertex operators are inserted. Since there is no canonical way
of choosing these local coordinates the result for the off-shell amplitude is ambiguous.

The suggestion made in \cite{1311.1257,1401.7014} 
was to go ahead and compute the renormalized masses
and S-matrix elements using these off-shell amplitudes despite the latter's dependence
on spurious data, and then prove that the renormalized masses and the S-matrix 
elements computed this way do not depend on the spurious data. 
Refs.\cite{1311.1257,1401.7014} were able to
establish the latter result provided we restrict the choice of local coordinates to
within a special class -- those satisfying the requirement of gluing compatibility.
This means that if we are near a boundary of the moduli space where the 
punctured Riemann
surface $\Sigma$ used for computing an amplitude can be represented by two 
separate punctured
Riemann surfaces $\Sigma_1$ and $\Sigma_2$
glued at one each of their punctures using the standard plumbing fixture
prescription,
then the choice of local coordinates at the external punctures of $\Sigma$ 
must agree with those induced from the choice of local coordinates at the
punctures of $\Sigma_1$ and $\Sigma_2$. As long as we restrict the choice of local
coordinates at the punctures within this class, the results for the physical 
quantities were shown to be independent of the choice of local coordinates.

For
bosonic string theory we also have an underlying 
string field theory\cite{wittensft,9206084}. Off-shell
amplitudes computed from this field theory in the Siegel gauge fall within the general
class of off-shell amplitudes described in \cite{nelson}, and automatically come with
a set of gluing compatible coordinate system\cite{0708.2591}. 
Thus the renormalized masses 
and S-matrix elements computed from string field theory would also agree with the
ones computed with a general system of gluing compatible local coordinate
system.

The discussion above should have an immediate generalization to supersymmetric
string theories. The computation of on-shell amplitudes in this theory has undergone
much clarification in recent 
years\cite{1209.5461,Belopolsky,9706033,dp,Witten,donagi-witten} where the notion
of local coordinate systems on Riemann surfaces is replaced by local superconformal
coordinates on super-Riemann surfaces and the final result for the amplitude is
expressed as an integral over supermoduli spaces of super-Riemann surfaces instead
of ordinary moduli spaces of Riemann surfaces.
Generalizing the results of \cite{1311.1257,1401.7014} we would expect that  
the definition of an off-shell amplitude will now depend on the
choice of local super-conformal coordinate system at the punctures, and  
that as long as the choice of the local superconformal coordinate system at the
punctures is gluing compatible, the result for physical quantities should be 
independent of the choice of these coordinate systems. However the complete
set of rules for off-shell amplitudes in superstring theory have not been laid out,
although \cite{Belopolsky,9706033} go a long way. 
Another route to defining off-shell amplitudes in superstring theory would be to develop
a superstring field theory. 
Despite considerable 
progress\cite{wittenssft,9202087,9503099,0109100,0406212,0409018,1201.1761,
1303.2323,1312.1677,1312.2948,1312.7197,1403.0940}
this has not yet been fully achieved.
This prevents us from carrying out 
explicit computation of renormalized masses and S-matrix elements in string theory
except at low orders ({\it e.g.} at one loop order 
two point amplitude with {\it on-shell} external states is enough
to compute the renormalized mass).

The goal of this paper is to give a definition of off-shell amplitude in superstring theory
which can be used for practical computation. However instead of using superconformal 
formalism where the spurious data resides in the choice of local superconformal 
coordinates at the punctures, we shall use the 
formalism involving picture changing  operators\cite{FMS,Verlinde:1987sd} where
the spurious data resides in the choice of local {\it bosonic} coordinate system at
the punctures and the locations of the picture changing operators. 
Thus in this formalism the off-shell amplitudes are expressed as integrals over the moduli
spaces of ordinary punctured Riemann surfaces, with the integrand being appropriate
correlations functions of off-shell vertex operators, ghost fields and picture changing
operators.

It has been known since \cite{Verlinde:1987sd} that the choice of locations of the picture
changing operators corresponds to a choice of gauge for the gravitino field. It has also
been known from  the work of
\cite{donagi-witten} that it is not possible to make a global choice of gauge for the gravitino
field -- we must work with different gauge choice in different parts of the moduli space. This
breakdown of global gauge choice for the gravitino shows up in the picture changing formalism as
spurious singularities of the integration measure appearing in a real codimension two
subspace of the moduli space. We give a procedure for dealing with these singularities
by introducing the notion of `vertical integration' -- integrating along a direction in which the
location of the picture changing operators vary keeping the moduli fixed. 
The off-shell amplitude defined this way is ambiguous, but this ambiguity is at the same
level as the one associated with the choice of locations of the picture changing operators and does
not affect the renormalized masses or S-matrix elements.

The rest of the paper is organised as follows. In \S\ref{s1} we review the construction
of off-shell amplitudes in bosonic string theory, following closely the work of
\cite{nelson,9206084}. In \S\ref{s2} we generalize this construction to off-shell 
NS sector
amplitudes in superstring
theory using picture changing operators. We allow  the locations
of the picture changing operators to
vary as we change the moduli of the Riemann surface.
In this case
we need to take into account the fact 
that the correlation function for computing the integration measure requires insertion of
additional operators related to the picture changing operators by a set of descent
equations\cite{Verlinde:1987sd,9202087}. 
In \S\ref{snonhol} we discuss the origin of the spurious poles in the superstring
integration measure and our method of dealing with them using the notion of 
vertical integration. In \S\ref{sspecial} we use this formalism to show that the renormalized masses
and S-matrix elements of special states are independent of the choice of the locations
of the picture changing operators even though the off-shell amplitudes do 
depend on them.
In \S\ref{sramond} we extend our prescription to
amplitudes involving Ramond sector external
states. This turns out to be more subtle than those involving NS sector external
states and we suggest a way to 
deal with these subtleties by giving up manifest symmetry of the off-shell
amplitude under the permutations of the external states. 
This will lead to a sensible approach if the S-matrix elements 
can be shown to have the permutation symmetry, but this has not been proved.
Finally in \S\ref{sfi}
we illustrate the utility of our method even for on-shell amplitudes 
by applying it to compute the effects of   Fayet-Iliopoulos terms
in SO(32) heterotic string theory 
compactified on Calabi-Yau 3-folds\cite{DSW} . 
This computation
has been done earlier in different 
formalism\cite{ADS,DIS,greenseiberg,AtickS,1304.2832,1403.5494,1404.5346}, 
and our analysis using picture changing
operator yields results in agreement with the earlier results.

We end this introduction with a word on convention. As 
emphasized in \cite{1311.1257,1401.7014},
an off-shell amplitude $\Gamma^{(n)}_{a_1\cdots a_n}(k_1,\cdots k_n)$  in
string theory -- where
$a_i$ denotes the quantum numbers and $k_i$ denotes the momentum of the
$i$-th external state -- do not compute the analog of the off-shell Green's function
$G^{(n)}_{a_1\cdots a_n}(k_1,\cdots k_n)$ in a quantum field theory. Instead the
two are related as
\be \label{es1.1}
\Gamma^{(n)}_{a_1\cdots a_n}(k_1,\cdots k_n) = G^{(n)}_{a_1\cdots a_n}(k_1,\cdots k_n)
\, \prod_{i=1}^n (k_i^2 + m_i^2)
\ee
where $m_i$ is the {\it tree level} mass of the $i$-th external state. Throughout 
this paper this is what we shall analyze. Of course once we have computed $\Gamma^{(n)}$,
it is easy to find $G^{(n)}$ using \refb{es1.1}.

\sectiono{Off-shell amplitudes in the bosonic string theory} \label{s1}

In this section we shall review the construction of off-shell amplitudes in bosonic string 
theory\cite{nelson,9206084} following closely the conventions of \cite{9206084}.
Let $\MM_{g,n}$ denote the moduli space of genus $g$ Riemann surface with $n$
punctures, $\PP_{g,n}$ denote the moduli space of genus $g$ Riemann surface with 
$n$ punctures with some choice of local coordinates around each puncture and
$\wh\PP_{g,n}$ denote the quotient of $\PP_{g,n}$ by independent phase rotation of the
local coordinate around each puncture.
Both $\PP_{g,n}$ and $\wh\PP_{g,n}$
are infinite dimensional spaces. We also denote by $\MM_g$ the moduli
space of genus $g$ Riemann surface without the punctures. Then we have the
natural projection
\be
\PP_{g,n} \to \wh\PP_{g,n} \to \MM_{g,n} \to \MM_g\, ,
\ee
which corresponds to forgetting about some part of the data at each step.
In fact we can regard $\PP_{g,n}$ to be a fiber bundle over the base $\wh \PP_{g,n}$
with the phases of the local coordinates acting as the fiber directions, $\wh\PP_{g,n}$
as the fiber bundle over the base $\MM_{g,n}$ with the choice of the local
coordinates at the punctures up to phases as the fiber directions, and 
$\MM_{g,n}$ as the fiber bundle over the base $\MM_g$ with the locations of
the punctures as the fiber directions.

\subsection{Schiffer variation} \label{s2.1}

Let $\Sigma$ denote an element of $\PP_{g,n}$, i.e.\ a Riemann surface of genus $g$
and $n$ punctures and some specific choice of local coordinates around each puncture.
For given $\Sigma$, let
$w_a$ be the choice of local coordinate around the $a$-th puncture  
with the puncture situated at $w_a=0$.
We shall assume that the coordinates
$\{w_a\}$ have been chosen (possibly by scaling them with small numbers) so that
$w_a$ is a valid coordinate system on $\Sigma$
for $|w_a|\le 1$. We
denote by $D_a$ the disk $|w_a|< 1$. 
It will also be convenient to choose some fixed coordinate system on 
$\Sigma-\cup_a D_a$. 
A concrete way to do this is as follows. We can cut
$\Sigma - \cup_a D_a$ along $3g-3+2n$  homotopically non-trivial circles
to divide $\Sigma - \cup_a D_a$ into $2g-2+n$ disjoint parts, each with the topology
of a sphere with three holes. 
We can then label the $i$-th part $\sigma_i$ by a complex coordinate
$z_i$ in which $\sigma_i$ takes the form of a complex plane with three
holes cut out of it. We shall denote by $z$ the collection of the coordinates
$\{z_i\}$. 
At the boundary circle separating two such components
$\sigma_i$ and $\sigma_j$, 
the coordinates $z_i$ and $z_j$ are related by some functional relation
\be \label{ecomprel}
z_i = f_{ij}(z_j)
\ee
where $f_{ij}(z_j)$ is an analytic function that maps the common circle between
$\sigma_i$ and $\sigma_j$ from the $z_j$ plane to the $z_i$ plane in a
one to one fashion but could have singularities
elsewhere.
Furthermore the coordinates $w_a$ labelling the local
coordinates on the punctures are also related to the coordinate $z$ of one
of the
components $\sigma_i$ -- 
that shares the boundary circle $|w_a|=1$ -- by a functional 
relation  of the form 
\be \label{ex1}
z = f_a (w_a)\, ,
\ee
where $f_a$ maps the circle $|w_a|=1$ to the corresponding circle in the $z$-plane
in a one to one fashion 
but could have
singularities both inside $D_a$ as well as on $\Sigma - D_a$.
Note that by an abuse of notation, in \refb{ex1} 
we have labelled the coordinate of
$\sigma_i$ as $z$. Since there is always a unique $\sigma_i$ that shares 
a boundary with $D_a$, this will not cause any confusion.  
The information about the moduli of the
Riemann surface as well as the local coordinate system around the punctures
is then contained in the transition functions $f_{ij}$ and $f_a$, although 
they are not all independent. For example an infinitesimal 
coordinate transformation 
of the form $z_i\to z_i + \eps v(z_i)$ where $v(z_i)$ is a non-singular
vector field on $\sigma_i$  will change the $f_{ij}(z_j)$'s and
possibly some $f_a(w_a)$ if the $i$-th component shares a boundary with
$D_a$, but these changes do not change the moduli of the Riemann
surface or the local coordinates around the punctures.

We'll need to study the tangent space of $\PP_{g,n}$ associated with deformations of
the punctured Riemann surface and/or the choice of local coordinates around the 
punctures. There are various ways of describing this tangent space {\it e.g.} by
infinitesimal deformations of the various functions $f_{ij}(z_j)$ and
$f_a(w_a)$, but one convenient
way of doing this is  via the Schiffer variation. 
The idea of Schiffer
variation is that locally we can generate the full set of deformations in $\PP_{g,n}$
by deforming the functions $f_a(w_a)$ keeping the $f_{ij}(z_j)$'s fixed.
We shall now describe how it can be used to define a tangent to $\PP_{g,n}$.
Let us consider a deformation in $\PP_{g,n}$ labelled by an infinitesimal
parameter $\eps$ and let $f_a^\eps$ be the deformed form of $f_a$.
We introduce the coordinate system $w_a^\eps$ via the relations
\be \label{ex2}
z = f_a^\eps(w_a^\eps)\, .
\ee
We can combine \refb{ex1} and \refb{ex2} to get a relation between $w_a^\eps$
and $w_a$ of the form
\be \label{e1}
w_a^\eps = (f_a^{\eps})^{-1} (f_a(w_a)) = w_a + \eps v^{(a)}(w_a)\, ,
\ee
for some vector field $v^{(a)}(w_a)$ that is non-singular around the $|w_a|=1$
curve but can have singularities away from it. Thus we can use the vector field
$v^{(a)}(w_a)$ to describe a tangent vector of $\PP_{g,n}$.
We can also use \refb{ex1}-\refb{e1}
to write
\be \label{ex3}
f_a^\eps(w_a) = f_a(w_a) -\eps\, 
v^{(a)}(z), \quad v^{(a)}(z) \equiv f_a'(w_a) v^{(a)}(w_a)
\, .
\ee
By an abuse of notation we have used the same symbol 
$v^{(a)}(z)$ with changed argument to represent the
vector field $v^{(a)}(w_a)$ written in the 
$z$ coordinate system.

In order to be more general let us consider such vector fields around each 
puncture and consider a deformation of the type given in \refb{e1} around
each puncture. Together they describe a deformation of $\PP_{g,n}$ and hence
a tangent vector of $\PP_{g,n}$ labelled by $\vec v = (v^{(1)}, \cdots v^{(n)})$.
It is easy to verify that if $\delta_{\vec v}$ denotes the tangent vector of $\PP_{g,n}$
generated by $\vec v$, then we have
\be \label{evcom}
[\delta_{\vec v_1}, \delta_{\vec v_2}] = \delta_{[\vec v_2, \vec v_1]}, \quad
[\vec v_2, \vec v_1]^{(a)} \equiv \left(
v_2^{(a)}(z) \p_z v_1^{(a)}(z)  -v_1^{(a)}(z) \p_z v_2^{(a)}(z)
\right) \, .
\ee
Now we have the following general results (see {\it e.g.} section 7 of \cite{9206084}
for proofs of these results):
\begin{enumerate}
\item 
Let $v(z)$ be a globally defined 
vector field on $\Sigma$ that is holomorphic everywhere except possibly
at the punctures.
If $v^{(a)}(z)$ is the restriction of $v(z)$ to $\p D_a$,
then the deformations generated by $\vec v(z)
\equiv (v^{(1)}(z),\cdots v^{(n)}(z))$ can be removed by coordinate
redefinition on $\Sigma-\cup_a D_a$, i.e. a redefinition of the coordinates
$z_k$ on $\sigma_k$. Thus $\vec v(z)$ describes a vanishing tangent vector on 
$\PP_{g,n}$.\footnote{Such a vector field will generate a deformation in a bigger
space that contains information about not only the local coordinates around the
punctures but also the coordinate system $z_k$ on the $\sigma_k$'s.} 
This  also works in the reverse direction, i.e.\ if the $n$-tuple of
vector fields $(v^{(1)},\cdots v^{(n)})$ 
 fail
to extend holomorphically into $\Sigma$ as a globally defined vector field away 
from the punctures then $\vec v(z)$ does describe a non-trivial deformation
on $\PP_{g,n}$.
\item If $\vec v(z)$ does not extend holomorphically into $\Sigma - \cup_a D_a$,
but extends holomorphically into the $D_a$'s  and vanish at the punctures,
then it describes the same point in $\MM_{g,n}$ but deforms the choice of local
coordinate system around the punctures.
\item If $\vec v(z)$ does not extend holomorphically into $\Sigma - \cup_a D_a$,
but extends holomorphically into the $D_a$'s  and  does not
vanish at the punctures,
then it describes the same point in $\MM_{g}$ but moves one or more of the 
punctures.
\item If $\vec v(z)$ does not extend holomorphically into $\Sigma - \cup_a D_a$,
and has poles at one or more punctures, then it describes a deformation on
$\MM_g$, i.e.\ changes the moduli of the underlying Riemann surface. Furthermore
the complete set of complex deformations of $\MM_g$ can be obtained by choosing a
set of $3g-3$ such vector fields with poles of order $1,\cdots 3g-3$ at any
of the punctures.
\end{enumerate}
In the following we shall continue to denote by $z$ some fixed coordinate
system on $\Sigma -\cup_a D_a$ represented by the collection of the
$z_i$'s, and by $w_a$ the local coordinates around the
punctures.

\subsection{Surface states}

A surface state 
$\langle\Sigma|$ associated with a Riemann surface  $\Sigma$
with $n$ punctures 
is a state in the dual space of the
$n$-fold tensor
product of the Hilbert space $\HH$ of the underlying CFT. It describes the state that is created
on the boundaries of $D_a$ by performing the functional integral over the fields of the
CFT on $\Sigma - \cup_a D_a$. 
More precisely, if we consider a state $|\Psi_1\rangle\otimes\cdots\otimes|\Psi_n\rangle$
in $\HH^{\otimes n}$, then
\be
\langle\Sigma|(|\Psi_1\rangle\otimes\cdots\otimes|\Psi_n)\rangle
\ee
describes the $n$-point correlation function on $\Sigma$ with the vertex operator for
$|\Psi_a\rangle$ inserted at the $a$-th puncture using the local coordinate
system $w_a$ around that puncture. Thus $\langle\Sigma|$ depends not only
on the moduli labelling $\MM_{g,n}$ but also on the choice of local coordinates around
the punctures. 
$\langle\Sigma|$ satisfies the identity
\be \label{eqbsigma}
\langle \Sigma| \sum_{r=1}^n Q_B^{(a)}=0\, ,
\ee
where $Q_B^{(a)}$ is the BRST operator acting on the Hilbert space of states at
the $a$-th puncture. 
Furthermore they also satisfy the identity (see {\it e.g.} \cite{9206084}):
\ben \label{edeltav}
\delta_{\vec v} \langle \Sigma | &=& -\langle \Sigma|  T(\vec v), \nonumber \\
\quad T(\vec v) &\equiv& \left(\sum_a  \ointop dw_a v^{(a)}(w_a) T^{(a)}(w_a) 
+ \sum_a \ointop d\bar w_a \bar v^{(a)}(\bar w_a) \bar T^{(a)}(\bar w_a)\right)
\nonumber \\
&=& \left(\sum_a  \ointop dz v^{(a)}(z) T^{(a)}(z) 
+ \sum_a \ointop d\bar z \bar v^{(a)}(\bar z) \bar T^{(a)}(\bar z)\right)\, ,
\nonumber \\
\een
where $\delta_{\vec v}$ is the tangent vector of $\PP_{g,n}$ associated with the
Schiffer variation  induced by the vector fields $\vec v$, $T^{(a)}$,
$\bar T^{(a)}$ are the
stress tensor components acting on the Hilbert space of the $a$-th puncture,
and
the integration contour over $w_a$ ($\bar w_a$) 
runs in the anti-clockwise (clockwise) direction around each puncture
and includes the usual $1/2\pi i$ normalization factors so that $\ointop dw/w=
\ointop d\bar w/\bar w=1$.  
In going from the second to the third line we have used the fact that
$v^{(a)}(z) T^{(a)}(z)$ transforms as a one form under  a coordinate transformation.
Using \refb{edeltav}
and the Virasoro commutation relations it is easy to verify that
\be
[T(\vec v_1), T(\vec v_2)] = T([\vec v_2,\vec v_1])\, .
\ee
Note that $\vec v$ and $i\vec v$ describe independent deformations. Equivalently
we can take $\vec v$ and $\vec {\bar v}$ to be independent.

\subsection{Integration measure on $\PP_{g,n}$}

We now describe the construction of a $p$-form on $\PP_{g,n}$ that can be
integrated over a $p$-dimensional subspace -- henceforth refered to as an
integration cycle. 
Let $|\Phi\rangle$ denote some element of $\HH^{\otimes n}$. Now
by definition a $p$-form should generate a number when contracted with $p$
tangent vectors of $\PP_{g,n}$ and this number should be anti-symmetric under
the exchange of any pair of tangent vectors. Since tangent vectors are
labelled by the $n$-tuple of vector fields $\vec v$, what we are looking for is a
multilinear function of $p$ such $n$-tuple of vector fields. Let $V_1,\cdots V_p$
be $p$ tangent vectors of $\PP_{g,n}$ and let $\vec v_1,\cdots \vec v_p$ be the
corresponding $n$-tuple vector fields. 
First we introduce an operator values $p$-form $B_p$, whose contraction with the
tangent vectors $V_1,\cdots V_p$ is given by
\be \label{edefB}
B_p[V_1,\cdots V_p]  = b(\vec v_1) \cdots b(\vec v_p) \, 
\ee
where $b(\vec v)$ is defined in the same way as $T(\vec v)$:
\ben \label{edefbv}
b(\vec v) &\equiv& \left(\sum_a  \ointop dw_a v^{(a)}(w_a) b^{(a)}(w_a) 
+ \sum_a  \ointop d\bar w_a \bar v^{(a)}(\bar w_a) \bar b^{(a)}(\bar w_a)\right)
\nonumber \\
&\equiv& \left(\sum_a  \ointop dz v^{(a)}(z) b^{(a)}(z) 
+ \sum_a  \ointop d\bar z \bar v^{(a)}(\bar z) \bar b^{(a)}(\bar z)\right)\, ,
\een
$b$, $\bar b$ being the anti-ghost fields. 
$B_p$ is clearly anti-symmetric under the exchange of
a pair of $\vec v_i$'s. 
Then we define the $p$-form $\Omega^{(g,n)}_p$ as\cite{9206084}:
\be \label{epform}
\Omega^{(g,n)}_p(|\Phi\rangle)
= (2\pi i)^{-(3g-3+n)}\, \langle \Sigma| B_p |\Phi\rangle\, .
\ee
Ghost number conservation on the genus $g$ Riemann surface
tells us that if $|\Phi\rangle$ carries total ghost number $n_\Phi$ then we
must have
\be \label{eghno}
n_\Phi -p = 6 - 6g\, ,
\ee
in order for $\Omega^{(g,n)}_p(|\Phi\rangle)$ to be non-zero.

Now it follows from our previous discussion that 
if there is a globally defined holomorphic vector field on the whole of
$\Sigma- \cup_a D_a$, then adding to each $\vec v_i$ an arbitrary multiple $c_i$
of $v$ describes the same set of tangent vectors in $\PP_{g,n}$. One can show that this
addition does not change the value of $\Omega^{(g,n)}_p(|\Phi\rangle)$ given in \refb{epform}.
The proof of this  uses the fact that such a 
deformations adds to $b(\vec v_i)$ a term $c_i \ointop v(z) b(z) dz$ where the integration
contour winds around all the punctures. We can now deform the contour in the
interior of $\Sigma$ and contract it to a point showing that the corresponding
contribution vanishes.

This shows that $\Omega^{(g,n)}_p$ indeed describes a $p$-form in $\PP_{g,n}$
and not in a bigger space that also keeps track of possible addition of globally defined
vector fields to the $v^{(a)}_i$'s.

\subsection{Restriction to $\wh\PP_{g,n}$}

So far we have worked with states in the general Hilbert space $\HH$ of matter-ghost
CFT. From now on we shall work with a restricted Hilbert space $\HH_0$ defined
via the condition
\be \label{econd}
|\Psi\rangle \in \HH_0\quad {\rm if} \quad
(b_0-\bar b_0)|\Psi\rangle = 0,  \quad (L_0-\bar L_0)|\Psi\rangle=0
\, ,
\ee
and take $|\Phi\rangle$ to be an element of $\HH_0^{\otimes n}$. Fot later
use
we shall also introduce the subspace $\HH_1$ containing
off-shell states of ghost number two in the Siegel gauge
\be \label{ephysicaloffshell}
|\Psi\rangle \in \HH_1\quad {\rm if} \quad |\Psi\rangle \in \HH_0, 
\quad (b_0+\bar b_0)|\Psi\rangle = 0\, , \quad
\hbox{ghost number} \, (|\Psi\rangle)=2\, .
\ee
The physical states which will appear as external states in S-matrix computation will
be of this type. 

One can show that for states satisfying \refb{econd} the following properties hold:
\begin{enumerate}
\item The $p$-form given in \refb{epform}, contracted with a tangent vector of
$\PP_{g,n}$ whose projection onto $\wh \PP_{g,n}$ vanishes, vanishes.
This follows from the fact that such a tangent vector represents a deformation in
which local coordinates at the punctures change by phases. Using \refb{edefbv} we
see that contracting such a tangent vector with the $p$-form will insert into the
correlation function
a $b(\vec v)$ that is a linear combination of $
\ointop w_a dw_a b^{(a)}(w_a) - \ointop \bar w_a d\bar w_a \bar b^{(a)}(\bar 
w_a) =
b_0^{(a)}-\bar b_0^{(a)}$. This
vanishes by \refb{econd}.
\item The $p$-form given in \refb{epform} remains unchanged if we move in 
$\PP_{g,n}$ along a direction that leaves its projection into $\wh P_{g,n}$
unchanged. Since such a deformation corresponds to changing the local coordinates
at the punctures by phases, we see  from \refb{edeltav}
that they correspond to insertions of a linear combination of 
$\ointop w_a dw_a T^{(a)}(w_a) - \ointop \bar w_a d\bar w_a 
\bar T^{(a)}(\bar  w_a) =L_0^{(a)} -
\bar L_0^{(a)}$. This vanishes by \refb{econd}.
\end{enumerate}
This essentially tells us that the $p$-form defined in \refb{epform} can be regarded as
a $p$-form on $\wh\PP_{g,n}$. 

\subsection{BRST identity} 

We shall now describe an important identity that is used for proving many properties of
the off-shell amplitude. Let us denote by $|\Phi\rangle$ 
a state in $\HH_0^{\otimes n}$. 
Then we
have the identity
\be \label{epformqb}
\Omega^{(g,n)}_p\left(Q_B |\Phi\rangle\right) 
= (-1)^p \, d\Omega^{(g,n)}_{p-1} \left(|\Phi\rangle\right)\, ,
\ee
where $Q_B=\sum_{a=1}^n Q_B^{(a)}$, $Q_B^{(a)}$ being the BRST operator
acting on the $a$-th copy of $\HH$.
Since this is an important identity that needs to be generalized for superstring
theories, we shall review its proof\cite{9206084}. 
For this let $\wh V_1,\cdots \wh V_p$ be a set of $p$ tangent vectors of $\wh\PP_{g,n}$,
and $\Omega^{(g,n)}_p(\wh V_1, \cdots \wh V_p)$ be the contraction of these $p$
tangent vectors with $\Omega^{(g,n)}_p$. Furthermore let $\vec v_1,\cdots \vec v_p$ be
the $n$-tuple vector fields associated with the tangent vectors $\wh V_1,\cdots \wh V_p$.
Then by definition:
\ben \label{e114}
d\Omega^{(g,n)}_{p-1}(\wh V_1, \cdots \wh V_p)
&=& \sum_{i=1}^p (-1)^{i+1} \wh V_i \, \Omega^{(g,n)}_{p-1} (\wh V_1, \cdots, \not \wh V_i
\cdots, \wh V_p) \nonumber \\
&+& \sum_{1\le i<j\le p} (-1)^{i+j} \Omega^{(g,n)}_{p-1} ([\wh V_i, \wh V_j], \wh V_1, \cdots, 
\not \wh V_i,
\cdots \not \wh V_j,
\cdots ,\wh V_p) \, ,
\een
where $\not ~$ indicates that the corresponding entry is deleted from the list and
the $\wh V_i$ in the first term on the right hand side has to be regarded as a
differential operator  involving derivative with respect to the coordinates of
$\wh\PP_{g,n}$.
Now
using  \refb{evcom}, \refb{edefB}, \refb{epform} and \refb{e114} 
we can translate \refb{epformqb} to
\ben \label{epform2}
\langle\Sigma| b(\vec v_1) \cdots b(\vec v_p) Q_B |\Phi\rangle 
&=& 
\sum_{i=1}^p (-1)^{p+i+1} \delta_{\vec v_i} 
\langle\Sigma| 
b(\vec v_1) \cdots \not b(\vec v_i) \cdots b(\vec v_p)|\Phi\rangle
\nonumber \\
&& +\sum_{1\le i<j\le p} (-1)^{p+i+j}
\langle\Sigma| b([\vec v_i, \vec v_j] )
b(\vec v_1) \cdots \not b(\vec v_i) \cdots \not b(\vec v_j) 
\cdots b(\vec v_p)|\Phi\rangle  \nonumber \\
&=& 
\sum_{i=1}^p (-1)^{p+i} 
\langle\Sigma| T(\vec v_i)
b(\vec v_1) \cdots \not b(\vec v_i) \cdots b(\vec v_p)|\Phi\rangle
\nonumber \\
&& +\sum_{1\le i<j\le p} (-1)^{p+i+j} 
\langle\Sigma| b([\vec v_i, \vec v_j] )  b(\vec v_1) \cdots \not b(\vec v_i) \cdots \not b(\vec v_j) 
\cdots b(\vec v_p)|\Phi\rangle  \, .
\nonumber \\
\een
To prove this equation we begin with the left hand side and
move $Q_B$ to the extreme left  picking up commutators on the way. When $Q_B$ acts
on $\langle\Sigma|$ the result vanishes by eq.\refb{eqbsigma}. So we only need
to worry about the (anti-)commutators. Using the fact that
\be
\{Q_B, b_{\vec v}\} = T_{\vec v}\, , 
\ee
we can express this into a sum of $p$ terms where in the $i$-th term the $b(\vec v_i)$
is replaced by $T(\vec v_i)$ and we pick an extra factor of $(-1)^{p-i}$. Next we move the
$T(\vec v_i)$ in the $i$-th term to the extreme left thereby generating the first set
of terms on the right hand side of \refb{epform2}. However 
in that process we pick up commutators
with $b(\vec v_j)$'s using the relation
\be
[T(\vec v_i), b(\vec v_j)] = - b([\vec v_i, \vec v_j])\, ,
\ee
and then move the $b([\vec v_i, \vec v_j])$ factor to the extreme left. This generates
an extra factor of $(-1)^{j-1}$. Finally exchanging the labels $i$ and $j$
we recover the second set of terms on the right hand side
of \refb{epform2}.


\subsection{General parametrization of tangent vectors} \label{s1gen}

Even though the Schiffer variations are able to describe  arbitrary tangent vectors in
$\wh \PP_{g,n}$, and are the most convenient ones for deforming the local 
coordinate system around the punctures and the locations of the punctures on a
fixed Riemann surface, they are not always the most convenient way of
describing the variation of
the moduli of the Riemann surface itself.
A more general description of a tangent vector can be given 
by deforming the functions $f_{ij}(z_j)$ introduced in \S\ref{s2.1}.
In this case by following the same
procedure as in \refb{ex3} we can 
introduce the relations
\be
z_i = f_{ij}^\eps (z_j^\eps), \quad z_j^\eps = (f_{ij}^\eps)^{-1} (f_{ij}(z_j))
\equiv z_j + \eps v(z_j)\, ,
\ee
where
$v(z_j)$ is a vector field on the Riemann surface that is analytic inside an
annulus containing the common boundary circle between $\sigma_i$ and
$\sigma_j$. Then the contraction of $\Omega^{(g,n)}_p$ with the corresponding
tangent vector is given by inserting into the correlation function a factor of
\be 
b(v) = \left(\ointop dz_j v(z_j) b(z_j) 
+ \ointop d\bar z_j \bar v(\bar z_j) \bar b(\bar z_j)\right)\, 
\ee
with the integration contour over $z_j$ ($\bar z_j$)
running along the circle forming the common boundary of
$\sigma_i$ and $\sigma_j$ keeping the $\sigma_j$ component to its left (right). This is
the generalization of the statement that the contour of integration over
$w_a$ ($\bar w_a$) in \refb{edefbv} was 
anti-cockwise (clockwise)
in the $w_a$ plane.

The simplest example of this is the plumbing fixture relation:
\be  \label{epp1}
zw = q\, ,
\ee
where $q$ is a complex parameter labelling the `plumbing fixture variable'. 
The $z$ coordinate system is used in the region $|z|\ge |q|^{1/2}$ and the $w$
coordinate system is used in the region $|w|\ge |q|^{1/2}$.
If we
regards $q$ and $\bar q$ as independent variables and consider the tangent vector
$\p/\p q$ then the corresponding vector field $v(z)$ computed from \refb{ex3} is
given by $-q^{-1}z$. Thus the contraction of $\Omega^{(g,n)}_p$ with such a vector will insert
\be 
-q^{-1}\ointop dz \, z\, b(z)  \, 
\ee
into the correlation function.
The contour runs anti-clockwise around $z=0$
along $|z|=|q|^{1/2}$.
By appropriate deformation of the integration contour this definition of $\Omega^{(g,n)}_p$
can be shown to be equivalent to the one given in terms of Schiffer variation.

\subsection{Off-shell amplitude and gluing compatibility} \label{soff}

So far we have described the construction of natural $p$-forms on $\wh\PP_{g,n}$ for a
given set of external states in $\HH^{\otimes n}$.
If we restrict the external states at the punctures
by requiring each of them to have
ghost number 2, so that the corresponding state $|\Psi_1\rangle \otimes \cdots
\otimes |\Psi_n\rangle$ carries total ghost number $2n$, then
\refb{eghno} tells us that \refb{epform}
vanishes unless
\be
p = 6g - 6 + 2n\, .
\ee
This is exactly the correct dimension of the moduli space of genus $g$ Riemann
surfaces. However for off-shell external states 
the $6g-6+2n$-form defined in \refb{epform} does not descend down
to a $6g-6+2n$-form on $\MM_{g,n}$ since it depends on the choice of local coordinate
system at the punctures and has non-vanishing contraction with tangent vectors which
correspond to deformation of the local coordinates without any deformation of
$\MM_{g,n}$. Thus the best we can do is to regard 
$\wh\PP_{g,n}$
as a fiber bundle over $\MM_{g,n}$ and integrate this $(6g-6+2n)$-form
over a section of the fiber bundle. This defines the off-shell string amplitude for
external states $|\Psi_1\rangle,\cdots, |\Psi_n\rangle$. The
result depends on the choice of the section, reflecting the fact that the off-shell
amplitudes depend on the choice of local coordinate system around the puncture.
However the physical quantities like the renormalized masses and S-matrix elements
are independent of the choice of the section\cite{1311.1257,1401.7014}.

As discussed in \cite{1311.1257,1401.7014}, 
for consistent off-shell
amplitudes we need to impose on the choice of this section the requirement
of gluing compatibility. This says that near a boundary of the moduli space when a
genus $g$ surface with $n$-punctures degenerates into a genus $g_1$ surface with
$n_1$ punctures and a genus $g_2$ surface with $n_2$ punctures with
$g=g_1+g_2$ and $n=n_1+n_2-2$, the choice of local coordinates on the
original surface must be taken to be those induced from the local coordinates
at the punctures on the two surfaces into
which it degenerates. More precisely suppose that 
$u_1,\cdots u_{n_1}$ denote the local coordinates around the $n_1$ punctures
of the genus $g_1$ Riemann surface and 
$v_1,\cdots v_{n_2}$ denote the local coordinates around the $n_2$ punctures
of the genus $g_2$ Riemann surface. Suppose further that we construct the
genus $g_1+g_2$ Riemann surface by gluing the $a$-th puncture of the first surface
and the $b$-th puncture of the second surface using 
the plumbing fixture relation:
\be \label{egluing}
u_a v_b = e^{-s+i\theta}, \quad 0\le s<\infty, \quad 0\le \theta<2\pi\, .
\ee
This will automatically give a choice of local coordinates on the genus $g_1+g_2$
Riemann surface with $n_1+n_2-2$ punctures. The requirement is that {\it for
all Riemann
surfaces of genus $g=g_1+g_2$ with $n=n_1+n_2-2$ punctures which can be
constructed this way, the local coordinates at the punctures must be taken to be
the ones that is induced from the choices $(u_1,\cdots \not\nobreak 
\hskip -5pt u_a, \cdots u_{n_1},
v_1, \cdots \not\nobreak \hskip -5pt v_b, \cdots v_{n_2})$.}
The off-shell Siegel gauge amplitudes constructed
from closed string field theory automatically induces such gluing compatible
local coordinate system\cite{0708.2591}.

A more explicit description of this  condition is as follows.
Let us
describe the first Riemann surface as a collection of different
components $\{\sigma^{(1)}_k\}$ and $D^{(1)}_1,\cdots D^{(1)}_{n_1}$ as in 
\S\ref{s2.1} and the
second Riemann surface as a collection of different
components $\{\sigma^{(2)}_k\}$ and $D^{(2)}_1,\cdots D^{(2)}_{n_2}$.
The choice of the sections in $\wh\PP_{g_1,n_1}$ and $\wh\PP_{g_2,n_2}$
correspond to specific relations between the coordinate systems on these
different components on their common boundary circles.
Now one of the components $\sigma^{(1)}_k$, which has common boundary 
with $D^{(1)}_a$, is glued to one of the components $\sigma^{(2)}_k$,
having common boundary with $D^{(2)}_b$, according to \refb{egluing} to
form the Riemann surface of genus $g$ and $n$ punctures. 
For such Riemann surfaces we can 
label the coordinates of $\MM_{g,n}$ by 
the cooordinates
of $\MM_{g_1,n_1}$, coordinates of $\MM_{g_2,n_2}$, $s$ and $\theta$.
On the other hand
the coordinate of $\PP_{g,n}$ can be described by 
specifying the relationship
between the coordinate systems  on $\{\sigma^{(1)}_k\}$, $\{\sigma^{(2)}_k\}$,
$D^{(1)}_1,\cdots \not \hskip -5pt D^{(1)}_a,\cdots D^{(1)}_{n_1}$  and
$D^{(2)}_1,\cdots \not \hskip -5pt D^{(2)}_b,\cdots D^{(2)}_{n_2}$ on their
overlap circles.
Then the gluing compatibility
condition requires that the section in $\wh\PP_{g,n}$ should be chosen such that
the relationships between the coordinates of $\{\sigma^{(1)}_k\}$ and
$D^{(1)}_1,\cdots \not \hskip -5pt D^{(1)}_a,\cdots D^{(1)}_{n_1}$
depend only on a subset of the base coordinates -- those
labelling $\MM_{g_1,n_1}$ --  but not on the coordinates of
$\MM_{g_2,n_2}$ or $s$, $\theta$.
Furthermore the dependence of these relations 
on the coordinates of $\MM_{g_1,n_1}$
must be the one induced from the choice
of the section in $\wh\PP_{g_1,n_1}$. Similarly the relationships between the
coordinate systems on $\sigma^{(2)}_k$ and
$D^{(2)}_1,\cdots \not \hskip -5pt D^{(2)}_b,\cdots D^{(2)}_{n_2}$
depend only on the coordinates of $\MM_{g_2,n_2}$ according to the choice
of the section in $\wh\PP_{g_2,n_2}$.

Let us denote by $S_1$ and $S_2$ a pair of sections on $\wh\PP_{g_1,n_1}$ and $\wh\PP_{g_2,n_2}$
and let $S$ be the $(6g-6+2n)$ dimensional 
subspace of $\wh\PP_{g,n}=\PP_{g_1+g_2,n_1+n_2-2}$ containing the family of
Riemann surfaces obtained by plumbing fixture of the Riemann surfaces associated with
the sections $S_1$ and $S_2$. 
Then the tangent space of $S$ can be labelled by $\p/\p s$, $\p/\p\theta$ and
the tangent vectors of $S_1$ and $S_2$. It follows that the $b$-ghost insertions needed for
computing the contraction of $\Omega^{(g,n)}_{6g-g+2n}$ with these tangent vectors
automatically factorize into
\be \label{ebfactor}
-i\, B^{(1)}_{6g_1 - 6 + 2 n_1}\,  b_0^+ b_0^- \, B^{(2)}_{6 g_2 - 6 + 2 n_2}\, ,
\ee
where the superscripts $(1)$ and $(2)$ refer to the two Riemann surfaces, and
\be
b_0^\pm \equiv (b_0 \pm \bar b_0), 
\quad b_0 \equiv 
\ointop {d u_a \, u_a} b(u_a), \quad \bar b_0 \equiv \ointop {d \bar u_a\, \bar u_a}\,
\bar b(\bar u_a) \, .
\ee
In \refb{ebfactor} $B^{(1)}$ and $B^{(2)}$ represent the effect of contraction of $\Omega^{(g,n)}_{6g-6+2n}$ with
the tangent vectors of $S_1$ and $S_2$, $b_0^+$ represents the effect of
contraction of $\Omega^{(g,n)}_{6g-6+2n}$ with the tangent vector $\p/\p s$ and 
$-i\, b_0^-$ represents the effect of
contraction of $\Omega^{(g,n)}_{6g-6+2n}$ with the tangent vector $\p/\p \theta$.
Furthermore,
it follows from \refb{ebfactor}, and the factorization property of correlation functions in
conformal field theories
on Riemann surfaces, that the full integration measure $\Omega^{(g,n)}_{6g-6+2n}(|\Phi\rangle)$,
restricted to $S$, 
also factorizes. 
We shall begin with a more general factorization formula and then restrict to the case of
interest.
If $|\Phi\rangle\in \HH_0^{\otimes n}$ 
has the form $|\Phi_1\rangle\otimes |\Phi_2\rangle$ where $|\Phi_1\rangle
\in \HH_0^{n_1-1}$ 
denotes the states at the external punctures of the first Riemann surface and
$|\Phi_2\rangle
\in \HH_0^{n_2-1}$ 
denotes the states at the external punctures of the second Riemann surface, 
and if $N_1$ and $N_2$ denote total ghost numbers carried by $|\Phi_1\rangle$ and
$|\Phi_2\rangle$, 
then
we have, 
\ben \label{emeasurefactorA}
\Omega^{(g,n)}_{p}(|\Phi\rangle) |_S
&=&  {1\over 2\pi} \sum_{0\le p_1\le 6g_1-6+2n_1, \, 0\le p_2\le
6g_2-6+2 n_2\atop p_1+p_2=p-2}
\sum_{i,j} \langle \vp_i^c | b_0^+ b_0^-
e^{-s(L_0+\bar L_0)} e^{i\theta(L_0-\bar L_0)} |\vp_j^c\rangle
 \nonumber \\
&& (-1)^{p_1p_2+N_1+p_1 + 1} \, \, ds \wedge d\theta \wedge \Omega^{(g_1,n_1)}_{p_1}(|\Phi_1\rangle
\otimes |\vp_i\rangle)|_{S_1}   \wedge  \, \Omega^{(g_2,n_2)}_{p_2}(|\vp_j\rangle
\otimes |\Phi_2\rangle)|_{S_2} 
\nonumber \\ 
\een
where the 
subscript $S$ on the left hand side denotes that this relation is valid for $\Omega^{(g,n)}_{p}$
restricted to the section $S$ and the subscripts $S_1$ and $S_2$ on the right 
denotes similar restrictions on $\Omega^{(g_1,n_1)}_{p_1}$ and 
$\Omega^{(g_2,n_2)}_{p_2}$.
The sum over $i,j$ runs over all states in $\HH_0$ 
and $\langle\vp_i^c|$ is the conjugate state
of $|\vp_i\rangle$ satisfying
\be \label{einner}
\langle \vp_i^c |\vp_j\rangle = \delta_{ij}, \quad \langle \vp_j |\vp_i^c\rangle = 
(-1)^{n_{\vp_i}} \delta_{ij}, \quad \sum_i |\vp_i\rangle \langle  \vp_i^c | = 
(-1)^{n_{\vp_i}}  \sum_i |\vp_j^c\rangle \langle  \vp_j | = {\bf 1}
\, ,
\ee
where $\langle \vp_i|$ is the BPZ conjugate of $|\vp_i\rangle$
and $n_{\vp_i}$ is its ghost number. It follows that
$|\vp_i^c\rangle$ does not belong to $\HH_0$, but has the form 
$(c_0-\bar c_0) |\chi_i\rangle$
for some state $|\chi_i\rangle \in \HH_0$.
In \refb{emeasurefactorA} 
one factor of $-1$ comes from the combination of $-i$ in \refb{ebfactor} and
the normalization factor $(2\pi i)^{-3g+3-n}$ in \refb{epform}. 
Other sign factors come from rearranging the $b$-ghost insertions and external
operator insertions in proper order so as to admit the interpretation given on the right
hand side of \refb{emeasurefactorA}. In particular
a factor of $(-1)^{p_2 N_1}$
arise from the need to move the $p_2$ number of $b$-ghost insertions 
through $|\Phi_1\rangle$ so that they sit next to the external vertex operators inserted
on $\Sigma_2$. Another factor of 
$(-1)^{N_2+p_2}$ comes from noting that on the right hand side we have once used
the last identity of \refb{einner} and that the $|\vp_j\rangle$ involved in this sum
has $n_{\vp_j}=6-6g_2-N_2+p_2$.  Finally a factor of
$(-1)^{(p_2+N_2)p_2}$ comes from moving 
$\vp_j$ through the $p_2$ ghost interstions to sit next to the states $|\Phi_2\rangle$.
Together they give the net factor of 
\be\label{enetfactor}
(-1)^{p_2 N_1 +(N_2+p_2) (p_2+1) +1}\, .
\ee
Using ghost charge conservation we see that in order to get a non-vanishing contribution to 
the right hand side of \refb{emeasurefactorA} we need $(-1)^{N_2+p_2}=(-1)^{N_1+p_1}$.
This reduces \refb{enetfactor} to $(-1)^{p_1p_2+N_1+p_1+1}$ as given in
\refb{emeasurefactorA}. 
Eventually integration over $\theta$ imposes a 
projection $\delta_{L_0,\bar L_0}$ on $|\vp_j^c\rangle$, and cancels the 
multiplicative factor of $1/2\pi$.

Another useful formula expresses $\Omega^{(g,n)}_p|_S$ restricted to the boundary
$s=\Lambda$ for some large number $\Lambda$. 
We have to follow the same logic as before except that there are no $ds$ factor in the
wedge product and no $b_0^+$ insertion in the correlator. 
The result is 
\ben \label{emeasurefactorB}
\Omega^{(g,n)}_{p}(|\Phi\rangle) |_{S; s=\Lambda}
&=& {1\over 2\pi} \sum_{0\le p_1\le 6g_1-6+2n_1, \, 0\le p_2\le
6g_2-6+2 n_2\atop p_1+p_2=p-1}
\sum_{i,j} \langle \vp_i^c | b_0^-
e^{-\Lambda (L_0+\bar L_0)} e^{i\theta(L_0-\bar L_0)} |\vp_j^c\rangle
 \nonumber \\
&& (-1)^{p_1p_2+p_2}  \, d\theta \wedge \Omega^{(g_1,n_1)}_{p_1}(|\Phi_1\rangle
\otimes |\vp_i\rangle)|_{S_1} \, \wedge  \, \Omega^{(g_2,n_2)}_{p_2}(|\vp_j\rangle
\otimes |\Phi_2\rangle)|_{S_2} \, .
\nonumber \\ 
\een
The overall sign is calculated as follows.
We begin with the configuration
where $d\theta$ sits to the extreme left in the wedge product and $b_0^-$ sits to the 
extreme left in the correlation function. 
Following the same manipulation as before
we get 
the sign factor given in \refb{enetfactor}, but now there is an 
extra factor of $(-1)^{N_1+p_1}$ in order to move the $b_0^-$
from the extreme left through the $b$-insertions associated with $\Omega^{(g_1,n_1)}_{p_1}$
and $|\Phi_1\rangle$. (This factor was absent in the previous case since what was moved
is $b_0^+ b_0^-$.) Now using the fact that ghost charge conservations demands
that $(-1)^{N_1+p_1}=(-1)^{N_2+p_2+1}$ we get the sign factor given in
\refb{emeasurefactorB}.

When each of the states at the external punctures describe Siegel gauge off-shell
states carrying ghost number two, i.e.\ belong to the subspace $\HH_1$,
then $N_1$ and $N_2$ are even. If we further restrict to
the case $p_i=6 g_i - 6 + 2 n_i$, we get from \refb{emeasurefactorA}
\ben \label{emeasurefactor}
\Omega^{(g,n)}_{6g-6+2n}(|\Phi\rangle) |_S
&=& -{1\over 2\pi} \sum_{i,j} ds \wedge d\theta \wedge  \Omega^{(g_1,n_1)}_{6g_1-6+2n_1}(|\Phi_1\rangle
\otimes |\vp_i\rangle)|_{S_1} \, \wedge  \, \Omega^{(g_2,n_2)}_{6g_2-6+2n_2}(|\vp_j\rangle
\otimes |\Phi_2\rangle)|_{S_2} 
 \nonumber \\ &&
\times \langle \vp_i^c | b_0^+ b_0^-
e^{-s(L_0+\bar L_0)} e^{i\theta(L_0-\bar L_0)} |\vp_j^c\rangle
\een
Since in this case $|\Phi_1\rangle$ has total ghost number
$2(n_1-1)$ and $|\Phi_2\rangle$ has total ghost number
$2(n_2-1)$, it follows from the ghost number conservation law given in 
\refb{eghno} that $|\vp_i\rangle$ and $|\vp_j\rangle$ must also carry ghost
number two. Furthermore the appearance of $b_0^+ b_0^-=2\,\bar  b_0\,
b_0$ in the second
line of \refb{emeasurefactor} implies that the basis states $|\vp_i^c\rangle$,
$|\vp_j^c\rangle$ can be taken to be annihilated by $c_0,\bar c_0$ and hence
the conjugate basis states $|\vp_i\rangle$,
$|\vp_j\rangle$ should be annihilated by $b_0$, $\bar b_0$. Thus $|\vp_i\rangle$
actually belongs to the space $\HH_1$ described in 
\refb{ephysicaloffshell}. By normalizing the basis states $|\vp_i\rangle$ to
satisfy
\be
\langle \vp_i| c_0^- c_0^+|\vp_j\rangle =
\delta_{ij}\, , \quad c_0^\pm \equiv {1\over  2 } (c_0 \pm \bar c_0)\, ,
\ee
we can ensure that
\be
\langle \vp^c_i | b_0^+ b_0^-| \vp^c_j\rangle = \delta_{ij}\, .
\ee
Thus we can
drop the second line of \refb{emeasurefactor}, 
insert the operator
$e^{-s(L_0+\bar L_0)} e^{i\theta(L_0-\bar L_0)}$ in front of 
$|\vp_i\rangle$ in the first line\cite{1311.1257} and
 set $j=i$ in the sum.

\sectiono{Off-shell NS sector amplitudes in superstring theory} \label{s2}

In this section we shall construct off-shell amplitudes for NS sector external states 
in heterotic string theory by generalizing the procedure described
in \cite{9202087}.  Ramond sector states require special treatment and will be discussed
in \S\ref{sramond}.
Generalization to type II string theory is straightforward and
will be discussed briefly in \S\ref{ssuper}.
Also in this section we shall assume
that  it
is possible to choose a gauge for the gravitino globally over the whole moduli space.
In this case
we can
integrate out the fermionic coordinates of the supermoduli space and express the
amplitudes as integrals over the  moduli space of ordinary Riemann surfaces
at the cost of inserting a set of picture changing
operators into the world-sheet correlation functions\cite{Verlinde:1987sd}. 
As emphasized in \cite{1209.5461}, this assumption fails at sufficiently high
genus. 
In such cases the measure in the moduli space, constructed using the picture
changing operators, has spurious singularities in codimension two subspaces 
of the moduli space where the gauge choice for the gravitino breaks 
down\cite{Verlinde:1987sd}.
In \S\ref{snonhol} we shall address  how to deal with these
spurious singularities.

\subsection{Superconformal ghost system and off-shell string states}

A classical background in heterotic string theory is based on a two dimensional
superconformal field theory with supersymmetry in the right-moving (holomorphic)
sector of the world-sheet. Besides the matter superconformal field theory with
central charge (26,15), it also has anti-commuting $b$, $c$, $\bar b$, $\bar c$
ghosts and commuting $\beta, \gamma$ ghosts with total central charge $(-26,-15)$.
The $(\beta,\gamma)$ system can be `bosonized' as
\be \label{eboserule}
\gamma = \eta\, e^{\phi}, \quad \beta= \p\xi \, e^{-\phi}, \quad \delta(\gamma)
= e^{-\phi}, \quad \delta(\beta) = e^\phi\, ,
\ee
where $\xi, \eta$ are  fermions and $\phi$ is a scalar with background charge.
We shall use the standard
convention in which the (ghost number, picture number, GSO) assignments of various fields
are:
\ben
&& c, \bar c: (1,0,+), \quad b, \bar b: (-1, 0,+), \quad \gamma: (1,0,-), \quad \beta:(-1,0,-),
\nonumber \\
&& \xi: (-1,1,+), \quad \eta: (1,-1,+), \quad
e^{q\phi}: (0, q, (-1)^q)\, .
\een
$e^{\pm\phi}$ are fermionic operators. The operator products of $b$, $c$,
$\xi$, $\eta$ and 
$e^{q\phi}$ operators take the form
\be
c(z) b(w) =(z-w)^{-1}+\cdots, \quad
\xi(z)\eta(w) = (z-w)^{-1}+\cdots, \quad e^{q_1\phi(z)} e^{q_2\phi(w)} =
(z-w)^{-q_1q_2} e^{(q_1+q_2)\phi(w)}+ \cdots \, .
\ee
With this $\gamma(z)\beta(w)\sim -(z-w)^{-1}$. This has an additional $-$ sign compared
to the standard convention used {\it e.g.} in \cite{1209.5461}, but we have chosen to stick to the
bosonization rules \refb{eboserule} of \cite{Verlinde:1987sd}. 
The BRST charge is given by
\be
Q_B = \ointop dz j_B(z) + \ointop d\bar z \bar j_B(z)\, ,
\ee
where
\be
\bar j_B(\bar z) = \bar c(\bar z) \bar T_m(\bar z)
+\bar b(\bar z) \bar c(\bar z) \bar\p \bar c(\bar z)\, ,
\ee
\be \label{ebrstcurrent}
j_B(z) =c(z) (T_{m}(z) + T_{\beta,\gamma}(z) )+ \gamma (z) T_F(z) + b(z) c(z) \p c(z) 
-{1\over 4} \gamma(z)^2 b(z)\, .
\ee
Here $\bar T_m(\bar z)$ is the anti-holomorphic part of the matter stress tensor,
$T_{m}(z)$ is the holomorphic part of the matter stress tensor, $T_{\beta,\gamma}(z)$
is the stress tensor of the $(\beta,\gamma)$ system and $T_F(z)$ is the 
world-sheet supersymmetry current in the matter sector. The signs of various
terms in \refb{ebrstcurrent} are consistent with our conventions, as can be seen
{\it e.g.} from the fact that the components of the
total super stress tensor computed from the
(anti-)commutator of $Q_B$ with $b(z)$, $\beta(z)$ and $\bar b(\bar z)$ satisfy
the correct operator product relations.
Finally the picture changing operator $\VVV$
is given by\cite{FMS,Verlinde:1987sd}
\be \label{epicture}
\VVV(z) = \{Q_B, \xi(z)\} = c \partial \xi + 
e^\phi T_F - {1\over 4} \p\eta e^{2\phi} b
- {1\over 4} \p\left(\eta e^{2\phi} b\right)\, .
\ee
This is a BRST invariant dimension zero primary operator 
and carries picture number $1$.

We shall define the subspace $\HH_0$ of  off-shell 
states in the matter ghost conformal field theory as in \refb{econd} 
with some additional restrictions:
\be \label{econdsup}
|\Psi\rangle \in \HH_0\quad {\rm if} \quad
(b_0-\bar b_0)|\Psi\rangle = 0,  \quad (L_0-\bar L_0)|\Psi\rangle=0, \quad
\eta_0|\Psi\rangle=0,
\quad \hbox{picture number}\, (|\Psi\rangle)=-1\, .
\ee
The $\eta_0|\Psi\rangle=0$ condition tells us that we are working in 
the small Hilbert space\cite{FMS}.
It will also be useful to define a subspace $\HH_1$ containing
off-shell states of ghost number two in the Siegel gauge
\be \label{ephysicaloffshellsup}
|\Psi\rangle \in \HH_1\quad {\rm if} \quad |\Psi\rangle \in \HH_0, 
\quad (b_0+\bar b_0)|\Psi\rangle = 0\, , \quad \hbox{ghost number} \, (|\Psi\rangle)=2\, .
\ee

\subsection{The integration measure} 

For simplicity we shall describe the construction of off-shell amplitudes in
the heterotic string theory but the generalization to the case of type II strings is
straightforward. 
On a genus $g$ Riemann surface, in order to get a non-vanishing correlation
function the total picture number of all the operators must add up to $2g-2$.
The naive guess would be that the 
construction of the off-shell amplitude 
would proceed in a manner identical to that in the 
case of bosonic string theory except that in the construction of the $p$-form in the moduli
space the surface state $\langle \Sigma|$ 
should be replaced by
\be 
\langle \Sigma| K\, ,
\ee
where $K$ is the product of $2g-2+n$ picture changing operators. 
In this case the picture number carried by the states in $\HH_0$ inserted at
the punctures and the picture changing operators add up to $2g-2$ as required.
We can in fact 
generalize this a bit by writing
\be \label{ekinsert}
K =\sum_\alpha A^{(\alpha)} \VVV(z_1^{(\alpha)}) \VVV(z_2^{(\alpha)})  \cdots
\VVV(z_{2g-2+n}^{(\alpha)}) \, ,
\ee
where the sum over $\alpha$ in \refb{ekinsert}
runs over arbitrary number of values,
$A^{(\alpha)}$ are arbitrary real numbers satisfying $\sum_\alpha A^{(\alpha)}=1$
and $z_1^{(\alpha)},\cdots z_{2g-2+n}^{(\alpha)}$ are the locations of the picture
changing operators for the $\alpha$-th term. For definiteness we shall assume
that the coordinates $z_i^{(\alpha)}$ lie on $\Sigma-\cup_a D_a$
and 
are measured in the fixed $z$ coordinate system introduced in
\S\ref{s2.1}. More precisely if a picture changing operator
is located on the component $\sigma_k$ then its location is measured in the
coordinate system $z_k$ on $\sigma_k$ that we introduced in \S\ref{s2.1}.

With this definition most of the identities satisfied by the off-shell bosonic string theory
amplitude generalizes to the heterotic string theory provided we continue to impose
the conditions \refb{econdsup}. There is however one caveat.
In proving the analog of \refb{epformqb} we have to assume that the constants
$A^{(\alpha)}$ as well as the locations $z_i^{(\alpha)}$ of the picture changing 
operators remain fixed as we move in $\wh\PP_{g,n}$ using Schiffer variation. 
Otherwise in \refb{e114} the tangent vector $\wh V_i$ acting on the first term on
the right hand side will give additional contributions containing derivatives of
$A^{(\alpha)}$ and $z_i^{(\alpha)}$ with respect to the coordinates on 
$\wh\PP_{g,n}$. While it is certainly possible to keep $A^{(\alpha)}$ 
and $z_i^{(\alpha)}$ fixed
locally, various global issues may prevent us from keeping them fixed over the
entire section in $\wh\PP_{g,n}$ over which we integrate. 
Thus we need to allow $A^{(\alpha)}$ 
and / or $z_i^{(\alpha)}$ to depend on the coordinates of $\wh\PP_{g,n}$.
For simplicity we shall take the $A^{(\alpha)}$'s to be fixed moduli independent
constants and allow the $z^{(\alpha)}_i$'s to be moduli dependent 
-- this is not necessary but will suffice for our analysis.
A minimal remedy will then be to include extra terms in $\Omega^{(g,n)}_p$
that can account for moduli dependence of 
the $z_i^{(\alpha)}$'s\cite{Verlinde:1987sd,9202087}. However we shall develop
a slightly more general formalism that will be useful for dealing with the spurious
poles in \S\ref{snonhol}.

This general formalism involves 
extending $\wh\PP_{g,n}$ to a larger space $\wt\PP_{g,n}$ by appending
the data on the locations $(z_1,\cdots z_{2g-2+n})$ of $(2g-2+n)$
picture changing operators  to $\wh\PP_{g,n}$. Thus
$\wt\PP_{g,n}$ can be regarded
as a fiber bundle over the base $\wh\PP_{g,n}$, with 
$z^{}_i$'s acting as fiber coordinates. 
The tangent vectors of $\wt \PP_{g,n}$ can be labelled by
\begin{enumerate}
\item Schiffer variations generated by $n$-tuple of vector fields 
$(v^{(1)}(z),\cdots v^{(n)}(z))$ keeping the locations of the picture changing
operators fixed in the $z$ coordinate system,
\item
${\p/\p z^{}_i}$ for every  $i$. These move the picture changing operators
keeping fixed the Riemann surface, the punctures and the coordinate system
on the Riemann surface.
\end{enumerate}
A general  choice of picture changing operators like the one given in \refb{ekinsert}
with moduli independent $A^{(\alpha)}$
can be regarded as a weighted average of several sections of $\wt\PP_{g,n}$
labelled by $\alpha$. Since
eventually we shall be interested in integrating forms over these sections, the integral
of a form over such weighted averages can be interpreted as the weighted average
of the integrals over different sections.

We now define operator valued $r$-forms $K^{(r)}$
along the fiber as follows:
\be \label{ekr0}
K^{(0)} =  \VVV(z_1^{}) \VVV(z_2^{})  \cdots
\VVV(z_{2g-2+n}^{})\, , 
\ee
\ben \label{ekvalue}
K^{(r)} &=& \bigg[\left(\VVV(z^{}_1) - \p \xi(z^{}_1) d z^{}_1\right)
\wedge \left(\VVV(z^{}_2) - \p \xi(z^{}_2) d z^{}_2\right)\wedge \cdots
\nonumber \\ &&
\wedge \left(\VVV(z^{}_{2g-2+n}) - \p \xi(z^{}_{2g-2+n}) d z^{}_{2g-2+n}
\right)
\bigg]^{(r)}\, ,
\een
where the superscript $(r)$ on the right hand side indicates that we need to pick the
$r$ form from the expansion of the terms inside the square bracket. More
explicitly, we have
\ben \label{ekvalueexplicit}
K^{(1)} &=&  \sum_{i=1}^{2g-2+n}
S_i^{}
\prod_{j=1\atop j\ne i}^{2g-2+n}  \VVV(z_j^{}) \, ,
\nonumber \\ 
K^{(2)} &=&   \sum_{i,j=1\atop i<j}^{2g-2+n} S_i^{}\wedge S_j^{}
 \prod_{k=1\atop k\ne i,j}^{2g-2+n}  \VVV(z_k^{}) \, , \nonumber \\
\cdots &=& \cdots \nonumber \\
K^{(2g-2+n)} &=& 
 S_1^{}\wedge S_2^{} \wedge \cdots
\wedge S_{2g-2+n}^{}\, ,
\een
where 
\be \label{edefsi}
S_i^{} \equiv - d z_i^{} \p\xi(z_i^{}) \, .
\ee
$K^{(r)}$'s satisfy the 
`descent relations'\cite{9202087}
\be \label{ekr}
d_F K^{(r)} = (-1)^{r+1} [K^{(r+1)},Q_B\}\, \quad \hbox{for} \quad 1\le r \le 2g-2+n, 
\quad d_F K^{(2g-2+n)}=0\, ,
\ee
where $d_F$ denotes exterior derivative along the fiber direction 
labelled by $\{z^{}_i\}$:
\be \label{edefdf}
d_F K^{(r)} \equiv  \sum_i d z^{}_i 
\wedge {\p\over \p  z^{}_i} K^{(r)}
\, .
\ee
The symbol $[~\}$ stands for commutator if $K^{(r+1)}$ is grassmann even and
anti-commutator if $K^{(r+1)}$ is grassmann odd.

Next we define
\be \label{edefomp}
\Omega^{(g,n)}_p(|\Phi\rangle) = (2\pi i)^{-(3g-3+n)}\,  \langle \Sigma| \BB_p |\Phi\rangle\, ,
\quad \BB_p\equiv \sum_{r=0\atop
r\le 2g-2+n}^{p} K^{(r)}\wedge B_{p-r}\, ,
\ee
where $B_p$ has been defined in \refb{edefB}.  
Since the notation is somewhat abstract, we shall now 
clarify the meaning of \refb{edefomp}. Let us consider $p$
tangent vectors $\{V_1+ U_1, \cdots V_p+U_p\}$ of $\wt\PP_{g,n}$
where each $V_k$ is
associated with a Schiffer variation generated by the $n$-tuple of vector fields
$\vec v_k(z)$ {\it keeping the coordinates $z^{}_i$'s fixed}
and each $U_k$ is a vector field of the form 
$\sum_{i} u_{k;i} {\p / \p z^{}_i}$ that generates 
shift of the location of the picture changing operators keeping the 
moduli of the Riemann surface, the punctures as well as the coordinate
system on the Riemann surface fixed. Then
\ben \label{eexpomp}
\Omega^{(g,n)}_p(|\Phi\rangle)[V_1+U_1, \cdots V_p+U_p] 
&=&  (2\pi i)^{-(3g-3+n)}\,   \bigg\langle \Sigma| \sum_{r=0\atop r\le 2g-2+n}^{p} \sum_{\GG_r} (-1)^{\bf P} 
K^{(r)}[\{U_i; i\in \GG_r\}]\nonumber \\ && \quad
\wedge B_{p-r}[\{ V_j; j\in \GG_r^c\}]|\Phi\bigg\rangle\, ,
\een
where the sum over $\GG_r$ runs over all subsets of length $r$ of $1,\cdots p$,
$\GG_r^c$ denotes the complement of $\GG_r$ and $(-1)^{\bf P}$ is the sign that is picked
up while rearranging $1,\cdots p$ to the arrangement $\{i\in \GG_r\}, \{j\in \GG_r^c\}$.
$[~]$ on the left hand side 
denotes contraction with the arguments inside the square bracket as usual.

In words these rules may be stated as follows.
\begin{enumerate}
\item $\Omega^{(g,n)}_0(|\Phi\rangle)$ is given by the correlation function of the
vertex operators describing the state $|\Phi\rangle$, inserted using the local
coordinate system associated with the point in $\wt\PP_{g,n}$ we are at, with
additional insertion of $K^{(0)}$ into the correlation function.
\item $\Omega^{(g,n)}_p$ is defined in terms of $\Omega^{(g,n)}_0$ as follows. For every
contraction with a tangent vector $\p/\p z^{}_i$ we replace the $\VVV(z^{}_i)$ term by
$-\p\xi(z^{}_i)$. On the other hand for every contraction of $\Omega^{(g,n)}_p$
with a tangent vector of $\wt\PP_{g,n}$ associated with Schiffer variation by the
$n$-tuple of vector fields $\vec v$, we
insert into the correlation function a $b(\vec v)$.
\end{enumerate}

Generalization of
the ghost number conservation equation \refb{eghno} tells us that if
$|\Phi\rangle$ has total ghost number $n_\Phi$  and total picture number $p_\Phi$
then for $\Omega^{(g,n)}_p$ to be non-zero we must have
\be \label{eghnopicno}
n_\Phi -p = 6 - 6g, \quad p_\Phi = -n\, .
\ee
The second condition is automatically satisfied if $|\Phi\rangle\in\HH_0$ with
$\HH_0$ defined as in \refb{econdsup}.

\subsection{Properties of the integration measure}

First we shall verify that the property mentioned below \refb{eghno} and
the two properties mentioned below \refb{ephysicaloffshell} hold
once we restrict the string states to satisfy \refb{econdsup}.
The proof of the properties mentioned below \refb{ephysicaloffshell} 
are identical to that in the case of
bosonic string theory.
This justifies our regarding $\Omega^{(g,n)}_p$ as a form in $\wt\PP_{g,n}$ rather
than in a larger space that we get by appending the picture changing data to
$\PP_{g,n}$.
The proof of the property mentioned below \refb{eghno} 
however is more subtle.
Let $v(z)$ denote a globally defined vector
field on $\Sigma - \cup_a D_a$ and $\vec v = (v^{(1)}, \cdots v^{(n)})$ denote the
collection of vectors obtained from the restriction of $v(z)$ on $\p D_a$. 
This generates a deformation of the local coordinates around the punctures
according to \refb{ex1}-\refb{ex3},
but this can be undone by a change in the coordinate system $z$ in
$\Sigma- \cup_a D_a$ with $z+\eps v(z)$ as the new coordinate. 
This is the reason why earlier this generated 
a vanishing tangent vector of $\PP_{g,n}$ and hence also
of $\wh\PP_{g,n}$. But in superstring theory, the change in the $z$ coordinates,
required to undo the deformation of the local coordinates, 
will move the location 
$z^{}_i$ of the picture changing
operators by $\eps v(z^{}_i)$. Thus we expect that the tangent vector of
$\wt \PP_{g,n}$ associated with the vector field $v$ will not vanish but should be
equal to
\be \label{edefu}
U = \sum_i v(z^{}_i) {\p \over \p z^{}_i}\, .
\ee
Equivalently $V-U$ should vanish as a tangent vector of $\wt\PP_{g,n}$.
We shall now try to verify that the contraction of $\Omega^{(g,n)}_p$ with such a 
tangent vector vanishes. 
Let $V$ denote the
tangent vector of $\wt \PP_{g,n}$ associated with the vector field $v(z)$.
Denoting this contraction of $\Omega^{(g,n)}_p$ with $V$ 
by $\Omega^{(g,n)}_p[V]$ and using \refb{edefomp} we get
\be \label{eg1}
\Omega^{(g,n)}_p[V] =  (2\pi i)^{-(3g-3+n)} \, \langle\Sigma|\sum_{r=0}^{2g-2+n} 
(-1)^r \, K^{(r)}\wedge B_{p-r}[v] |\Phi\rangle\, .
\ee
where
\be
B_{p-r}[v] \equiv b(v) B_{p-r-1}, \quad
b(v) \equiv \ointop v(z) b(z) dz + \ointop \bar v(\bar z) \bar b(\bar z) d\bar z\, .
\ee
We can now
move $b(v)$ to the left of the $K^{(r)}$ using the expression \refb{ekvalue} for the
$K^{(r)}$'s and the anti-commutation relations:
\be \label{eg2}
[\VVV(z), b(v)] = - v(z)\p\xi(z), \quad \{S_i, b(v)\} = 0\, .
\ee
Once $b(v)$ moves to the left we can contract the integration contour by deforming it
into $\Sigma$ and the contribution vanishes. Thus the 
net contribution to the right hand side of
\refb{eg1} comes from the commutators and can be expressed as
\be \label{eg5}
- (2\pi i)^{-(3g-3+n)}\Big\langle\Sigma\Big
|\sum_{\ell = 1}^{2g-2+n} v(z^{}_\ell)
\p \xi(z_\ell^{}) \sum_{r=0}^{2g-2+n} 
\sum_{i_1, \cdots i_r=1\atop i_1< i_2< \cdots <i_r, \, 
i_s\ne \ell}^{2g-2+n} S^{}_{i_1}\wedge \cdots \wedge S^{}_{i_r}
\prod_{k=1\atop k\ne \ell, i_1, \cdots i_r}^{2g-2+n} \VVV(z^{}_k)
B_{p-r-1}\Big|\Phi\Big\rangle\, .
\ee
Using \refb{ekvalueexplicit} this can be interpreted as
\be
(2\pi i)^{-(3g-3+n)}\Big\langle \Sigma\Big| \sum_{r=0}^{2g - 2 + n}  K^{(r+1)}[U] \wedge B_{p-r-1} |\Phi\rangle
=\Omega^{(g,n)}_p[U]
\ee
where $U$ has been defined in \refb{edefu}. This shows that 
$\Omega^{(g,n)}_p[U]=\Omega^{(g,n)}_p[V]$,  precisely as expected.


Next we shall show that
$\Omega^{(g,n)}_p$ satisfies the
identity \refb{epformqb} with $d$ now denoting the exterior derivative in
$\wt\PP_{g,n}$.
For this we write
\ben \label{emeasure}
(2\pi i)^{3g-3+n}\Omega^{(g,n)}_p(Q_B|\Phi\rangle) &=& \Big\langle \Sigma\Big| \sum_{r=0}^{2g-2+n} K^{(r)}\wedge 
[B_{p-r}, Q_B\}\Big|\Phi\Big\rangle \nonumber \\
 && + \Big\langle \Sigma\Big| \sum_{r=0}^{2g-2+n} 
(-1)^{p-r}
[K^{(r)} , Q_B\}
\wedge  B_{p-r}\Big|\Phi\Big\rangle \, ,
\een
using the relation $\langle\Sigma|Q_B=0$. Now using the same manipulations as
in the case of bosonic string theory, the first term on the right hand side
can be interpreted as
\be \label{eone}
\sum_{r=0}^{2g-2+n} (-1)^{p-r} (-1)^r 
d_T \langle \Sigma| K^{(r)} \wedge B_{p-r-1}|\Phi\rangle \, ,
\ee
where $d_T\equiv d - d_F$ in the `tangential exterior derivative along
the base $\wh\PP_{g,n}$' 
defined so that its contraction 
with a tangent associated with Schiffer variation
{\it keeping $z^{}_i$'s fixed} is
given as in \refb{e114} while its contraction with $\p/\p z^{}_i$ 
vanishes.\footnote{Neither $d_T$ nor $d_F$ are good operators in the sense
that neither of them satisfies the constraint that their contraction with  
$V$ associated with Schiffer variation by a vector field $v(z)$ that
is globally defined on $\Sigma -\cup_a D_a$,
and
$U$ given in
\refb{edefu},
are equal. But $d_T+d_F$, which is
the full exterior derivative operator in $\wt\PP_{g,n}$, has this property.}
The $(-1)^{p-r}$ in \refb{eone} 
is the result of the $(-1)^p$ factor in \refb{epformqb}, while the
$(-1)^r$ factor is the result of passing $d_T$ through the $r$-form 
$K^{(r)}$. On the other hand, using \refb{ekr} we can express the second term
on the right hand side of \refb{emeasure}
as
\be \label{etwo}
\sum_{r=0}^{2g-2+n} (-1)^{p-r} (-1)^r d_F
\langle \Sigma| K^{(r-1)} \wedge B_{p-r}|\Phi\rangle
= (-1)^p
\sum_{r=0}^{2g-2+n} d_F
\langle \Sigma| K^{(r)} \wedge B_{p-r-1}|\Phi\rangle \, .
\ee
In going from the left hand side to the right hand side of this equation we have
made an $r\to r+1$ shift and used 
that the $r=0$ term on the left hand side and $r=2g-2+n$ terms on the right hand side
vanishes.
Adding \refb{eone} and
\refb{etwo} and using the fact that 
$d_F+d_T$ represents the total exterior derivative $d$ on $\wt\PP_{g,n}$,
we arrive at the
equation
\be \label{etotal}
\Omega^{(g,n)}_p(Q_B|\Phi\rangle) = (-1)^p d \Omega^{(g,n)}_{p-1}(|\Phi\rangle)\, .
\ee

\subsection{General parametrization of $\wt\PP_{g,n}$} \label{s2gen}

We can also generalize the above prescription to the more general 
labelling of the tangent vectors of $\wh\PP_{g,n}$ as 
described in \S\ref{s1gen}. In this formalism
we describe a Riemann surface as different components $\sigma_k$ and the
coordinate disks $D_a$, each with its own coordinate
system, glued together at their boundary circles. We can move in $\wh\PP_{g,n}$
by changing the 
functional relationship between the coordinates of two components sharing a common
boundary circle. Thus a particular deformation of one of these functions,
encoded in a vector field
defined around the common boundary,
will describe
a tangent vector of $\wh\PP_{g,n}$. 
The contraction of the
$p$-form $\Omega^{(g,n)}_p$ in bosonic string theory with such a tangent vector
involves inserting into the correlator contour integrals of $b$ and $\bar b$ along
the common boundary circles of two components, 
weighted by the vector field $v(z)$ and $\bar v(\bar z)$ 
associated with this deformation. 

In heterotic or
type II string theory, we need to insert picture changing operators in the correlation
function to get a proper integration measure. 
As already described earlier, if a particular picture changing operator
is located on the component
$\sigma_k$ then its coordinate $z^{}_i$ 
is measured in the $z_k$ coordinate system. 
Then $\wt \PP_{g,n}$ can be constructed as a fiber bundle over $\wh\PP_{g,n}$ with
$z^{}_i$'s as the fiber coordinates. The tangent vectors along the fiber 
are linear combinations of 
$\p/\p z^{}_i$'s and generate shift of the locations of the picture changing operators, keeping fixed
the Riemann surface, the punctures and the coordinate systems $\{z_k\}$ and $\{w_a\}$ 
on the different
components $\{\sigma_k\}$ and $\{D_a\}$ of the Riemann
surface. $K^{(r)}$'s are now defined as in 
\refb{ekvalue}.
The tangent vectors along the base $\wh\PP_{g,n}$, described in the last paragraph,
are  lifted to 
$\wt \PP_{g,n}$ by identifying them as deformations that generate the same 
deformations along the base $\wh\PP_{g,n}$, and keeps, for every $i$, 
the coordinate
$z_i^{}$ of the $i$-th picture changing operator fixed.
With these modifications $\Omega^{(g,n)}_p$ is defined in the same way as in
\refb{edefomp}, except that, as described in the last paragraph, 
the $b(v)=
\ointop v(z) b(z) + \ointop \bar v(\bar z) \bar b (\bar z)$'s are now more general
objects which use vector fields
$v(z)$ describing the changes in the functional relationship between different
components and the contour integral runs along the boundary circle separating two
such components.

Now it is clear that the location of the boundary between two adjacent components
$\sigma_i$ and $\sigma_j$ is somewhat arbitrary and can be shifted without changing
the relation between $z_i$ and $z_j$. But due to this shift a picture changing operator
initially located in $\sigma_i$ may move to $\sigma_j$ or vice versa. Since this is
change is not physical, the integration measure should not change. While the
proof of this is straightforward, we shall illustrate this only by an example.
Let us consider the case described in \refb{epp1} and
pretend that $q$ is the only modulus of $\wh\PP_{g,n}$ that we are
interested in. Suppose further that we have only one picture changing
operator that we want to insert at the point $z_1$ in the $z$ coordinate system.
Then $\wt \PP_{g,n}$ will be labelled by $(q,z_1)$ and the 1-form 
$\Omega^{(g,n)}_1$ will be associated with the 
insertion\footnote{For $\Omega^{(g,n)}_0$ and $\Omega^{(g,n)}_2$ the analysis is trivial.}
\be \label{ezi}
-dq \, q^{-1} \VVV(z_1) \ointop dz \, z\, b(z)  - dz_1 \p \xi(z_1) \, ,
\ee
where it is understood that $z_1$ is placed away from the origin
relative to the integration contour in the $z$-plane (see Fig.~\ref{fmove}(a)).
On the other hand if we want to insert the picture changing operator in the $w$
coordinate system, then $\Omega^{(g,n)}_1$ is  associated with the insertion
\be \label{ewi}
-dq \, q^{-1} \VVV(w_1) \ointop dw \, w\, b(w)  - dw_1 \p \xi(w_1) \, ,
\ee
with $w_1$ being away from the origin relative to 
the integration contour in the $w$ plane.

\begin{figure}

\begin{center}

\figmove

\end{center}

\caption{Pictorial representation of the integration contours for eqs.\refb{ezi} and \refb{ezialt}.
\label{fmove}}

\end{figure}

 Let us suppose that in the two cases
 the picture changing operators are located at the same physical position. 
 This can be achieved by moving the (artificial) boundary between the two 
 components labelled by $z$ and $w$ across the location of the picture
 changing operator.
 In that case we should get identical results using \refb{ezi} and \refb{ewi}. 
 Let us test this.
 First noting that $b$ is a primary of dimension 2, $\p\xi$ is a 
 primary of dimension 1 and $\VVV$ is a primary of dimension 0, we have
 \be
 \VVV(w) = \VVV(z), \quad
 \p\xi(w) = (\p z/\p w) \p\xi(z) = - q^{-1} \, z^2 \p\xi(z), \quad 
 b(w) = (\p z/\p w)^2 b(z) = q^{-2} \, z^4 b(z), 
 \ee
 where the argument of an operator is also used to denote the coordinate system
 in which it is inserted. Thus \refb{ewi} can be written as
 \be \label{ezialt}
 -dq \, q^{-1} {\ointop} dz \, z\, b(z) \, \VVV(z_1)  + q^{-1} z_1^2  dw_1 \p \xi(z_1) \, .
\ee
In writing the above equation we have taken into account the fact that an anti-clockwise
contour in $w$ plane produces a clockwise contour in the $z$ plane and hence costs
an extra $-$ sign when we make this into an anticlockwise contour
$\ointop$. In the first term we have placed $\VVV(z_1)$ on the right of
the contour integral,  signifying the fact that in \refb{ezialt}
the point $z_1$ is towards the origin in the $z$ plane relative to the integration 
contour  (see Fig.~\ref{fmove}(b))
since in the $w$-plane it was away from the origin. Using
the relation $w_1=q/z_1$, we now write, as a differential form in $\wt\PP_{g,n}$
labelled by $q$ and $z_1$,
\be 
dw_1 = - q z_1^{-2} dz_1 + z_1^{-1} dq\, .
\ee
Substituting this into \refb{ezialt} we get
\be \label{ezialt2}
 -dq \, q^{-1} {\ointop} dz \, z\, b(z) \, \VVV(z_1)  - dz_1 \p \xi(z_1) 
 + q^{-1} z_1 \, dq \p\xi(z_1)\, ,
\ee
where the integration contour is still as shown in Fig.~\ref{fmove}(b).
In order to compare with \refb{ezi} we  
now move the contour of integration in the first term through the point $z_1$
so that $z_1$ is situated away from the origin relative to 
the integration contour. In that process we pick up
a residue of the form $-dq \, q^{-1} z_1 \p\xi(z_1)$ which precisely cancels the
last term in \refb{ezialt2}. Thus we get
\be
-dq \, q^{-1} \VVV(z_1) {\ointop} dz \, z\, b(z)  - dz_1 \p \xi(z_1) \, .
\ee
This precisely agrees with \refb{ezi}.

\subsection{Off-shell amplitude of NS sector fields}  \label{soffshell}

For defining off-shell amplitudes in superstring theory
we shall need to regard $\wt\PP_{g,n}$ as a
fiber bundle over the base $\MM_{g,n}$ and integrate $\Omega^{(g,n)}_{6g-6+2n}$ on
a section -- or more generally on the formal weighted average of several sections
as in \refb{ekinsert} -- of this fiber bundle. 
This means that for every point in $\MM_{g,n}$ we need to make a choice
of local coordinate system around each puncture and the locations of the
picture changing operators, and integrate $\Omega^{(g,n)}_{6g-6+2n}$ on the subspace
of $\wt\PP_{g,n}$ that it defines. 
The section can be arbitrary subject to 
the requirement of gluing compatibility. This requires first of all that
the choice of local coordinates must be subject to the same kind of
constraints as given in \S\ref{soff}. However we must also put
constraint on the $K^{(r)}$'s. 
It requires that when we take a Riemann surface of genus $g_1+g_2$
and $n_1+n_2-2$ punctures, constructed from the plumbing fixture of a 
Riemann
surface of genus $g_1$ and $n_1$ punctures with another Reimann surface
of genus $g_2$ and $n_2$ punctures, then we must have
\be \label{epic1}
K^{(r)}_{g_1+g_2, n_1+n_2-2} = \sum_{s=0}^r K^{(s)}_{g_1, n_1} \wedge
K^{(r-s)}_{g_2, n_2}\, ,
\ee
where $K^{(s)}_{g,n}$ denotes the choice of $K^{(s)}$ on the genus $g$
Riemann surface with $n$ punctures.
Since all the $K^{(r)}$'s for $r\ge 1$ are determined in terms of $K^{(0)}$,
it is enough to satisfy this equation for $r=0$. In that case it takes the simple
form
\be \label{epic2}
K^{(0)}_{g_1+g_2, n_1+n_2-2} = K^{(0)}_{g_1, n_1}  \,
K^{(0)}_{g_2, n_2}\, .
\ee 
This in particular requires that the $2g-2+n$ picture changing operators on the
glued Riemann surface are distributed such that $2g_1-2+n_1$ of them
lie on the first 
surface and $2g_2-2+n_2$ of them
lie on the second surface\cite{1209.5461}. A systematic
procedure for constructing such gluing compatible sections will be discussed in
\S\ref{sgluing}. For such Riemann surfaces we see from 
\refb{epic1} and \refb{ebfactor} that after contraction with the
tangent vectors of the section, $\BB_{6g-6+2n}$ defined in \refb{edefomp} factors as
\be \label{ebbfactor}
-i\, \BB^{(1)}_{6g_1 - 6 + 2 n_1}\,  b_0^+ b_0^- \, \BB^{(2)}_{6 g_2 - 6 + 2 n_2}\, .
\ee
This leads to an exact analog of \refb{emeasurefactorA}, \refb{emeasurefactorB}
for general external states in $\HH_0$ inserted at the punctures:
\ben \label{emeasurefactorAsup}
\Omega^{(g,n)}_{p}(|\Phi\rangle) |_S
&=&  {1\over 2\pi} \sum_{0\le p_1\le 6g_1-6+2n_1, \, 0\le p_2\le
6g_2-6+2 n_2\atop p_1+p_2=p-2}
\sum_{i,j} \langle \vp_i^c | b_0^+ b_0^-
e^{-s(L_0+\bar L_0)} e^{i\theta(L_0-\bar L_0)} |\vp_j^c\rangle
 \nonumber \\
&& (-1)^{p_1p_2+N_1+p_1 + 1}  ds \wedge d\theta \wedge \Omega^{(g_1,n_1)}_{p_1}(|\Phi_1\rangle
\otimes |\vp_i\rangle)|_{S_1} \, \wedge  \, \Omega^{(g_2,n_2)}_{p_2}(|\vp_j\rangle
\otimes |\Phi_2\rangle)|_{S_2} \, ,
 \nonumber \\ 
\een
\ben \label{emeasurefactorBsup}
\Omega^{(g,n)}_{p}(|\Phi\rangle) |_{S; s=\Lambda}
&=& {1\over 2\pi} \sum_{0\le p_1\le 6g_1-6+2n_1, \, 0\le p_2\le
6g_2-6+2 n_2\atop p_1+p_2=p-1}
\sum_{i,j} \langle \vp_i^c | b_0^-
e^{-\Lambda (L_0+\bar L_0)} e^{i\theta(L_0-\bar L_0)} |\vp_j^c\rangle
 \nonumber \\
&& (-1)^{p_1p_2+p_2}  \, d\theta \wedge \Omega^{(g_1,n_1)}_{p_1}(|\Phi_1\rangle
\otimes |\vp_i\rangle)|_{S_1} \, \wedge  \, \Omega^{(g_2,n_2)}_{p_2}(|\vp_j\rangle
\otimes |\Phi_2\rangle)|_{S_2} \, .
\nonumber \\ 
\een
If we restrict the external states to be in $\HH_1$ and take $p_i=6g_i-6+2n_i$
for $i=1,2$ we get
\ben \label{emeasurefactorsup}
\Omega^{(g,n)}_{6g-6+2n}(|\Phi\rangle) |_S
&=&-{1\over 2\pi}  \sum_{i,j} ds \wedge d\theta \wedge \Omega^{(g_1,n_1)}_{6g_1-6+2n_1}(|\Phi_1\rangle
\otimes |\vp_i\rangle)|_{S_1} \, \wedge  \, \Omega^{(g_2,n_2)}_{6g_2-6+2n_2}( |\vp_j\rangle
\times |\Phi_2\rangle) |_{S_2}
  \nonumber \\ &&
\times \langle \vp_i^c | b_0^+  b_0^- 
e^{-s(L_0+\bar L_0)} e^{i\theta(L_0-\bar L_0)} |\vp_j^c\rangle\, .
\een
Since in this case each external state at the punctures carries ghost number two and picture number $-1$,
it follows from \refb{eghnopicno} that $|\vp_i\rangle$, 
$|\vp_j\rangle$ carry ghost number
two and picture number $-1$. Furthermore due to the $b_0^+ b_0^-$ factor in the second
line of \refb{emeasurefactorsup}, $|\vp_i\rangle$ and $|\vp_j\rangle$ are annihilated
by $b_0$ and $\bar b_0$. Thus they correspond to states in $\HH_1$
defined in \refb{ephysicaloffshellsup}.

Note that if we choose the section in such a way that the locations of the picture changing
operators remain fixed in the chosen coordinate system on the Riemann
surface, then the tangents to the section have vanishing contraction with the
$K^{(r)}$'s for $r\ge 1$. Thus we can set all the $K^{(r)}$'s other than $K^{(0)}$ to
zero and get back the usual formalism in which we insert the picture changing 
operators at fixed points on the Riemann surface. In general however we shall allow
the locations of the picture changing operators to vary as we move along the
base $\MM_{g,n}$. We shall also not impose any holomorphicity condition on the
section, and allow the locations of the picture changing operators to depend
non-holomorphically on the coordinates of $\MM_{g,n}$.

\subsection{Vertical integration} \label{svertical}

If we are only interested in integrating $\Omega^{(g,n)}_{6g-6+2n}$ over a
section, we could from the beginning express this as an integral over $\MM_{g,n}$
by regarding the $z^{}_i$'s and the local coordinates
as functions of the coordinates of $\MM_{g,n}$. 
This will entail replacing the $dz^{}_i$'s by
$(\p z^{}_i / \p t_k) dt_k$ from the beginning,
where $t_k$ are the coordinates of 
$\MM_{g,n}$. However the advantage of regarding the $\Omega^{(g,n)}_p$'s as $p$-forms
on $\wt\PP_{g,n}$ is that we can integrate $\Omega^{(g,n)}_p$
over any $p$-dimensional subspace
of $\wt\PP_{g,n}$. In particular we can use this to carry out `vertical integration'
i.e.\ integration over fibers keeping the base point in $\MM_{g,n}$ fixed. This is
necessary for example if we want to choose the integration cycle such that parts
of it involves moving the picture changing operators keeping the point in $\MM_{g,n}$
fixed.

\begin{figure}
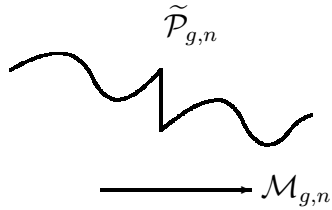

\begin{center}
\figvertical
\end{center}
\caption{Pictorial representation of a subspace of $\wt\PP_{g,n}$ containing a 
vertical segment that contains a tangent vector along the fiber.
\label{figvertical}}

\end{figure}

A special case of this that will be of interest to us is as follows. 
Suppose that on a codimension one 
subspace $\KK$ of $\MM_{g,n}$, we turn the integration cycle in the 
vertical direction so that at every point in $\KK$ we change the 
location $z^{}_i$ of a particular picture changing operator from its initial value
$u$ 
to some final value $v$ keeping fixed the local coordinates at the punctures and
the locations of the other picture changing operators. 
This has been depicted in Fig.~\ref{figvertical}.
Both $u$ and $v$, as well as the local coordinates at the punctures and the
locations of other picture changing operators can of course vary along
$\KK$. In this case  we can label the vertical part of the integration cycle 
by the coordinates of
$\KK$ and the coordinate $z^{}_i$ along the fiber, and we can integrate 
$\Omega^{(g,n)}_{6g-6+2n}$
over this part on the integration cycle. 
This integration can be performed by 
first integrating along the fiber labelled by $z^{}_i$
and then integrating the result along $\KK$.
Integration along the fiber will give
\be 
\int_{u}^{v} dz^{}_i \, \Omega^{(g,n)}_{6g-6+2n}\left[{\p\over \p z^{}_i}\right]\, .
\ee
Now it follows from \refb{ekvalue}-\refb{eexpomp} that
contraction with $\p/\p z^{}_i$ effectively 
replaces the $\VVV(z^{}_i) - \p\xi (z^{}_i)  d z^{}_i$ term in
\refb{ekvalue} by 
$-\p \xi(z^{}_i)$. 
Since the rest of the
operators entering the definition of $\Omega^{(g,n)}_p$ 
have no dependence on $z^{}_i$, the
integration along the fiber direction 
produces an insertion of
\be 
-\int_u^v dz^{}_i \,\p\xi(z^{}_i) = (\xi(u) - \xi(v))\, ,
\ee
into the correlation function. 
This replaces the $\VVV(z^{}_i) - \p\xi (z^{}_i)  d z^{}_i$ term in
\refb{ekvalue}. Note in particular that the result depends only on the initial and final
values of $z^{}_i$ and not on the contour in the $z^{}_i$-plane
along which we integrate. At the same time it must be noted that this operator belongs to
the small Hilbert space since it represents the difference in $\xi$ at two points.
The rest of the insertions into the correlation
function can be determined from the contraction of $\Omega^{(g,n)}_{6g-7+2n}$  with the tangent
vectors of the image of $\KK$ in $\wt\PP_{g,n}$.

Note that the  above result is valid only if  we vary
the location of only one picture changing operator. If we want to use the above
result to move the
locations of several picture changing operators then this can be done by
making the vertical segment composed of several parts, and in each part we
vary the location of only one picture changing operator.\footnote{If this breaks some
symmetry {\it e.g.} permutation symmetry among external punctures, we can
always average over various possibilities. This will increase the number of
$A^{(\alpha)}$'s.}
The result will depend on the
order in which we vary the locations of the picture changing operators, but this can
be addressed in the same way as the effect of a general variation in the choice of 
integration cycle as
described in \S\ref{spicture}.

For reasons that will become clear in \S\ref{snonhol}, we need to define the
superstring amplitude by integration over integration cycles containing such
vertical segments. In all subsequent discussion we shall be working with such 
general choice of integration cycle.

\subsection{Construction of gluing compatible integration cycles} \label{sgluing}

An algorithm for constructing gluing compatible sections in bosonic string theory has been 
described in \S3.2, \S3.3 of \cite{1401.7014}. We can follow a similar procedure for constructing
gluing compatible integration cycles in superstring theory and use it to divide the contributions
to a given off-shell amplitude into one particle irreducible (1PI) and one particle
reducible (1PR) parts. First of all on three punctured spheres and all one punctured
tori parametrized by the torus modular parameter $\tau$ we make some specific choice
of the local
coordinates at the punctures subject to the symmetry that permutes the 
three punctures. A class of choices can be found in \cite{kugo,saadi,9206084}. 
We also fix the
location of the picture changing operator  consistent with this symmetry -- this may
require averaging over more than one set of choices as represented by the sum over
$\alpha$ in \refb{ekinsert}. Also for reasons to be clear later we shall take the locations
of the picture changing operators to be in the region $|w_a|>1$ for each $a$ where 
$w_a$ denotes the local coordinate around the $a$-th puncture.  This of course 
presupposes that the $|w_a|<1$ regions are sufficiently small so that they do not cover the
whole surface but this can always be done by scaling the $w_a$'s by a sufficiently
small number $\lambda$, and we shall only work with such choices of local coordinates.
We declare the
three punctured sphere and all one punctured tori to be 1PI Riemann surfaces.
We can now glue pairs of these 1PI surfaces using the plumbing fixture relations
to construct four punctured spheres and two punctured tori\footnote{We can also
construct zero punctured genus two Riemann surfaces by gluing a pair of one
punctured tori, but these play no role in our analysis.} and choose the local
coordinates at the punctures and the locations of the picture changing operators
to be those induced from the original Riemann surfaces which are being glued.
Since the picture changing operators were taken to be in the region $|w_a|>1$ for
each puncture including the ones that are being used for gluing, it follows that
after gluing they are at distinct points on the final Riemann surface for all $s$ and
$\theta$ as long as we take $s\ge 0$. 
During this construction we treat the external punctures as distinct so that {\it e.g.}
while gluing two three punctured spheres to get a four punctured sphere we
get separate contributions from $s$, $t$ and $u$ channel diagrams. 
We call
the family of Riemann surfaces obtained this way
1PR Riemann surfaces and declare the rest of the
four punctured spheres and two punctured tori as 1PI Riemann surfaces. 
We now make some choice of local coordinates and picture changing 
operators (possibly including vertical segments) on the
new 1PI Riemann surfaces consistent with  permutation symmetry,
the $|w_a|>1$ conditions for the locations of the picture changing operators and
the requirement that both the choice of local coordinates and the locations of the
picture changing operators smoothly match onto those on 1PR Riemann surfaces on
the codimension one subspace of the moduli space that forms the common boundary
of the moduli space of 1PI Riemann surfaces and the moduli space of 1PR Riemann
surfaces.
We
continue this process by constructing, for any given value of $g$ and $n$,
all 1PR Riemann surfaces by gluing two or more 1PI Riemann surfaces with
lower $g$ and / or $n$.
At each stage the Riemann surfaces which cannot be obtained by gluing 
two or more 1PI surfaces of lower genera / lower number of punctures are declared
to be 1PI.

Once the division into 1PI and 1PR Riemann surfaces have been made, the 1PI
off-shell amplitudes are defined by restricting the integral of $\Omega^{(g,n)}_{6g-6+2n}$
to run over 1PI Riemann surfaces only. The rest of the contributions to the amplitude are
defined to be 1PR.
Clearly this division depends on the choice of local coordinates
on the 1PI surfaces, but 
the analysis of \cite{1311.1257,1401.7014} shows that the
physical
renormalized masses and S-matrix elements are independent of the choice
of local coordinates.\footnote{For some choice of local coordinates on 1PI surfaces
it may happen that a given 1PR Riemann surface may appear more than once as a
result of gluing {\it e.g.} once from $s$-channel diagram and once from $t$-channel
diagram in the case of four punctured sphere. In that case the definition of 1PI family
of surfaces will require us to subtract this contribution. This can be avoided by
scaling the choice of local coordinates of the original 1PI surfaces by some small
number $\lambda$ which reduces the size of the moduli space covered by the 1PR
family of Riemann surfaces. A definite choice of local coordinate system that
avoids this can be found in \cite{9206084}.}

\subsection{Infrared regulator} \label{sinfra}

The off-shell amplitudes can have infrared divergences from separating type 
degenerations represented by the $s\to\infty$ limit of \refb{egluing}. These
can be divided into two kinds -- generic degenerations where the momentum
flowing through the punctures being glued is general off-shell momentum and
special degenerations where the momentum flowing through the punctures being glued is
forced to be zero or on-shell\cite{1307.5124}. In the case of generic degenerations -- which
also include all non-separating type degenerations -- we regulate the
divergence as $s\to\infty$ by making an analytic continuation $s\to i\, s$ and
include a damping factor $e^{-\eps s}$ in the integral\cite{1307.5124}. 
On the other hand for special
degeneration we restrict the $s$ integral by some upper cut-off $\Lambda$\cite{1209.5461}. 

Degenerations which glue two Riemann surfaces each of which carries two or more
external punctures are always treated as generic.
Degenerations which glue two Riemann surfaces of which one has only one external
puncture (other than the puncture that is being glued) correspond to mass and
wave-function renormalization. If the mass is renormalized then we'll have to work
with off-shell amplitudes where we
take the external momentum carried by the particle to be at a generic off-shell 
value\cite{1311.1257,1401.7014} 
and hence we are forced to treat this as a generic degeneration. On the other hand 
if the mass is not renormalized then we can keep the momentum carried by the 
state on-shell and treat this as a special degeneration. However we can also work
with off-shell external momentum and treat ths as a generic degeneration and take
the momentum on-shell at the end of the computation. Presumably both methods 
will lead to the same result for the wave-function renormalization factor
at the end but a formal proof of this has not been given. In any case since the
wave-function renormalization factor is not a physical observable and, in particular,
depends on the choice of local coordinates, this is not a pressing issue.

Separating type degenerations in which one of the Riemann surfaces has no external
puncture and the other carries all external punctures are always treated as special, since
the momentum flowing through the punctures that are glued is forced to be zero.

\subsection{Effect of changing the locations of the picture changing operators} 
\label{spicture}

\begin{figure}

\begin{center}
\figrboundary
\end{center}

\vskip -1.3in 

\caption{Illustration of \refb{erboundary} and \refb{echangeinf}
} \label{frboundary}

\end{figure}

Once we have chosen a gluing  compatible integration cycle, we need 
to follow the procedure of
\cite{1311.1257,1401.7014} to show that even though 
the off-shell amplitudes depend on the
choice of local coordinate system at the punctures, physical quantities like
the renormalized mass or the S-matrix elements are independent of this choice.
A new question that arises for superstrings is: 
how does the amplitude depend on the choice of the
locations of picture changing operators? 
We can address this question together with the old question: how does the 
amplitude depend on the choice of local coordinates? Both correspond to a change
in the integration cycle of $\wt\PP_{g,n}$
on which we integrate to find the amplitude.
If $S$
and $\wh S$ are two integration cycles and $R$ is the region bounded by them then 
we have (see Fig.~\ref{frboundary})
\be \label{erboundary}
\p R = \wh S - S + (\p R)'\, ,
\ee
where $(\p R)'$ is the intersection of $R$ with the boundary of
$\wt\PP_{g,n}$ containing degenerate Riemann surfaces.
This gives
\be \label{exx23}
\int_{\wh S} \Omega^{(g,n)}_{6g-6+2n}(|\Phi\rangle) - \int_{S} \Omega^{(g,n)}_{6g-6+2n} (|\Phi\rangle)
= \int_{R} d\Omega^{(g,n)}_{6g-6+2n}(|\Phi\rangle) - \int_{(\p R)'} \Omega^{(g,n)}_{6g-6+2n}(|\Phi\rangle)
\, .
\ee
Using \refb{etotal} we can express the right hand
side of \refb{exx23} as
\be \label{echange}
- \int_{R} \Omega^{(g,n)}_{6g-5+2n}(Q_B|\Phi\rangle)
- \int_{(\p R)'} \Omega^{(g,n)}_{6g-6+2n}(|\Phi\rangle) \, .
\ee
It is useful to write down the result for
infinitesimal change which can be
parametrized 
by some infinitesimal vector 
$V_f$ of the tangent space of $\PP_{g,n}$ at every point on the original
integration cycle $S$, labelling the displacement between $S$ and $\wh S$.
Clearly $V_f$ is defined only
up to the addition of a tangent vector of $S$, but this will not affect the final
result.  
In that case \refb{echange} takes the 
form
\be \label{echangeinf}
- \int_{S} \Omega^{(g,n)}_{6g-5+2n}(Q_B|\Phi\rangle)[V_f]
+ \int_{\p S}  \Omega^{(g,n)}_{6g-6+2n}(|\Phi\rangle)[V_f]
\, ,
\ee
where $\p S$ denotes the intersection
of $S$ with the boundary of $\wt\PP_{g,n}$ containing degenerate Riemann
surfaces.

From now on we shall focus 
on the effect of changing the locations of the picture changing operators, but the
arguments given below can be easily generalized to give an 
alternative analysis of the effect of the change in the local coordinate system, leading to
the same results as in \cite{1311.1257,1401.7014}.
First
let us consider the first term in \refb{echangeinf}. 
For a gluing compatible choice of coordinate system, we can follow the procedure
reviewed in \S\ref{sgluing} to break up the integral over $S$ 
as sum of 1PI contributions, two 1PI contributions joined by a propagator, three 1PI
contributions joined by two propagators etc. It will be useful for our analysis to express
the 1PR contributions  in terms of the
constituent 1PI contributions. The basic identity that allows us to do this 
can be derived by analyzing  
a region of the moduli space where the Riemann surface $\Sigma$ 
is constructed by gluing two Riemann surfaces $\Sigma_1$ and $\Sigma_2$ by
plumbing fixture.
In this case $V_f$ can be expressed as $V_f^{(1)}+V_f^{(2)}$ where $V_f^{(1)}$
captures the effect of changing the locations of the picture changing operators on
$\Sigma_1$ and $V_f^{(2)}$ denotes the effect of changing the locations of the
picture changing operators on $\Sigma_2$. In this case  using a sllight generalization of
\refb{emeasurefactorAsup} to include contraction with the vector $V_f$ field and that 
\be
Q_B(|\Phi_1\rangle\otimes |\Phi_2\rangle)= 
(Q_B|\Phi_1\rangle)\otimes |\Phi_2\rangle +
|\Phi_1\rangle\otimes Q_B|\Phi_2\rangle\, ,
\ee
where $|\Phi_1\rangle$ and $|\Phi_2\rangle$ denote states inserted at the 
external punctures 
of $\Sigma_1$ and $\Sigma_2$ respectively,
one can 
show that when restricted to the integration cycle $S$, we have
\ben \label{emeasurefactorvF}
&& \Omega^{(g,n)}_{6g-5+2n}(Q_B|\Phi\rangle)[V_f] |_S \nonumber \\
&=& -{1\over 2\pi} \bigg[\sum_{i,j} \Omega^{(g_1,n_1)}_{6g_1-5+2n_1}(Q_B|\Phi_1\rangle
\otimes |\vp_i\rangle)[V_f^{(1)}]_{S_1} 
\, \wedge  \, \Omega^{(g_2,n_2)}_{6g_2-6+2n_2}( |\vp_j\rangle\otimes |\Phi_2\rangle
)_{S_2} \nonumber \\ &&
\qquad \wedge\, ds \wedge d\theta  
\times \langle \vp_i^c | b_0^+ b_0^- 
e^{-s(L_0+\bar L_0)} e^{i\theta(L_0-\bar L_0)} |\vp_j^c\rangle
\nonumber \\ &&
- \sum_{i,j} \Omega^{(g_1,n_1)}_{6g_1-6+2n_1}(Q_B|\Phi_1\rangle
\otimes |\vp_i\rangle)_{S_1}\, \wedge  \, \Omega^{(g_2,n_2)}_{6g_2-5+2n_2}( |\vp_j\rangle
\otimes|\Phi_2\rangle
)[V_f^{(2)}]_{S_2}  \nonumber \\ &&
\qquad \wedge\, ds \wedge d\theta  
\times \langle \vp_i^c | b_0^+ b_0^- 
e^{-s(L_0+\bar L_0)} e^{i\theta(L_0-\bar L_0)} |\vp_j^c\rangle
\nonumber \\ &&
-  \sum_{i,j} \Omega^{(g_1,n_1)}_{6g_1-5+2n_1}(|\Phi_1\rangle
\otimes |\vp_i\rangle)[V_f^{(1)}]_{S_1} \, \wedge  \, \Omega^{(g_2,n_2)}_{6g_2-6+2n_2}(
 |\vp_j\rangle\otimes Q_B|\Phi_2\rangle
)_{S_2} \nonumber \\ &&
\qquad \wedge\, ds \wedge d\theta  
\times \langle \vp_i^c | b_0^+ b_0^- 
e^{-s(L_0+\bar L_0)} e^{i\theta(L_0-\bar L_0)} |\vp_j^c\rangle
\nonumber \\  &&
+  \sum_{i,j} \Omega^{(g_1,n_1)}_{6g_1-6+2n_1}(|\Phi_1\rangle
\otimes |\vp_i\rangle)_{S_1}\, \wedge  \, \Omega^{(g_2,n_2)}_{6g_2-5+2n_2}(
|\vp_j\rangle \otimes Q_B|\Phi_2\rangle
)[V_f^{(2)}]_{S_2}  \nonumber \\ &&
\qquad \wedge\, ds \wedge d\theta  
\times \langle \vp_i^c | b_0^+ b_0^- 
e^{-s(L_0+\bar L_0)} e^{i\theta(L_0-\bar L_0)} |\vp_j^c\rangle\bigg]\, .
\een

If $\Sigma_1$ and $\Sigma_2$ are 1PI contributions then the right hand side of
\refb{emeasurefactorvF} is already expressed in terms of 1PI contributions. If
$\Sigma_1$ and/or $\Sigma_2$ are obtained as a result of gluing 1PI
Riemann surfaces
with lower genus and/or lower number of punctures then we need to again express 
the right hand side in terms of the amplitudes on these Riemann surfaces by making 
repeated use of \refb{emeasurefactorAsup} and
\refb{emeasurefactorvF}. At the end we get the result in
terms of 1PI amplitudes.

Let us now analyze the contribution from the second term in \refb{echangeinf} that
involves an integral over $\p S$. We shall assume that the 
integration cycle $S$ (as well as the
deformed integration cycle generated by the vector field $V_f$) has been chosen
in a modular invariant fashion so that when we regard the base space $\MM_{g,n}$
as the fundamental domain in the Teichmuller space, the contributions from the
apparent boundaries, whose different components are related to each other by
modular transformations, cancel. Under this assumption, 
 the relevant component of $\p S$ arises from separating type 
 degenerations of the Riemann surfaces\cite{1209.5461}.
Near $\p S$ the Riemann surface $\Sigma$ is described as
the result of gluing two Riemann surfaces using the plumbing fixture relation
\refb{egluing}, with the degeneration corresponding to the $s\to\infty$ limit. 
As discussed in \S\ref{sinfra}, we can divide these into two kinds -- the generic
degeneration where the momentum flowing through the degenerating punctures is 
a generic off-shell momentum and the special degeneration in which the momentum
flowing through the node is either zero or satisfies a classical on-shell condition.
For a generic degeneration we make an analytic continuation $s\to i s$ 
and introduce a damping factor $e^{-\eps s}$ in the integral. As a result
the $s$ integral is convergent and there are no boundary contributions from the
$s\to \infty$ end due to the damping factor. The same argument can be used to rule out
boundary contributions from non-separating type degenerations.
On the other hand for special degenerations we shall regulate the infrared divergence
by putting a sharp cut-off $s\le \Lambda$ on the $s$ integral
and let $\theta$ run over the full range $0\le \theta<2\pi$. 
This boundary contribution can be computed using a slight generalization of
\refb{emeasurefactorBsup} and gives
\ben \label{omegaboundary}
&& \Omega^{(g,n)}_{6g-6+2n}(|\Phi\rangle)[V_f] |_{S;s=\Lambda} \nonumber \\
&=&   -{1\over 2\pi} \, \bigg[-\, d\theta \wedge \sum_{i,j} \Omega^{(g_1,n_1)}_{6g_1-5+2n_1}(|\Phi_1\rangle
\otimes |\vp_i\rangle)[V_f^{(1)}]|_{S_1} \, \wedge  \, \Omega^{(g_2,n_2)}_{6g_2-6+2n_2}(
|\vp_j\rangle\otimes |\Phi_2\rangle
) |_{S_2}
\nonumber \\ &&
\times \langle \vp_i^c | b_0^- 
e^{-\Lambda (L_0+\bar L_0)} e^{i\theta(L_0-\bar L_0)} |\vp_j^c\rangle
\nonumber \\ &&
+\, \,  d\theta \wedge \sum_{i,j} \Omega^{(g_1,n_1)}_{6g_1-6+2n_1}(|\Phi_1\rangle
\otimes |\vp_j\rangle)|_{S_1}\, \wedge  \, \Omega^{(g_2,n_2)}_{6g_2-5+2n_2}(|\vp_i\rangle
\otimes |\Phi_2\rangle
)[V_f^{(2)}] |_{S_2}
 \nonumber \\ &&
\times \langle \vp_j^c | b_0^- 
e^{-\Lambda (L_0+\bar L_0)} e^{i\theta(L_0-\bar L_0)} |\vp_i^c\rangle\bigg]\, .
\een
Note that in the second term in \refb{omegaboundary}
we have exchanged the roles of $|\vp_i\rangle$ and
$|\vp_j\rangle$ compared to the convention used in \refb{emeasurefactorvF}.
This will be useful later.

\subsection{Extension to type II strings} \label{ssuper}

The extension of this formalism to type II string theory is in principle straightforward.
Now \refb{edefomp} will have to be replaced by 
\be \label{edefompsuper}
\Omega^{(g,n)}_p(|\Phi\rangle) = (2\pi i)^{-(3g-3+n)}\, \langle \Sigma| \BB_p |\Phi\rangle\, ,
\quad \BB_p\equiv \sum_{r,s=0\atop
r,s\le 2g-2+n}^{p} K^{(r)}\wedge \bar K^{(s)} \wedge B_{p-r-s}\, ,
\ee
where $\bar K^{(s)}$ involving the left-moving (anti-holomorphic) fields is defined 
in the same way as $K^{(r)}$. While choosing an integration cycle,
in general the weight factors $\bar A^{(\alpha)}$ and
the locations $\bar z^{(\alpha)}_i$ of the anti-holomorphic picture changing operators can be 
chosen to be independent of their holomorphic counterparts.
The rest of the analysis will proceed as in the case of heterotic string theory with the
understanding that
the vector field
describing change of locations of the picture changing operators
will now correspond to a vector of the form 
$ \sum_{i} \delta z^{}_i {\p / \p z^{}_i}
+ \sum_{i} \delta \bar z^{}_i {\p / \p \bar z^{}_i}$.

\sectiono{Dealing with spurious poles}
\label{snonhol}

Appearance of the second term on the right hand side of \refb{echangeinf}
shows that 
when we change the location of the picture changing operators, the integration
measure in $\MM_{g,n}$ changes by a total derivative term
even for on-shell amplitudes for which $Q_B|\Phi\rangle=0$. 
This is related to the fact that when we regard the amplitude as the result of an
integral over the supermoduli space, we have to make a choice of the integration
cycle that determines the even nilpotent part of the bosonic 
moduli. 
For example if we have a supermoduli space with one bosonic coordinate
$x$ and a pair of fermionic coordinates $\xi,\eta$, then for defining an
integral of the form
\be \label{ex.1}
\int dx d\xi d\eta f(x,\xi, \eta)\, ,
\ee
we need to pick an `integration cycle'
\be \label{ex.2}
x = u + h(u) \xi \eta
\ee
where $u$ is an ordinary bosonic variable and $h(u)$ is some arbitary function
labelling the even nilpotent part of $x$. We then
substitute \refb{ex.2} into \refb{ex.1} and integrate over $u,\xi,\eta$. One finds that 
after $\xi,\eta$ integration we are left with an integrand that is a function of $u$, but
it depends on the function $h(u)$ through a total derivative term.

The consequence of this ambiguity on integration over the supermoduli space
has been discussed extensively in 
\cite{ARS,global,catoptric,Witten,1209.5461,1304.2832}. We shall not review this here,
but only mention that
this makes the
computation of the amplitude in fermionic string theory
tricky under certain circumstances -- namely when the
supermoduli space is not holomorphically projected\cite{donagi-witten}.
Resolution of this subtlety has been described in \cite{1209.5461} by expressing the 
amplitude as integral over the supermoduli space.
In  the formalism involving
picture changing operators related subtleties arise in the form of 
spurious poles\cite{Verlinde:1987sd}
-- poles in the integrand which depend on the locations of the picture changing operators.
Since the dependence on the locations of the picture changing operators is a total
derivative in the moduli space, one would naively expect that these singularities are
`fake' as their locations can be moved around by adding total derivative terms in the
integrand over the moduli space. Nevertheless we have to find a systematic way of
dealing with these singularities. This will be the main goal of this section.
In the last paragraph of \S\ref{savoid} we shall briefly discuss the relation between
our approach and the one suggested in \cite{1209.5461}.

\subsection{Spurious poles}

We shal now briefly review the origin of the spurious poles\cite{Verlinde:1987sd}. 
They arise from the correlation functions of the $e^{q\phi}$, $\eta$
and $\xi$ operators. Regarded as a function of the locations of these operators,
these correlation functions have zeroes and poles controlled by the operator product
expansions of these fields but also have poles at points where no two operators
are coincident. 
This is seen by explicitly writing down the expression for 
an arbitrary correlator of these fields on a genus $g$ Riemann 
surface. Up to an overall normalization it takes the 
form\cite{Verlinde:1987sd,lechtenfeld,morozov}
\ben \label{espurious}
&& \Big \langle \prod_{i=1}^{n+1} \xi(x_i) \prod_{j=1}^n \eta(y_i) 
\prod_{k=1}^m e^{q_k \phi(z_k)} 
\Big\rangle_\delta
\nonumber \\
&=& {\prod_{j=1}^n \vt[\delta](-\vec y_j + \sum \vec x -\sum \vec y + \sum q \, \vec z - 2\vec\Delta)\over
\prod_{i=1}^{n+1} \vt[\delta](-\vec x_i + \sum \vec x -\sum \vec y + \sum q \, \vec z - 2\vec\Delta)}
{\prod_{i<i'} E(x_i, x_{i'}) \, \prod_{j<j'} E(y_j, y_{j'})\over
\prod_{i,j} E(x_i, y_j) \, \prod_{k<\ell} E(z_k, z_\ell)^{q_k q_\ell} 
\prod_k \sigma(z_k)^{2 q_k} 
}\, , \nonumber \\
&& \sum_{k=1}^m q_k = 2(g-1)\, .
\een 
In this equation $\delta$ stands for the spin structure, $\vt$ denotes the
genus $g$ theta functions, $E(x,y)$ denotes the prime form, $\sigma(z)$ is
a ${1\over 2} g$ differential representing the conformal anomaly of the ghost
system and $\vec\Delta$ is the Riemann class characterizing the divisor of zeroes
of the theta function. A detailed explanation of all of these quantities can be
found in \cite{Verlinde:1987sd,VerlindeChiral}. 
$\sum \vec x$, $\sum \vec y$ and $\sum q\, \vec z$ denote respectively $\sum_{i=1}^{n+1} \vec x_i$, 
$\sum_{j=1}^n \vec y_j$ and $\sum_{k=1}^m q_k \vec z_k$ with
\be
\vec x \equiv \int_p^x \vec \omega\, ,
\ee
where $\vec \omega$ is a $g$-dimensional vector of holomorphic one forms
on the Riemann surface and $p$ is an arbitrary point on the Riemann surface 
(with dependence on $p$ compensated by $p$-dependence of $\vec\Delta$).

Note that on the left hand side of \refb{espurious} we have one more
$\xi$ compared to $\eta$. This reflects that the correlation function has been
written in the `large Hilbert space' in which  on
any Riemann surface there is a
$\xi$ zero mode
that needs to be soaked up. In all operators that we use for
computing amplitudes -- vertex operators of external states, picture changing operators,
BRST operators etc.\ -- $\xi$ always appears in the combination $\p\xi$.  As these
do not carry any zero mode of $\xi$, we need to insert an explicit factors of
$\xi(z_0)$ in the correlation function to soak up the $\xi$ zero mode. Since the
correlation function is independent of $z_0$, normally we work in the `small 
Hilbert space' where we do not display the $\xi(z_0)$ factor
in the correlation function. Indeed
if we take the derivative of \refb{espurious} with respect to
$n$ of the $n+1$ $x_i$'s, then the correlation function becomes manifestly independent
of the last remaining $x_i$. We can then drop this from the argument and get the
correlation function in the `small Hilbert space'. In our analysis we shall always work
in the small Hilbert space. In particular, even though in some expressions we may use the
operator $\xi$ without any derivative, they will always appear in a combination
$\xi(A)-\xi(B)$ so that it is really a shorthand for $\int_A^B \p\xi(z) dz$.

The prime form $E(x,y)$ has a simple zero at $x=y$ and various prime
forms in \refb{espurious} capture the zeroes and poles associated with the
short distance singularities / zeroes in the operator product of the various
fields. The zeroes of the
$\prod_{i=1}^{n+1} \vt[\delta](-\vec x_i + \sum \vec x -\sum \vec y + \sum q \, \vec z - 2\vec\Delta)$ factor
in the denominator are responsible for the spurious poles. 
With some effort one can see that when we use \refb{eboserule} and
\refb{espurious} to compute the correlation
functions of $\beta$ and $\gamma$ this denominator factor becomes independent of
the locations of $\beta$ and $\gamma$\cite{lechtenfeld,morozov}. 
Thus there are no spurious poles as functions of the
arguments of $\beta$ and $\gamma$ operators. Since the BRST current is constructed from
$\beta$ and $\gamma$ we see that there are no spurious poles in the
argument of the BRST current either. However in the expressions for picture
changing operators there are `bare' $\p\xi$, $\eta$ and $e^{q\phi}$ factors which cannot
be expressed as polynomials of (derivatives of) $\beta$ and $\gamma$, and hence the
correlators  will have spurious singularities as functions of the locations of the
picture changing operators.
Physically
these spurious 
singularities can be traced to the fact that the gauge choice for the
world-sheet gravitinos, which lead to some particular insertion of 
picture changing operators
on the world-sheet, fail to be a good choice of gauge precisely when the correlator
of these operators hits a spurious singularity. 

Let us now examine how the spurious poles affect the construction of off-shell
amplitudes. Let us suppose that we have made some gluing compatible
choice of integration cycle in $\wt \PP_{g,n}$ for defining these amplitudes.
This corresponds to specifying the
locations of the picture changing operators (and choice of local coordinates
at the punctures) as a function of the moduli labelling
$\MM_{g,n}$. At a generic point of $\MM_{g,n}$ the 
$\prod_{i=1}^{n+1} \vt[\delta](-\vec x_i + \sum \vec x -\sum \vec y + \sum q \, \vec z - 2\vec\Delta)$ factor
in the denominator of \refb{espurious} is not expected to vanish, and hence the
integration measure is non-singular at a generic point. 
However since we need to satisfy one complex equation for
this factor to vanish, we expect to encounter spurious poles over a real
codimension
two
subspace of the moduli space.

There are also other more obvious singularities which depend on the locations
of the picture changing operators, {\it e.g.} the ones arising from collision of a pair
of picture changing operators or the collision of a picture changing operator with 
a vertex operator at a puncture. All of these occur on codimension two subspaces 
of the moduli space and the method we shall describe will deal with all these
singularities in the same way.

\subsection{Dealing with the spurious poles}  \label{savoid}

\begin{figure}
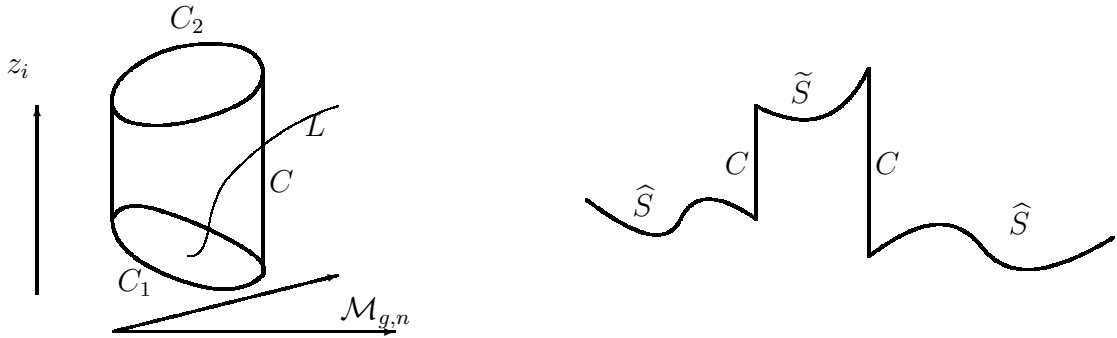


\vskip -.9in

\begin{center}
\hbox{\hskip .2in \figvertsec \hskip .8in  \figtwodsection}
\end{center}

\caption{A pictorial representation of integration along a vertical segment. The
two horizontal directions in the left hand side 
figure represent coordinates along $\MM_{g,n}$ transverse
to the codimension two subspace on which we have spurious pole and
the vertical direction labels the location $z^{}_i$ of a picture changing operator.
All other coordinates of $\wt\PP_{g,n}$
have been suppressed.
The integration cycle consists of 3-pieces -- a section $\wh S$
(which will be a two dimensional surface 
in this representation but not shown) whose inner boundary is the
curve $C_1$, the vertical cylindrical surface $C$ bounded by $C_1$ and $C_2$, 
and another section $\wt S$ (not shown) whose outer boundary is 
the curve $C_2$. 
The right hand figure shows the intersection of this integration
cycle with a vertical plane
where we see the three parts $\wh S$, $C$ and $\wt S$
of the integration cycle explicitly.
The thin curve marked $L$ in the left hand figure
describes the location of the spurious pole. 
As is clear from this figure, both the sections $\wh S$ and $\wt S$
can avoid the spurious pole. Integration over $C$ will encounter the
spurious pole, but as this is expressed as a difference between an integral along $C_1$
and an integral along $C_2$, this also avoids the spurious pole.
} \label{figvertsec}

\end{figure}

Our goal will be to describe a procedure for integrating through 
these spurious singularities.
Suppose we have a section of $\wt\PP_{g,n}$ which
encounters spurious singularities on a real codimension two subspace $\NN$ 
of the base
$\MM_{g,n}$.
Let us consider a tubular neighborhood $T$ surrounding $\NN$.
Outside $T$ the integrand has no spurious poles. Our prescription will be to
turn the integration cycle
along a `vertical direction'  -- in the sense described in \S\ref{svertical} --
as we reach the boundary of $T$ on the base $\MM_{g,n}$.
This corresponds to changing the location $z^{}_i$ of one of the
picture changing operators keeping fixed the locations of the other picture changing operators,
local coordinates around the punctures and the coordinates of $\MM_{g,n}$.
This is done for every point on $\p T$. Thus the vertical segment is a $6g-6+2n$
dimensional subspace of $\wt\PP_{g,n}$ labelled by the coordinates of $\p T$ and the
contour along which $z^{}_i$ varies.
The vertical segment is continued till the final arrangement of the picture changing operators
as a function of the coordinates on $\p T$
are such that there are no longer any spurious poles inside $T$. 
At that point we can turn the
section `horizontal' and integrate over the interior of $T$. This has been shown pictorially
in Fig.~\ref{figvertsec} where the projection of the curves $C_1$ and $C_2$ 
on $\MM_{g,n}$ correspond to $\p T$ and the projection of the interior of the
cylindrical region $C$ on $\MM_{g,n}$ corresponds to $T$.

Now naively one might expect this procedure to run into the spurious singularities
as we integrate along
the vertical direction. After all the location of the spurious poles in $\MM_{g,n}$ must
vary continuously as we change the locations of the picture changing operators, and
since in the initial configuration there are spurious poles in the interior of $T$ and 
in the final configuration
there are no spurious poles in the interior of $T$, the poles 
must pass through the boundary
of $T$ as we change the location $z^{}_i$ of the $i$-th picture changing
operator.  This suggests that the orbit of the spurious pole(s) must cross the
vertical segment. This has been shown by the thin line $L$ in Fig.~\ref{figvertsec}.
On the other hand the formalism of \S\ref{svertical} shows that integrating
along the vertical segment, corresponding to integrating $z^{}_i$
from $u$ to $v$ (say), requires us to replace the $\VVV(z^{}_i)
-\p \xi(z^{}_i) d z^{}_i$ factor in \refb{ekvalue}  by
$-\int_{u}^{v} \p \xi(z)  dz= (\xi(u)-\xi(v))$. 
Thus the result depends 
on only $u$ and $v$ and not on the path connecting $u$ to $v$. In particular
since for $\xi(u)$ insertion the spurious poles are in the interior of $T$ and for
$\xi(v)$ insertion the spurious poles are outside $T$, neither the contribution 
involving $\xi(u)$
nor the contribution involving $\xi(v)$ has any singularity on $\p T$. 
This shows that the result of the vertical integration is free
from any spurious singularity.

It is worth examining this in more detail. The point is that if instead of 
$\xi(v) - \xi(u)$ we use the expression $\int_{u}^{v} \p \xi(z)  dz$ then
somewhere along the contour $z$ reaches a point where the location of the
spurious pole reaches $\p T$ causing the integration measure to diverge.
However since the correlation function of $\xi(z)$ given in \refb{espurious} 
is single valued, the integral
$\int_{u}^{v} \p \xi(z)  dz$ can be carried out through this pole unambiguously,
leading to correlation function involving $(\xi(v) - \xi(u))$ 
which is manifestly free from any poles on $\p T$.
This gives a systematic procedure for dealing 
with spurious poles in the computation
of off-shell amplitudes.

\begin{figure}
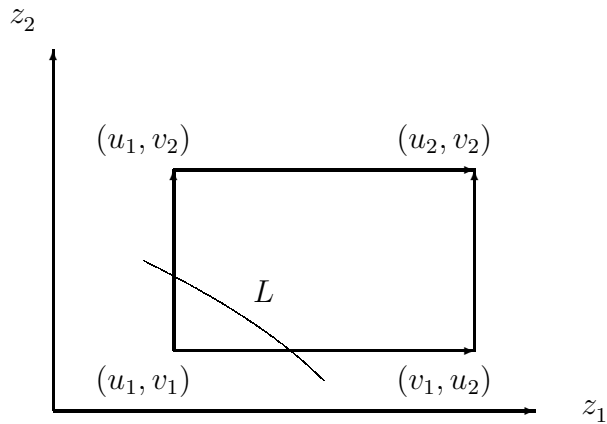


\begin{center}
\figsquare
\end{center}

\vskip -.5in

\caption{The two integration cycles in the $z^{}_1$-$z^{}_2$ plane.
\label{figsquare}
}

\end{figure}

This procedure is of course not completely
unambiguous since it depends on which
$z^{}_i$ we choose to vary to move the spurious singularity. To see
the effect of this, let us compare the results for two different vertical
segments of the integration cycle, in each of which the net
effect is to change the locations of two picture changing operators -- which we
shall take to be $z^{}_1$ and $z^{}_2$ -- from $(u_1, v_1)$
to $(u_2, v_2)$ in a way that moves the spurious singularity out. 
In the first case, once we
reach a point on $C_1$, we first change the location $z^{}_1$ from $u_1$
to $v_1$ to move the spurious singularity out, and then change $z^{}_2$
from $u_2$ to $v_2$.  In the second case we first change $z^{}_2$ from $u_2$
to $v_2$ to move the spurious singularity out, and then change $z^{}_1$
from $u_1$ to $v_1$. This has been shown in Fig.~\ref{figsquare} where we have
also displayed the movement of the spurious singularity by the thin line $L$. We now
calculate the difference between the integral of $\Omega^{(g,n)}_{6g-6+2n}$
over these two different cycles.
Formally this can be expressed as in \refb{echange} -- the only issue is whether
the two terms in \refb{echange} are free from the divergences
associated with the spurious poles. First we consider the first term  of \refb{echange}. 
It is clear that this can be obtained by
first integrating $\Omega^{(g,n)}_{6g-5+2n}(Q_B|\Phi\rangle)$  over the square in the 
$z^{}_1$-$z^{}_2$ plane shown in Fig.~\ref{figsquare} 
and then integrating the result over the
image of $\p T$ in $\wt\PP_{g,n}$ sans the directions labelled by
$z^{}_1$ and
$z^{}_2$. The integral in the 
$z^{}_1$-$z^{}_2$ plane
will correspond to the insertion of
\be \label{enov}
\int_{u_1}^{v_1}d z^{}_1  \int_{u_2}^{v_2}  d z^{}_2\,
\p \xi(z^{}_1) \, \p\xi(z^{}_2 )
= (\xi(u_1) - \xi(v_1)) \, (\xi(u_2) - \xi(v_2))\, .
\ee
Since there are no spurious singularities at the corner points of the square in
the $z^{}_1$-$z^{}_2$ plane this gives a finite result.
Similarly
the second term 
of \refb{echange} can be evaluated by first performing the integral of
$\Omega^{(g,n)}_{6g-6+2n}$ over the $z^{}_1$-$z^{}_2$ plane, leading to the 
insertion of \refb{enov} into the correlation function, and then integrating this
over the image in $\wt\PP_{g,n}$ of the intersection of $\p T$ with the 
compactification boundary. The latter arise from setting $s=\Lambda$ for some
plumbing fixture variable $s$.

This shows that the difference in the off-shell amplitude for two different choices
of vertical segment shown in Fig.~\ref{figsquare} can be expressed as 
\refb{echange} with finite expressions for both terms. However
since \refb{enov} is not infinitesimal, we do not have the analog of
\refb{echangeinf} which will be needed 
in
\S\ref{sspecial}
to prove that these contributions do not affect renormalized masses and
S-matrix elements. So we need to find a way to express this as a result of
successive infinitesimal changes. This can be done 
by taking a family of generalized integration cycles labelled by a 
continuous paramater $t$ such that the following conditions hold:
\begin{enumerate}
\item For each $t$ the integration cycle is a formal
weighted average of several integration cycles differing in their vertical segments.
\item For each $t$ the integration cycles satisfy the gluing compatibility condition
\refb{epic1}.
\item At $t=0$ and $t=1$ the integration cycles coincide with the original
integration cycles which we wanted to show are equivalent.
\end{enumerate}
The effect of infinitesimal change from $t$ to $t+\delta t$ will now involve
the insertion of terms like
\refb{enov} into the correlation functions as before after
carrying out the integral in the $z^{}_1$-$z^{}_2$ plane, but the
result will be multiplied by a factor of $\delta t$. 
Since this is an infinitesimal deformation, its effect can now 
be
analyzed as in \S\ref{sspecial} to show that these contributions do not affect the
renormalized masses and S-matrix elements.

More generally the rules for choosing picture changing operators 
will involve dividing up the moduli space into subregions, choose
the picture changing operators in each subregion as smooth functions of the moduli
ensuring that they do not encounter any spurious singularity, and at the boundary
of two such subregions use the prescription of vertical integration to 
interpolate between the two picture changing operator arrangements in the
two subregions. When two or more such boundaries meet there can be additional
subtleties since the vertical path across the different boundaries may not be
compatible, leaving a `gap' in the integration cycle in $\wt\PP_{g,n}$.\footnote{I wish
to thank E.~Witten for raising this point.} We then have to `fill this gap' by adding
new vertical components to the integration 
cycle in which two of the coordinates are along the `vertical direction'
as in \refb{enov}. In unpublished work with E.~Witten it has been shown that 
it is possible to find a consistent procedure for filling these gaps.

One also needs to check that this prescription for dealing with spurious
poles is consistent with gluing compatibility. 
Suppose we have chosen the integration cycles on
two families of 
1PI Riemann surfaces avoiding spurious poles. Now consider an 1PR contribution
obtained by joining the two families using plumbing fixture. 
The choice of integration cycles on the original
families of Riemann surfaces induces an integration cycle on the family of glued
Riemann surfaces. Does this automatically avoid the spurious poles? This can be
guaranteed by scaling the local coordinates at the punctures by a sufficiently
small number so that 1PR surfaces always describe Riemann surfaces close
to degeneration for the whole range $0\le s <\infty$. In this case the 
theta function on the glued surface,
responsible for the spurious poles, can be approximated by
products of theta functions on the original surfaces and hence the locations
of the spurious poles will be close to those on the original Riemann surfaces.
Thus as long as the original integration cycles were chosen to avoid the spurious 
poles on the original surface, the induced integration cycle will avoid the
spurious poles on the glued surface.

The prescription for avoiding the spurious poles given here
may be related to the one suggested in \cite{global}, i.e.\ taking the $z^{}_i$'s
to be holomorphic function of the moduli around the neighbourhood of the spurious
poles and then excluding a small tubular neighbourhood around the spurious poles
while carrying out the integral over $\MM_{g,n}$. However we shall not explore 
the relation here.

Finally we can also explore the connection between the prescription given here and
the suggestion of \cite{Verlinde:1988tx,1209.5461} of 
performing integration over the supermoduli space by
dividing it into open sets in each of which we can choose a good slice
for integration over the supermoduli, and then expressing the full result as sum of
contributions from such open sets using partition of unity. To see the
connection of this to the formalism employed here we note that in Fig.~\ref{figvertsec}
the two sections $\wh S$ and $\wt S$ can be regarded as two such open sets
and the extra term, coming from integration over the vertical segment, can be interpreted
as capturing the effect of decreasing the weight given to the choice of section
corresponding to $\wh S$ from 1 to 0 across $\p T$ and
at the same time increasing the weight given to the choice of section
corresponding to $\wt S$ from 0 to 1. Thus the prescription for dealing with the
spurious poles, as given here, seems to be in consonance with the one described in
\cite{Verlinde:1988tx,1209.5461}.

\sectiono{Mass renormalization and S-matrix of special states} \label{sspecial}

Using the method of \cite{1311.1257} one can show that the renormalized masses and 
S-matrix elements
of special
states are independent of the choice of local coordinates at the punctures as
long as we define the off-shell amplitudes following the procedure described in
\S\ref{s2}. 
We shall now demonstrate how the formalism
can be used to prove that the same physical quantities  
are also independent of the choice of locations of the picture changing 
operators.\footnote{We should 
add that the analysis of this and the next section has now been carried out in a much
more systematic manner in \cite{1411.7478,appear2} 
using the notion of one particle irreducible effective
string field theory.}

\subsection{Definition of special states} \label{sdefspecial}

We shall follow the notation of \cite{1311.1257} as closely as possible 
so that whenever possible
we can refer the reader to the analysis of \cite{1311.1257}. 
Let us consider a string theory on $R^{D,1}\times \KK$ where 
$\KK$ is a compact space and $R^{D,1}$ is $D+1$ dimensional Minkowski
space.
Then $SO(D)$ is the 
little group of a particle at rest. Let $G$ be the internal symmetry group of the
theory (if any).
By definition,  special states at mass level $m$ in the rest frame are described by
off-shell vertex operators with the following properties:
\begin{enumerate}
\item They have 
the form 
\be \label{evop}
W_\pm = c\, \bar c \, e^{-\phi} e^{\pm i k_0 X^0} V\, ,
\ee
where $V$'s are GSO odd superconformal primary operators of dimension
$(1+\alpha' m^2/4, 1/2+\alpha' m^2/4)$ constructed
from matter fields and transform in some set of irreducible representations 
$R_1,\cdots R_s$ of 
$SO(D)\times G$.
\item 
{\it All} zero momentum
matter vertex operators in the representation $R_i$
of $SO(D)\times G$ for $1\le i\le s$ have total conformal dimension 
\be \label{edefspecial}
h \ge  {3\over 2} +\alpha' {m^2\over 2}\, .
\ee
Furthermore if the equality in \refb{edefspecial} holds then the operator 
describes a special state when substituted into \refb{evop}.
\end{enumerate}
The on-shell
condition for the vertex operators $W_\pm$ given in \refb{evop}
is $k^2 = -m^2$, showing that
$m$ is the tree level mass of the state. 
We shall denote by $n_p$ the total number of special states with a given tree level
mass $m$. In general this may include states from more than one irreducible representation
of $SO(D)\times G$ i.e.\ $s$ can be larger than 1.
Special states are generalization of states lying on the leading Regge trajectory.

\begin{figure}
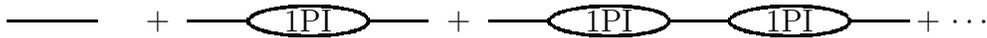

\begin{center}
\figanew
\end{center}

\vskip -1in

\caption{Pictorial representation of the full two point Greens function 
as a sum of 1PI
amplitudes connected by propagators.
\label{fnewa}
}
\end{figure}

\subsection{Renormalized mass}

Now we consider the net contribution 
to the two point Green's function of two arbitrary off-shell 
external states 
$|\Psi_1\rangle, |\Psi_2\rangle\in \HH_0$ satisfying the 
additional restrictions that $b_0^+|\Psi_i\rangle=0$
and that they carry momentum $\pm k=\pm (k^0,\vec 0)$.
Using \refb{emeasurefactorAsup} repeatedly, this can be expressed 
as sum of 1PI contributions glued together by
propagators as
shown in Fig.~\ref{fnewa}. This leads
to the following expression for the quantum corrected propagator 
\be \label{epropag}
\Pi=\Delta + \Delta \wh \FF \Delta + \Delta \wh \FF \Delta \wh\FF \Delta + \cdots
= \Delta + \Delta\FF\Delta \, ,
\ee
\be \label{effexp}
\FF \equiv \wh\FF + \wh\FF \Delta \wh\FF + \wh\FF \Delta \wh\FF\Delta \wh\FF + \cdots
= (1 - \wh\FF\Delta )^{-1}\wh\FF \, .
\ee
Here, as in \cite{1311.1257},  $\wh\FF$ is the 1PI contribution to the 2-point amplitude 
which includes the $b_0^\pm$ factors appearing in \refb{emeasurefactorAsup}
and is regarded as an operator acting on  states in $\HH_0$
carrying momentum $k$ and
annihilated by $b_0^+$. 
$\Delta$ is the tree level propagator\footnote{This definition of $\Delta$ 
differs from the definition used
in \cite{1311.1257,1401.7014} by a normalization factor of 2. This has the effect that in the $\alpha'=1$
unit, $\Delta$ restricted to mass level $m$ states is given by $2/(k^2+m^2)$. However we can
avoid this annoying factor of $2$ if we choose $\alpha'=2$.}
\be \label{eDelta}
\Delta  = {1\over 2\pi} \int_0^{2\pi} d\theta 
\, \int_0^\infty ds\, e^{-s (L_0+\bar L_0) +i\theta(L_0-\bar L_0)}=
 \delta_{L_0, \bar L_0}\, \int_0^\infty ds\, e^{-s (L_0+\bar L_0)}\, .
\ee
$\FF$ is the full off-shell two point amplitude.
The off-shell two point amplitude of two external states $|\Psi_1\rangle$ and $|\Psi_2\rangle$ 
defined in \S\ref{soffshell}, or its 1PI counterpart defined in \S\ref{sgluing}, can be obtained
by taking the matrix element of ${1\over 2} c_0\bar c_0 \FF$ and 
${1\over 2} c_0\bar c_0 \wh\FF$ between the
states $\langle\Psi_1|$ and $|\Psi_2\rangle$ -- the 
${1\over 2} \bar c_0 c_0$ factor being needed
due to the fact that we have absorbed a $b_0^+ b_0^-$ factor in the definition of
$\FF$ and $\wh\FF$.
A useful property of the operators $\FF$, $\wh\FF$ and $\Delta$ is that they
preserve ghost number and picture number, i.e.\ acting on a state of ghost number
$g$ and picture number $p$ they give back a state of ghost number $g$ and picture 
number $p$. They also preserve $SO(D)\times G$ symmetry, i.e. acting on a state in
representation $R$ of $SO(D)\times G$, they give back states in the same 
representation. 

If $P$ denotes 
the projection operator into $n_p$ dimensional subspace of special states with
tree level mass $m$, then the off-shell two point Green's function of special states
is obtained by multiplying \refb{epropag} by $P$ from both sides. In $\alpha'=2$ unit this  
is given in terms of $\FF$ by
\be  \label{epropagator}
P \Pi(k) P = (k^2+m^2)^{-1} P + (k^2+m^2)^{-1} P \FF P (k^2+m^2)^{-1}\, .
\ee
We can try to evaluate this using \refb{effexp}. The poles in $k^2$ near $-m^2$
come
from the explicit factors of $(k^2+m^2)^{-1}$ in \refb{epropagator} and the
poles of $\Delta$ in \refb{effexp}.
The
only intermediate states which can generate  poles at $k^2=-m^2$ from the
$\Delta$ factors in \refb{effexp} are special states themselves
since due to \refb{edefspecial}
other states at the same mass level transform in different representations of 
$SO(D)\times G$ and hence cannot mix with the special states\cite{1311.1257}.
This allows us to `integrate out'
all states other than the special states in \refb{effexp} following the procedure described in
\cite{1311.1257}, resum the series, and express \refb{epropagator} as
\be \label{especprop}
P \Pi(k) P  =
Z^{1/2}(k) (k^2 + M_p^2)^{-1} Z^{1/2}(k)^T\, ,
\ee
where $Z^{1/2}(k)$ is an $n_p\times n_p$ matrix which has no poles near $k^2=-m^2$ 
and $M_p$ is an $n_p\times n_p$ diagonal matrix.
The eigenvalues of
$M_p$ give physical renormalized masses of the special states and
$Z^{1/2}(k)$ evaluated at $k^2=-M_p^2$ correspond to wave-function renormalization factors.

\subsection{Effect of changing the locations of picture changing operators} \label{spic}

We shall now consider the change in the two point amplitude $\FF$ under an
infinitesimal change in
the locations of picture changing operators. For this we shall use 
\refb{echangeinf}. 
For now we shall proceed ignoring the boundary contribution described in the second
term in \refb{echangeinf}, -- its effect will be discussed separately in \S\ref{sboundary}.
The first term has two types of contributions -- one where $Q_B$ operates on the
external vertex operator on the left and the other where $Q_B$ operates on the 
external vertex
operator on the right. 
Let us focus on the terms where $Q_B$ acts on the external vertex operator
on the left.
Even though eventually we shall be interested in taking
both external states to be special states, let us for now restrict only the external state
on the left to be one of the $n_p$ special states carrying momentum $(k^0, \vec k=0)$
described by a vertex operator of type $W_+$ given in \refb{evop}. 
Since $Q_B W_+$ is
proportional to $(k^2+m^2)$ we can take out a factor of $(k^2+m^2)$ -- which is needed
to cancel the external tree level propagator factor\cite{1311.1257} -- and call the
rest of the contribution
$\delta \HH$.  If $n_p$ is the number of special states
at a given mass level then $\delta\HH$ is $n_p\times \infty$ dimensional matrix.
Now by repeated use of \refb{emeasurefactorvF} and \refb{emeasurefactorAsup}
$\delta\HH$ can be decomposed into sum of products of 1PI contributions and
propagators as in \refb{effexp}, but the effect of contraction with $V_f$ in
\refb{echangeinf}
will be to replace 
$\Omega^{(g_i,n_i)}_{6g_i-6 + 2n_i}$ by $\Omega^{(g_i,n_i)}_{6g_i-5+2n_i}[V_f]$
in {\it one of the} 1PI components,
representing the effect of moving the picture changing operators on that particular
component of the Riemann surface.
In order to write down an expression like \refb{effexp} for $\delta\HH$
we shall first
define several new quantities:
\begin{figure}
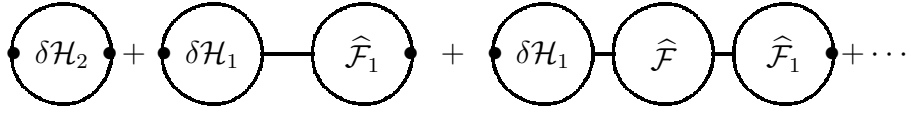

\begin{center}
\figatwo
\end{center}

\vskip -1in

\caption{Pictorial representation of the expression for $\delta \wh\HH$ given in
\refb{ehathh}. The black dots denote external states without propagator factors.
\label{figatwo}
}
\end{figure}

\begin{enumerate}
\item $\delta \HH_1$ denotes the 1PI contribution to the two point function with one
external state $(k^2+m^2)^{-1} Q_B W_+$ and the other external state arbitrary.
\item $\delta \HH_2$ will be defined in the same way as $\delta \HH_1$ but 
we replace $\Omega^{(g,n)}_p$ by $\Omega^{(g,n)}_{p+1}[V_f]$ in the integration measure.
\item $\wh\FF_1$ will be defined in the same way as $\wh \FF$, but 
we replace $\Omega^{(g,n)}_p$ by $\Omega^{(g,n)}_{p+1}[V_f]$ in the integration
measure \refb{edefomp}.
\item Next we define
\be \label{ehathh}
\delta \wh \HH = \delta \HH_2 + \delta \HH_1 \Delta \wh\FF_1 + \delta \HH_1 \Delta
\wh \FF \Delta \wh\FF_1 + \cdots = \delta \HH_2 + \delta\HH_1 \Delta (1 - \wh\FF \Delta)^{-1}
\wh\FF_1\, .
\ee
A pictorial representation of this has been shown in Fig.~\ref{figatwo}.
\end{enumerate}
In terms of these quantities we can now express $\delta \HH$ as
\be \label{efindh}
\delta \HH = \delta \wh \HH  + \delta \wh\HH \Delta \wh \FF + \delta \wh\HH \Delta \wh \FF
\Delta \wh \FF + \cdots = \delta \wh \HH (1 - \Delta \wh \FF)^{-1}
\ee
which has been pictorially represented in Fig.~\ref{figathree}. It is easy to see by
inspection that Fig.~\ref{figathree}, with $\delta\wh\HH$ defined via Fig.~\ref{figatwo},
sums over all possible contributions to $\delta\HH$.

\begin{figure}
\begin{center}
\figathree
\end{center}

\vskip -1in

\caption{Pictorial representation for the expression for $\delta\HH$ given in
\refb{efindh}.
\label{figathree}
}
\end{figure}

We shall now prove that $\delta\wh\HH$ has no poles at $k^2=-m^2$. For this
we examine each term on the right hand side of \refb{ehathh}. Since $\delta\HH_2$
gets contribution from 1PI amplitudes, it has no poles. In the second term we could 
in principle  get
a pole at $k^2=-m^2$ from a state of momentum $(k^0,\vec k=0)$
propagating in $\Delta$.
Let us denote the vertex operator of such a state by $e^{ik_0 X^0} \OO_g \wt V$
where $\OO_g$ is some ghost sector operator and $\wt V$ is a matter sector operator
with zero momentum. In order
to contribute to the pole at $k^2=-m^2$, 
$\OO_g  \wt V$ should have dimension $(\alpha'm^2/4, \alpha'm^2/4)$
since $e^{ik_0 X^0}$ has dimension $(\alpha'k^2/4, \alpha'k^2/4)$. Now
$\delta\HH_1$ has an insertion of $Q_B W_+$ on the left for some special
state vertex operator $W_+$. Since $Q_BW_+$ is a vertex operator of ghost
number 3 and picture number $-1$, by the conservation of ghost and picture number
the states that contribute to $\Delta$ must also have ghost number 3 and picture number
$-1$.  Since the states are annihilated by $b_0$ and $\bar b_0$, the 
minimum dimension ghost sector operator with ghost number 3 and
picture number $-1$ is $c\bar c \eta$ with total conformal weight $-1$.
Thus in order for $\OO_g\wt V$ to have dimension $(\alpha'm^2/4, \alpha'm^2/4)$, $\wt V$ 
must have
total dimension less than or equal to
\be \label{ebound}
 1+ \alpha'{m^2\over 4}  + \alpha'{m^2\over 4}  = {1} + \alpha'{m^2\over 2}\, .
\ee
But by $SO(D)\times G$ symmetry, $\wt V$ must belong to the same
representation as $W_+$ which is one of the $R_i$'s.
Hence its conformal weight has a lower bound
given in \refb{edefspecial}. We now see that \refb{ebound} is inconsistent with 
\refb{edefspecial}. Hence
there is no operator $\wt V$ that can contribute a pole to the second term on the
right hand side of \refb{ehathh}. 
Since $\wh \FF$ does not change the $SO(D)\times G$ transformation law of the
state, the same argument can be used to show that none of the  factors of $\Delta$
in the other terms on the right hand side of \refb{ehathh} can produce a pole at
$k^2=-m^2$. Thus $\delta \wh\HH$ is free from poles at $k^2=-m^2$.

We now note that \refb{effexp} and \refb{efindh} are identical to eqs.(3.19) of
\cite{1311.1257}, except that in \cite{1311.1257} $\delta$ was used to denote a change
induced by a change in the choice of local coordinate system. Also the
absence of poles in $\delta\wh\HH$ matches the similar property of
$\delta\wh\HH$ in \cite{1311.1257}.
Thus from here on one can repeat arguments identical to those given in
\cite{1311.1257} to show that the change $\delta\HH$  in the two point function 
given in
\refb{efindh} can be absorbed into a change in the wave-function renormalization factor
of the state associated with the special state vertex operator inserted on the left,
and
the renormalized masses do not change under a change of
the locations of the picture changing operators. Similar analysis would establish that
the effect of the terms in \refb{echangeinf}
in which $Q_B$ acts on the external state on the right can be absorbed into a 
wave-function renormalization of the external state on the right.

\subsection{S-matrix}

\begin{figure}
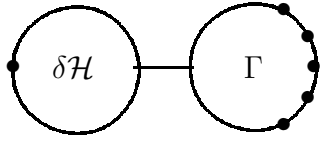

\begin{center}
\figy
\end{center}

\vskip -1in

\caption{A contribution to the change in the S-matrix under the change in the location of the
picture changing operators. \label{figy}}
\end{figure}

Similar argument can be used to show that the S-matrix of special states  
are also invariant under a change in the locations of the
picture changing operators. For this we use 
\refb{echangeinf} to
manipulate the change in the
$n$-point amplitude in a way similar
to that for the two point amplitude. 
As in the case of mass renormalization, we postpone discussion of 
the boundary terms arising from the second
term in \refb{echangeinf} to \S\ref{sboundary}.
So we only have to examine the first term. This can be expressed as a sum
of terms, in each of which  $Q_B$ acts on 
a particular external special state. 
If $W$ denotes the vertex operator of this state 
then the term involving $Q_BW$ can be manipulated in the same way as in
the case of mass renormalization, expressing it as sums of products of amplitudes 1PI in 
momentum $k$ and propagators carrying momentum $k$. The
resulting contribution 
can be expressed 
as the sum of diagrams shown in Fig.~\ref{figathree}
attached by a propagator $\Delta$ 
to an amplitude $\Gamma$ that is 1PI in momentum $k$ and carries all other
external states (see Fig.~\ref{figy})
plus a term without any pole near $k^2=-m^2$. The term without pole does not
contribute to the S-matrix.
As in \cite{1311.1257}, 
the contribution of Fig.~\ref{figy}
can be shown to cancel against the 
change in the
wave-function renormalization factor $Z(k)^{-1/2}$ that appears in the expression for the S-matrix,
establishing that the S-matrix is invariant under a change of the locations of the picture
changing operators.

\subsection{Boundary contributions} \label{sboundary}

In this section we shall study the effect of possible boundary contributions 
corresponding to the second term on the right hand side of \refb{echangeinf}.
The analysis of this section should be regarded as an iterative procedure
since we need to use some of the results of  \S\ref{sdecoupling} at lower genus / lower
number of punctures. Also this analysis is independent of the nature and number
of external states we have and holds for general external states.

As discussed above \refb{omegaboundary},
possible sources of this boundary contribution are from special
degenerations where a given Riemann surface degenerates to two Riemann
surfaces, with 
the momentum flowing through the
degenerating punctures forced to vanish or satisfy tree level on-shell condition for some state.
As long as we work in the off-shell formalism where 
the momenta of all the external states are generic and off-shell, the
only possible contribution involves the case where the momentum flowing through the degenerating
punctures vanishes.
This requires that one of the
component Riemann surfaces  has no punctures (except the one 
corresponding to the degenerating puncture) and
all the external states are inserted at the punctures of the other component. 

The relevant boundary contribution is given in \refb{omegaboundary}. Without any loss of
generality we can take the surface $\Sigma_1$ to be the one without any external puncture and
the surface $\Sigma_2$ to contain all the external punctures. 
Thus we have $n_1=1$.
Now ghost and picture number
conservation laws given in \refb{eghnopicno} tells us that in both terms on the right hand side of
\refb{omegaboundary} $|\vp_i\rangle$ must carry picture number $-1$ and ghost number 3
whereas $|\vp_j\rangle$ must have picture number $-1$ and ghost number 2. Furthermore
the presence of $b_0^-$ factors in the propagator tells us that both $|\vp_i^c\rangle$ and
$|\vp_j^c\rangle$ must be annihilated by $c_0^-$ and hence both $|\vp_i\rangle$ and
$|\vp_j\rangle$ must be annihilated by $b_0^-$. 
Both $|\vp_i\rangle$ and $|\vp_j\rangle$ must be GSO even since the rest of the operators inserted
in the correlations functions are GSO even. The integration over the angular variable
$\theta$ projects into states with $L_0=\bar L_0$.
Finally since eventually we shall take the upper
cut-off $\Lambda$ on the $s$-integral
to infinity, the non-vanishing contributions to the boundary terms come from
states with $L_0=\bar L_0\le 0$. Thus we must classify all possible candidates for $|\vp_i\rangle$
and $|\vp_j\rangle$ consistent with these properties. In fact the choices of $|\vp_i\rangle$ and
$|\vp_j\rangle$ are correlated due to the requirement that
$ \langle \vp_i^c | b_0^- 
|\vp_j^c\rangle$ should be non-zero in order to get a non-vanishing contribution. 
This is best implemented by choosing 
a basis of states 
such that for each $|\vp_j\rangle$ satisfying the restrictions given above,
there is a unique $|\vp_i\rangle$ for which
$\langle \vp_j^c|b_0^-|\vp_i^c\rangle$ is non-zero (or equivalently 
$\langle \vp_j|c_0^-|\vp_i\rangle$ is non-zero).
We shall now
list the possible choices of $|\vp_i\rangle$ and $|\vp_j\rangle$
in heterotic
string theory following \cite{catoptric}:
\begin{enumerate}
\item $\vp_j= c\bar c e^{-\phi} V, \quad \vp_i = (\p c + \bar \p \bar c) c\bar c e^{-\phi}
V$ where $V$ is a GSO
even matter sector vertex operator of dimension $(h+1/2, h)$ with $h\le 1/2$. 
Absence of tachyons in the
spectrum of physical states tells us that the only operators of this kind present in the theory
have dimension
(1,1/2), and the corresponding $\vp_j$ describes a vertex operator of a zero momentum
physical massless state. 
We shall assume that this represents a moduli field with vanishing potential -- 
the case where the
field has a potential can be analyzed following the procedure described in 
\cite{1404.6254}.
In this case 
\be \label{eterm1}
\int_{\MM_{g_1,n_1=1}} \Omega^{(g_1,1)}_{6g_1-6+2}( |\vp_j\rangle)|_{S_1}
\ee
represents a zero momentum tadpole of a physical massless state at genus $g_1$.
Assuming that the vacuum we are working with is stable, this zero momentum tadpole must vanish and
hence the second term on the right hand side of \refb{omegaboundary} vanishes after
integration over $\MM_{g_1, 1}$.
On the other hand
\be \label{eterm2}
\int_{\MM_{g_2,n_2}} \Omega^{(g_2,n_2)}_{6g_2-6+2n_2}(|\Phi_2\rangle
\otimes |\vp_j\rangle)|_{S_2}
\ee
represents the effect of inserting a zero momentum massless external state in the genus $g$ 
amplitude of the external states. 
This multiplied by any constant represents the effect of shifting the vacuum expectation
value of the corresponding field by that constant.
Thus for this choice of $|\vp_j\rangle$
the first term on the right hand side of
\refb{omegaboundary} can be interpreted as the 
result of shifting the expectation value of
this massless state by an amount proportional to\cite{catoptric,1209.5461}
\be \label{eterm3}
\int_{\MM_{g_1, 1}} \Omega^{(g_1,1)}_{6g_1-5+2}( |\vp_i\rangle)[V_f^{(1)}]|_{S_1}\, .
\ee
This is turn can be interpreted as a field redefinition of the corresponding 
scalar field.
\item  $\vp_j = {1\over 2} c\eta+ \bar c\bar\p^2 \bar c c\p\xi e^{-2\phi}$, $\vp_i =
(\p c + \bar \p \bar c) ({1\over 2} 
c\eta- \bar c\bar\p^2 \bar c c\p\xi e^{-2\phi})$. In this case
we have
\be \label{epsijpuregauge}
\vp_j = \{ Q_B, (\p c + \bar \p \bar c) c\p\xi e^{-2\phi}\}\, .
\ee
This is a pure gauge state. Thus by the result of \S\ref{sdecoupling} at a lower
genus / lower number of punctures,
\refb{eterm1} vanishes for this choice of $\vp_j$
whereas the effect of \refb{eterm2} 
can be absorbed into a wave-function renormalization of external states. 

\item $\vp_j = {1\over 2} c\eta- \bar c\bar\p^2 \bar c c\p\xi e^{-2\phi}$, $\vp_i  =
(\p c + \bar \p \bar c) ({1\over 2} c\eta+ \bar c\bar\p^2 \bar c c\p\xi e^{-2\phi})$.
In this case $|\vp_j\rangle$ is a BRST invariant state and represents zero
momentum dilaton in the $-1$ picture. 
Thus \refb{eterm1} now has the interpretation 
of a dilaton tadpole. In a consistent background this must vanish. On the other
hand \refb{eterm2} can be interpreted as the result of a zero momentum dilaton insertion
into the amplitude. Thus for this choice of $|\vp_j\rangle$
the first term on the right hand side of
\refb{omegaboundary} can be interpreted as the 
result of shifting the expectation value of
the dilaton by an amount proportional to \refb{eterm3}\cite{catoptric,1209.5461}.
Equivalently we can regard this as a field redefinition of the dilaton field.

\item $\vp_j = (\partial c +\bar\partial \bar c) c \p \xi e^{-2\phi} \bar c U$,
$\vp_i = c\eta \bar c U$ where $U$ is a dimension (1,0) GSO even
operator in the matter sector. 
If the matter CFT has a discrete symmetry which changes the sign of $U$ 
but leaves the super stress tensor invariant,
then the terms in \refb{omegaboundary} involving correlation function on
$\Sigma_1$, -- the $\Omega^{(g_1,1)}_{6g_1-5+2}( |\vp_i\rangle)[V_f^{(1)}]|_{S_1}$ factor 
in the first term and the $\Omega^{(g_1,1)}_{6g_1-6+2}( |\vp_j\rangle)|_{S_1}$ in the
second term --  vanishes. This is so {\it e.g.}\ in ten dimensional flat
space-time or toroidal compactification where $U=\p X^M$ for some compact 
or non-compact coordinate $X^M$, and there is always a symmetry that reverses the
sign of $X^M$ together possibly with some other $X^N$'s and their superpartners. 
Most known string compactifications have this property.
Henceforth we 
shall restrict our analysis to those theories in which the contribution from
this term to the right hand side of \refb{omegaboundary} vanishes.\footnote{At
genus one  the relevant correlator on the
torus in the matter CFT 
involves just the one point function of the $U(1)$ current $U$. 
Due to translational symmetry on the torus 
we can replace this by the contour integral
of $U$ along the $a$-cycle. In this case we can represent the correlator in the matter
sector as a trace over all fields weighted by the $U$-charge and 
$e^{2\pi i (\tau L_0 -\bar\tau\bar L_0)}$. This receives equal and opposite contributions
from the CPT conjugate states and hence always vanishes.}

\item $\vp_j = (\p c +\bar \p \bar c) c e^{-\phi} V$, 
$\vp_i = c e^{-\phi}  V \bar c \bar\p^2 \bar c$ where
$V$ is a dimension $(0,1/2)$ GSO odd matter operator. 
Again if the matter CFT has a discrete symmetry under which
$V\to -V$ but the super-stress-tensor remains invariant
then the terms in \refb{omegaboundary} involving correlation function on
$\Sigma_1$, -- the $\Omega^{(g_1,1)}_{6g_1-5+2}( |\vp_i\rangle)[V_f^{(1)}]|_{S_1}$ factor 
in the first term and the $\Omega^{(g_1,1)}_{6g_1-6+2}( |\vp_j\rangle)|_{S_1}$ in the
second term -- vanishes. This is so {\it e.g.} in ten dimensional flat
space-time or toroidal compactification where $V=\psi^M$ for some fermionic 
field $\psi^M$ which is the superpartner of some compact 
or non-compact coordinate $X^M$, and there is always a symmetry that reverses the
sign of $(\psi^M,X^M)$ together possibly with some other $X^N$'s and their superpartners. 
Again most known string compactifications have this property.
Henceforth we shall restrict our analysis to those theories for which this amplitude
vanishes.

\end{enumerate}

\subsection{Decoupling of pure gauge states} \label{sdecoupling}

\begin{figure}
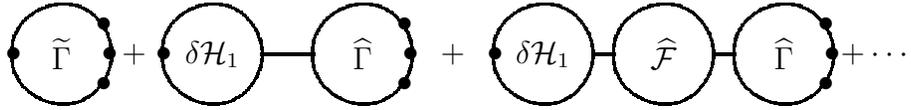

\begin{center}
\figatwonew
\end{center}

\vskip -1in

\caption{Result of organizing the amplitude with $Q_BW$ as an external state
into product of terms 1PI in momentum $k$ and propagators carrying momentum $k$.
The momentum $k$ enters through the external state represented by the left black dot
in each diagram.
\label{figatwonew}}

\end{figure}

An argument very similar to the one given in \S\ref{spic}
can be used to show the vanishing of the
S-matrix elements involving one or more pure 
gauge states of the form $Q_B|\Lambda\rangle$ 
and the rest of the states corresponding to special states. 
Our starting point is again \refb{etotal}. 
Up to 
boundary terms which arise from
the integration over total derivative term $d\Omega^{(g,n)}_{p-1}$ and can be treated as in
\S\ref{sboundary},
we can transfer the BRST operator 
from over $\Lambda$ to the other external states which we have assumed to be special states. 
Let us 
denote the
vertex operator of such a special state by $W$ and the momentum carried by it
by $k$ and analyze the term involving the insertion of $Q_BW$ to the amplitude.
We can decompose the amplitude into sum of products of terms which are
1PI in the momentum $k$ and propagators carrying momentum $k$. 
The resulting organization of the amplitude takes the form shown in
Fig.~\ref{figatwonew}, with $\wh\Gamma$ denoting a component that is
1PI in momentum $k$ and carries all other external states and $\wt\Gamma$
denoting a contribution to the full amplitude that is 1PI in momentum $k$ and carries
all external states. $\delta\HH_1$ is the same quantity that appears in  \S\ref{spic}.
The same argument as before now tells
us that none of these diagrams  have any poles at $k^2+m^2=0$ where $m$ is
the tree level mass of the special states, and hence after 
adding them we do not get a pole at $k^2=-m_p^2$. Thus the contribution from these
diagrams to the S-matrix vanishes.
The boundary terms can be analyzed as in \S\ref{sboundary} with $\Lambda$ insertion now
playing the role of contraction with $V_f$. There is an added simplification in that as
long as $\Lambda$ carries generic momentum, it can only be inserted on the surface
$\Sigma_2$ where the rest of the vertex operators are inserted. Thus the analog of the
terms in \refb{omegaboundary} involving $V_f^{(1)}$ will be absent.

\begin{figure}
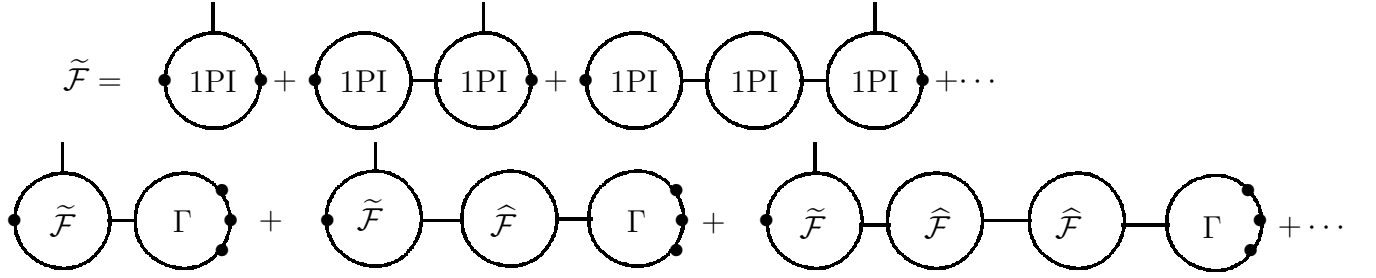


\vskip -.5in 

\figxa

\vskip -.8in

\figxb

\vskip -.8in

\caption{The first line gives the definition of $\wt\FF$ with the external left black dot
in each graph denoting the insertion of $Q_B W$ carrying momentum $k$
and the vertical external line on top denoting 
the insertion of the zero momentum vertex operator
$\Lambda=(\p c + \bar \p \bar c) c\p\xi e^{-2\phi}$ that appears in \refb{epsijpuregauge}.
It follows from the analysis similar to that for $\delta\wh\HH$
in \S\ref{spic} that $\wt\FF$ has no poles near $k^2+m^2=0$
despite the appearance of internal propagators carrying momentum $k$ from the second term
onwards.
The second line gives the diagrams that could generate poles near $k^2+m^2=0$
in the amplitude containing $Q_BW$, $\Lambda$ and other special states as external
legs.
Here $\Gamma$ is 1PI in momentum $k$ and carries all external states other than
$Q_BW$ and $\Lambda$. Since the net contribution of the second line can be interpreted as
the result of  multiplying the external propagator of the Green's function 
by $\wt\FF$ from the left, the effect of this is to change the wave-function
renormalization factor $Z^{1/2}$ of the special state described by 
the vertex operator $W$
by a factor proportional to $\wt\FF$.
\label{figxa}
}
\end{figure}

The same arguments can be used to show that the insertion of \refb{epsijpuregauge}
to an amplitude gives vanishing contribution to the S-matrix element, but a few 
additional terms need to be analyzed. First of all, since \refb{epsijpuregauge} carries
zero momentum, the term containing $Q_BW$ insertion to the amplitude can have
poles from diagrams  shown in the second line of
Fig.~\ref{figxa}. However as discussed in the
caption of this figure, the effect of these terms can be absorbed into a redefinition of the
wave-function renormalization factor of the special state described by the vertex operator $W$.
The second difference is that in the boundary terms, 
the operator $\Lambda = -(\p c + \bar \p \bar c) c\p\xi e^{-2\phi}$ that is left
after stripping off $Q_B$ from \refb{epsijpuregauge} can be inserted into $\Sigma_1$
as well as on $\Sigma_2$ 
and hence we have the analog of both terms that appear in \refb{omegaboundary}.
This however does not pose any difficulty since the boundary terms may be analyzed
in the same way as in \S\ref{sboundary}.

\subsection{Type II string theory}

The analysis for type II string theory proceeds in the same way except that the list
of operators $\vp_i$ and $\vp_j$ which can appear in the sum over states in
\refb{omegaboundary} are different. We shall now analyze their contributions.
\begin{enumerate}
\item $\vp_j= c\bar c e^{-\phi} e^{-\bar \phi} V, \quad \vp_i = 
(\p c + \bar \p \bar c) c\bar c e^{-\phi} e^{-\bar\phi}
V$ where $V$ is a left-GSO and right-GSO
odd matter sector vertex operator of dimension $(h, h)$ with $h\le 1/2$. 
Absence of tachyons in the theory tells us that the only possible value of
$h$ is 1/2 in which case $\vp_j$ describes the vertex operator of a physical
massless state. We can then follow the analysis used in the case of heterotic string
theory to show that as long as tadpoles of massless fields vanish, boundary
contributions involving these operators can be absorbed into a shift of the vacuum
expectation values of the massless fields.

\item  $\vp_j = {1\over 2} (c\eta \bar c \bar \p\bar \xi e^{-2\bar \phi}
+  \bar c \bar\eta c\p\xi e^{-2\phi})$, $\vp_i =
{1\over 2} (\p c + \bar \p \bar c) (c\eta \bar c \bar \p\bar \xi e^{-2\bar \phi}
+ \bar c \bar\eta c\p\xi e^{-2\phi})
$. 
In this case
we have
\be \label{epsijpuregaugeii}
\vp_j = \{ Q_B, (\p c + \bar \p \bar c) c\p\xi e^{-2\phi} \bar c \bar \p\bar \xi e^{-2\bar \phi}
\}\, .
\ee
This is a pure gauge state. Thus as in the case of heterotic string theory,
its effect can be absorbed into a wave-function renormalization of external states. 

\item    $\vp_j = {1\over 2} (c\eta \bar c \bar \p\bar \xi e^{-2\bar \phi}
-  \bar c \bar\eta c\p\xi e^{-2\phi})$, $\vp_i =
{1\over 2} (\p c + \bar \p \bar c) (c\eta \bar c \bar \p\bar \xi e^{-2\bar \phi}
- \bar c \bar\eta c\p\xi e^{-2\phi})
$.  In this case $|\vp_j\rangle$ is a BRST invariant state and represents zero
momentum dilaton in the $(-1,-1)$ picture. 
Thus as in the case of heterotic string theory, the effect of this term can be absorbed
into a shift in the expectation value of the zero momentum dilaton field.

\item $\vp_j = (\p c + \bar \p \bar c) c\p\xi e^{-2 \phi} \bar c e^{-\bar\phi} U$,
$\vp_i = c\eta \bar c e^{-\bar\phi} U$ where $U$ is a left GSO odd, right GSO even
dimension $(1/2,0)$ operator in the matter sector. In a unitary theory such an
operator is a superconformal primary.  As in the case of heterotic string theory,
if the theory has a discrete symmetry under which $U\to -U$ keeping the
super-stress tensor invariant, then the matrix element of this operator on $\Sigma_1$
will vanish. We shall restrict our analysis to the class of theories for which this
holds..

\item $\vp_j = (\p c + \bar \p \bar c) \bar c\bar \p\bar \xi e^{-2 \bar \phi}  c e^{-\phi} V$,
$\vp_i = \bar c\bar \eta  c e^{-\phi} V$ where $V$ is a left GSO even, right GSO odd
dimension $(0,1/2)$ operator in the matter sector. This case can be treated exactly
as the previous one with the roles of left and right moving sectors on the world-sheet
exchanged.

\end{enumerate}

\subsection{General states}

For external states which are not special, 
we need to choose a suitable basis for physical, unphysical and
pure gauge states, `diagonalize' the propagator at each mass level after `integrating out'
fields at other mass levels
and identify the
renormalized physical states and their masses. For bosonic string theory this procedure
has been described in detail in  
\cite{1401.7014}. We expect that
there should not be any surprises in generalizing the analysis of \cite{1401.7014} to heterotic
or type II string theory, but we shall postpone a detailed analysis of this
question to the future.

\sectiono{Ramond sector}  \label{sramond}

Let us now consider the case where the external vertex operators also include
(an even number of) Ramond punctures. If we take all the external Ramond 
punctures in the $-1/2$ picture then on a genus $g$ surface with $n_B$ NS punctures
and $2n_F$ Ramond punctures, we need a total of $2g-2 + n_B + n_F$ picture
changing operators. With this change we can define the off-shell amplitudes in
the same way as in \S\ref{s2}.

\subsection{The problem}

The problem however appears while choosing a gluing compatible assignment
of picture changing operators 
near a degeneration where a Ramond sector state propagates along the
tube connecting the Riemann surfaces. Consider for example the case where the
Riemann surface described above degenerates into a pair of Riemann surfaces
of genus $g_1$ and $g_2$ with the first one carrying $n_{B_1}$ NS punctures and
$2n_{F1}$ R-punctures and the second one carrying $n_{B2}$ NS-punctures and 
$2 n_{F2}$ R-punctures, satisfying
\be \label{ecount}
g=g_1+g_2, \quad n_B = n_{B1} + n_{B2}, \quad 2 n_F = 2 n_{F1} +2 n_{F2} - 2\, .
\ee
{}From the fact that the total number of external Ramond states decreases by 2 after
gluing we know that the vertex operator at the punctures being glued must be Ramond
vertex operators. However in this case since the picture numbers of the Ramond
vertex operators at these punctures must add up to $-2$, it is not possible to take
both of them in the $-1/2$ picture as in the case of external states. Indeed 
we can see that we run into an apparent contradiction if we 
take both in the $-1/2$ picture. In that case on the first Riemann surface we
shall have $2g_1-2 + n_{B1} + n_{F1}$ picture changing operators and on the
second Riemann surface we shall have $2g_2-2 + n_{B2} + n_{F2}$ picture
changing operators. Using \refb{ecount} their numbers add up to 
$2g + n_B + n_F - 3$. This is one less then the number of picture changing operators
which were inserted on the original surface. 

This shows that it is impossible to satisfy the gluing compatibility condition in its original form
which required that the arrangement of picture changing operators
on the glued surface must agree with the collection of picture changing operators on the
individual surfaces.

\subsection{The prescription}

We shall now suggest a possible procedure which is not elegant but practical.
This procedure essentially specifies 
the rules for building up the full amplitude from 1PI amplitudes.
\begin{enumerate}
\item
For computing 2-point amplitude of two Ramond states, we take one of them in the
$-1/2$ picture and the other one in the $-3/2$ picture. In this case near any
Ramond degeneration of this amplitude, we pick the vertex operator at the
degenerating puncture
that is closer to the $-1/2$ picture vertex operator in the $-3/2$ picture and the
other one in the $-1/2$ picture. This basically means that if the degeneration is
into a genus $g_1$ and a genus $g_2$ surface then we take $2g_1$ picture
changing operators to lie on the genus $g_1$ surface and $2 g_2$ picture
changing operators to lie on the genus $g_2$ surface.
\item For $n$-point amplitude with $n\ge 3$ we take all external Ramond sector states
in the $-1/2$ picture and adopt the following algorithm.
For any given set of $n$ external states, we 
declare all subsets of length $\le (n-1)/2$ as $\BBB$-type, all subsets of length
$\ge (n+1)/2$ as $\AAA$-type and (if $n$ is even) classify the subsets of length
$n/2$ arbitrarily as $\AAA$-type or $\BBB$-type, with the constraint that the complement
of an $\AAA$-type set is always $\BBB$-type and vice versa.
Now for any given Ramond 
degeneration of the original punctured Riemann surface into a 
pair of punctured Riemann surfaces,
one side will contain an $\AAA$-type set of external states    and the other side will
contain a complementary set that is of $\BBB$-type. Our prescription will be that on the
component that contains external states in the $\AAA$-type set we choose the state at
the degenerating puncture to be a picture number $-1/2$ state, while 
on the
component that contains external states in the $\BBB$-type set we choose the state at
the degenerating puncture to be a picture number $-3/2$ state. This effectively means
that the extra picture changing operator is inserted on the component
of the Riemann surface that contains external states in the $\BBB$-type set.

Clearly the division of the external states into $\AAA$ and $\BBB$-type is
arbitrary.
The important
point is that the rules for making such divisions, although arbitrary, must {\it depend 
only on how the external states are divided up into subsets
by the degeneration and
not on the genera or the moduli of the
Riemann surfaces involved in the degeneration}. This is necessary for the 
factorization property of the amplitude to be discussed in \S\ref{spropramond}.
\end{enumerate}

This procedure implies that the 1PI amplitudes which are glued together to form the
full amplitude can some time 
have all the
external states in the $-1/2$ picture, 
and other times one of their external states in the $-3/2$ picture. 
The fact that any subset of a $\BBB$-type set is always $\BBB$-type guarantees
that we do not get any 1PI amplitude with more than one $-3/2$ picture
vertex operator.
Gluing
compatibility then tells us that for any such 1PI amplitude, the arrangement
of picture changing operators on the corresponding Riemann surface
(and of course the choice of local coordinates at the
punctures) must be chosen in a way that only depends on the moduli relevant
to that 1PI amplitude and is independent of the rest of the
Riemann surfaces to which it is attached by plumbing fixture. 

\subsection{Analysis of propagator and S-matrix} \label{spropramond}

\begin{figure}
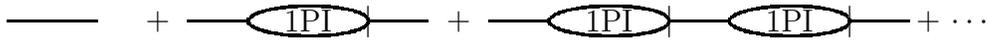

\begin{center}
\figa
\end{center}
\vskip -1in
\caption{Pictorial representation of eq.\refb{eprop}.  The horizontal lines represent
$\Delta$ and blobs marked 1PI represent $\wt\Gamma$. 
Note that this diagram is left-right asymmetric as in each 1PI component
the state on the left is in the $-1/2$
picture and the state on the right is in the $-3/2$ picture. The latter has been 
indicated by a $|$ at the vertex.
\label{fa}}
\end{figure}

We shall now briefly describe how this prescription helps us define propagators and
S-matrix elements. Let us start with the Ramond sector propagator where
one of the external vertex operators is in the $-1/2$ picture and the other one 
is in the
$-3/2$ picture. 
Let $k$ be the momentum carried by the $-1/2$ picture external vertex operator.
In this case the tree level propagator 
is
\be
\Delta = (L_0 + \bar L_0)^{-1} \delta_{L_0, \bar L_0}\, .
\ee
We now follow the procedure of \cite{1311.1257,1401.7014} 
to divide the higher genus amplitudes into
one particle reducible (1PR) and one particle irreducible (1PI) amplitudes.
Each 1PI component will have one external vertex
operator carrying momentum $k$ in the $-1/2$ picture and the other external vertex
operator carrying momentum $-k$ in the $-3/2$ picture.
If $\wt\Gamma$ denotes the 1PI amplitude then the full propagator can be written as
\be \label{eprop}
\Pi =\Delta + \Delta \wt\Gamma \Delta + \Delta \wt\Gamma \Delta \wt\Gamma \Delta
+\cdots = (\Delta^{-1} - \wt\Gamma)^{-1}\, .
\ee
This has been shown pictorially in Fig.~\ref{fa}. 
The poles of this give the renormalized mass$^2$. Following the procedure
described in \cite{1311.1257,1401.7014} one can further simplify 
this by integrating out states
at all mass levels except a particular level and reduce the problem to that of
diagonalization of a finite dimensional matrix, but we shall not discuss the details
here.

Note that the familiar $G_0$ in the numerator is missing from the tree level
propagator $\Delta$ that appears in 
\refb{eprop}. This is
related to choosing specific normalization of the basis states i.e. the choice
$\langle  -3/2, r| c_0 \bar c_0|-1/2,s\rangle = \delta_{rs}$ that has been used to
get the propagator $\Delta$. As a result the $G_0$ factor is
hidden inside $\wt\Gamma$ in \refb{eprop}.  At the end of this section we shall describe
how this factor can be recovered explicitly when the propagating state is a 
special state in the spirit described in  \S\ref{sdefspecial}.

Now consider the case of a general amplitude with external fermions with 
momenta $k_1,\cdots k_n$ and {\it tree level} masses $m_1,\cdots m_n$. One would
like such an amplitude to satisfy the following properties:
\begin{enumerate}
\item After multiplying the amplitude 
by a factor of $\Delta$ for each external leg to generate the off-shell 
Green's function as described in \refb{es1.1}, 
each amplitude should have an explicit factor of
the full propagator $\Pi(k_i)$ for each external leg. This will allow us to compute the
S-matrix element using the LSZ prescription.
\item If $k$ denotes the sum of a subset of the external momenta then the pole
of the S-matrix in the $k^2$ plane must come from the poles of $\Pi(k)$.
\end{enumerate}
Let us begin with the first property. We break up the amplitude into sums of products
of 1PI amplitudes as usual. 
Since subsets containing single external
fermions are always
$\BBB$-type,  the structure of self energy corrections associated with each
of the external Ramond sector states
will be the same as the one described in Fig.~\ref{fa} for the two point function.
In particular in any 1PI subamplitude inserted on the $i$-th external leg,
we always pick the $-3/2$ picture vertex
operator on the puncture 
carrying momentum $-k_i$ and $-1/2$ picture vertex operator on the puncture carrying
momentum $k_i$. Thus
we get precisely
the same factor \refb{eprop} for each external leg. 
From this we can compute the
on-shell S-matrix elements using the LSZ prescription as in 
\cite{1311.1257,1401.7014}.

\begin{figure}
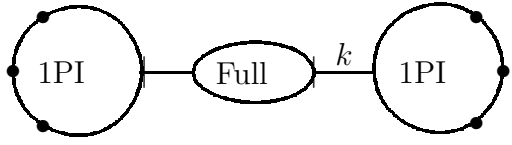

\begin{center}
\figb
\end{center}
\vskip -.5in
\caption{Pictorial representation of the second terms on the right hand sides of
eq.\refb{eGamma}. Here 1PI means sum of
contributions which are 1PI in the leg carrying momentum $k$, whereas  Full means sum
of all contributions to the 2-point function shown in Fig.~\ref{fa}. 
If external states on the left side belong to the $\BBB$-type set, then the internal state
on the left side inserted into the 1PI amplitude is in the $-3/2$ picture while the
internal state on the right inserted into the 1PI amplitude is in the $-1/2$ picture.
As in Fig.~\ref{fa}, the $-3/2$ picture state insertions are marked by $|$.
\label{fb}}
\end{figure}

A similar analysis can be used to prove the second property.
To find
the pole of the amplitude in $k^2$ we can express the off-shell amplitude $\Gamma$
as sums of 1PI and 1PR
contributions in legs carrying momentum $k$ as
\be \label{eGamma}
\Gamma = \wh\Gamma + \wh\Gamma_1^a  \Pi_{ab} \wh \Gamma_2^b 
 \, ,
\ee
where $\wh\Gamma$ represents contributions to $\Gamma$ which are 1PI in the
leg carrying momentum $k$ and $\wh\Gamma_1^a$, $\wh\Gamma_2^b$ are also
subamplitudes 1PI in momentum $k$. 
A pictorial representation of the second term on the right hand side of \refb{eGamma}
has been shown in Fig.~\ref{fb}. Between $\wh\Gamma_1^a$ and $\wh\Gamma_2^b$,
one of them carries external states in $\AAA$-type set and the other carries external states in
$\BBB$-type set. In the former the internal state will be inserted using $-1/2$ picture while
in the latter it will be inserted in the $-3/2$ picture. This makes it manifest that the poles
in the amplitude as a function of $k^2$ occur exactly at the poles of $\Pi$ given in
\refb{eprop}.

Following the analysis of \cite{1311.1257,1401.7014} and the
ones carried out here,
one may be able to prove that the renormalized masses
and S-matrix elements computed this way are independent of the detailed arrangement
of picture changing operators as well as of how we assign subsets of $1,\cdots n$
to be of $\AAA$-type and $\BBB$-type.  
A detailed analysis of this will be postponed to future work. However at this stage we
note that for special Ramond sector states -- in the spirit described in  \S\ref{sdefspecial}
-- one can recover the symmetry between the $\AAA$-type and $\BBB$-type sets as
follows. We simply require that in any degeneration limit 
the extra picture changing operator that is inserted on the part of the
amplitude with external states in the $\BBB$-type set approaches the degeneration node that
carries the $-3/2$ picture vertex operator.\footnote{This is equivalent to inserting the
picture changing operator on the propagator.} 
Since for special states all the relevant vertex operators
have the form $\bar c\, c\, e^{-3\phi/2} \tilde 
V_\alpha \, e^{ik\cdot X}(z)$ for some matter sector vertex 
operator $\tilde 
V_\alpha$, the only term in the picture changing operator 
that gives a non-zero contribution in
this limit is the $e^\phi\, T_F(w)$ term in $\VVV(w)$. 
Now for special states, there
are no matter sector operators carrying the same weight as $\tilde V_\alpha$ and
having conformal weight less than that of $\tilde V_\alpha$. Thus the maximum singularity
we can get from the operator product of $T_F(w)$ and $\tilde V_\alpha(z)$  
is $(w-z)^{-3/2}$, $3/2$ being
the conformal weight of $T_F$. This cancels with the $(w-z)^{3/2}$ factor from the
operator product of $e^\phi$ and $e^{-3\phi/2}$, producing a non-singular 
term.\footnote{The condition for being able to do this
is actually less stringent than a special state condition. The latter requires restriction on the
conformal weight in both the holomorphic and the 
antiholomorphic sectors, while here we only
need restriction on the holomorphic conformal weight.}
Its contribution is
\be \label{eopepre}
\lim_{w\to z} \,e^\phi T_F(w) \, \bar c\, c\, e^{-3\phi/2} \tilde V_\alpha \, e^{ik\cdot X}(z)
\propto \bar c\, c\, e^{-\phi/2} (\gamma^\mu k_\mu \pm M)_{\alpha}{}^{\beta}V_{\beta}(z) \, 
\ee
where $M$ is the tree level mass of the vertex operator and $V_{\beta}$ is another matter
sector operator that appears in the expression for the $-1/2$ picture vertex operator.
The $(\gamma^\mu k_\mu \pm M)_{\alpha}{}^{\beta}$ factor comes from the action  
on $\tilde V_\beta$ of the
$G_0$ term in the mode expansion of $T_F$.
This converts the $-3/2$ picture vertex operator to $-1/2$ picture vertex operator
and converts the propagator 
$\delta_{\alpha}{}^{\beta}/(k^2+M^2)$ to 
$(\gamma^\mu k_\mu \pm M)_{\alpha}{}^{\beta}
/(k^2+M^2)$ which is the correct Ramond sector propagator. Since now all Ramond 
degenerating 
nodes carry $-1/2$ picture vertex operators, we recover the symmetry between the
external states in the $\AAA$ and $\BBB$-type set.


\sectiono{Computation of Fayet-Iliopoulos terms using picture changing
operator} \label{sfi}

The procedure for defining off-shell amplitudes using picture changing operators,
as described above, can also be used for on-shell amplitudes. In this section we
shall describe how it can be used to compute the effect of
Fayet-Iliopoulos (FI) terms
in SO(32) heterotic string theory compactified on a Calabi-Yau
3-fold\cite{DSW,ADS,DIS,greenseiberg,AtickS,1304.2832,1403.5494,1404.5346}.

\subsection{Choice of locations of picture changing operators and local coordinate system}

We refer the reader to the original papers for the necessary background, and
focus here only on the computational aspect of the problem.
The problem involves computing an on-shell two
point function of two NS sector states at one loop order. The vertex operators
describing the states have the form
\be \label{evertex}
\bar c \, c \, e^{-\phi} \, V_1 \, e^{i k\cdot X} \quad \hbox{and} 
 \quad \bar c \, c \, e^{-\phi} \, V_2 \, e^{-i k\cdot X}, \quad k^2=0\, ,
\ee 
where  $X^\mu$
for $0\le\mu \le 3$ denote
the four non-compact target space-time coordinates
and $V_1,V_2$ are a pair of superconformal primaries of dimension 
(1,1/2) made of the degrees of freedom associated with the compact
directions.
Some of the
special properties of $V_i$ that we shall need will be reviewed
later as and when we need them.
In this section we shall work in the $\alpha'=1$ unit in which $X^\mu$ and its fermionic
partner $\psi^\mu$ have the following operator product:
\be \label{ematterope}
\p X^\mu(z) \p X^\nu(w) = -{\eta^{\mu\nu}\over 2 (z-w)^2}+\cdots, \quad
\psi^\mu (z) \psi^\nu(w) = -{\eta^{\mu\nu}\over 2 (z-w)}+\cdots\, ,
\ee
where $\cdots$ denote non-singular terms.
The matter energy momentum tensor $T(z)$ and its superpartner $T_F(z)$ have
the following form
\be 
T(z) = - \partial X^\mu \partial X^\nu \eta_{\mu\nu} + \psi_\mu \p \psi^\mu
+ T_{int}, \quad
T_F(z) = - \psi_\mu \p X^\mu + (T_F)_{int}\, ,
\ee
where the subscript $~_{int}$ denotes contributions from the compact directions.

For computing this amplitude we need to first compute
the correlation function on
the two punctured
torus with the two vertex operators given in \refb{evertex} 
inserted at the punctures, together with
appropriate insertion of ghost fields and the picture changing operators 
as described in \S\ref{s2}, and then integrate this over the moduli space of
two punctured tori. The latter will be parametrized by the coordinates of one
of the punctures (with the other one kept fixed) and the modular parameter
$\tau$ of the torus. Let $u$ be the coordinate system in which the torus is
described by the identification
\be 
u\equiv u+m + n\tau, \quad m,n\in \ZZZ\, .
\ee
Then we shall choose the punctures $P_1$ and $P_2$ 
to be at $u=0$ and $u=y$ and use $y$
and $\tau$ as complex coordinates of the moduli space. Furthermore we shall 
use
\be \label{ew1uw2}
w_1=u, \quad \hbox{and} \quad  w_2=u-y\, ,
\ee 
as the local coordinates around the puncture $P_1$ and $P_2$ (up to overall
phases).\footnote{This choice is not quite gluing compatible, as the latter requires that as we
take the degeneration limit $y\to 0$, the distance between the punctures should
remain fixed in the local coordinate system of each puncture. This can be achieved
by scaling the local coordinates by $1/y$. 
Following the analysis of \S\ref{s1} one can show that this will have two effects. 
First of all it will generate an extra factor of $|y|^{h_1+h_2}$ where $h_1$ and $h_2$
are the total conformal weights of the vertex operators inserted at the punctures.
Since for on-shell external vertex operators 
we have $h_1=0$, $h_2=0$, this effect disappears.
Second, since the relationship between
the local coordinates and the fixed coordinate $u$ vary by a scale factor as we vary
$y$,
we shall have additional insertions
in the correlation function involving
$b_0$ and $\bar b_0$ operators acting on external states. 
However since the external
states satisfy the Siegel gauge condition $b_0|\Psi\rangle = \bar b_0 |\Psi\rangle=0$,
this effect also disappears. 
Thus we can continue to use $u$ and $u-y$ as the local coordinates. \label{fo5}}

Let us now describe the choice of locations of the picture changing operators. We need two
of them. We shall choose one to be at a fixed location in the $u$ coordinate,
say at $u=u_1$ and the other one at a location $y=u_2 \equiv \alpha y$ for some
fixed constant $\alpha$. This has the property that in the degeneration limit when
$y\to 0$, if we regard the configuration as a three punctured sphere glued to a one
punctured torus (with $u/y$ as the coordinate system on the sphere and $u$ as the
coordinate system on the torus), the picture changing operator at $u_1$ lies
on the torus and the picture changing operator at $u_2$ lies on the sphere. This
is the correct prescription for gluing compatibility.  We could also have made a
non-holomorphic choice in which $u_2$ is taken to be a function of $y$ and $\bar y$. We
have not done this in order to keep the analysis simple, but the final result remains the
same even for this more general choice.

\begin{figure}
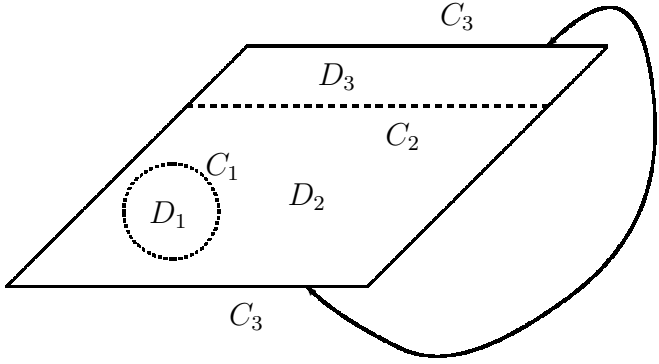

\figtorus
\caption{The covering of the torus by three different regions. \label{figtorus}
}
\end{figure}

Once we have made a choice of local coordinates and the location of
the picture changing operators, we have fixed the choice of the 
section in $\wt\PP_{1,2}$ on which we shall integrate. This means that the
tangent vectors $\p/\p\tau$, $\p/\p y$ and their barred counterparts now have definite
images in the tangent space of $\wt\PP_{1,2}$, and the integration measure
will be given by the contraction of $\Omega^{1,2}_4$ with these tangent vectors. This
in turn means that we should now be able to 
use \refb{eexpomp} to
write down explicitly the measure that needs to be integrated over $\MM_{1,2}$
to compute the relevant amplitude. 
We shall now do this explicitly. For this we shall follow the general
procedure described in \S\ref{s1gen} and \S\ref{s2gen} in which we divide the Riemann
surface into different components with different coordinate systems and the functional
relationship between the coordinates encode information about the moduli. We choose
three different coordinate systems on the torus: the coordinate $u$ introduced earlier
which is also the local coordinate $w_2$ around the puncture at $y=0$,
the local coordinate $w_1= u-y$ defined around the puncture at $y$, and a new
coordinate system $z$ defined as follows. Around the line $C_2$: Im~$u=a$ for some 
positive constant $a$ we have $z=u-\tau$ and along the line $C_3$: Im~$u=-b$ for
some negative constant $-b$ we have $z=u$.  Then the whole torus, whose fundamental
domain we shall take to be the region 
\be
-b\le {\rm Im} ~u< \tau_2-b, \quad 
-{1\over 2} + {\tau_1\over \tau_2} {\rm Im}~ u \le {\rm Re}~u< {1\over 2} + 
{\tau_1\over \tau_2} {\rm Im}~ u
\, ,
\ee
 is
covered by three regions.
Around the puncture at $y$ we identify a small disk $D_1$ inside which we use the
$w_1=u-y$ coordinate system. In the region
 \be \label{eureg}
D_2:  -b\le {\rm Im} ~u< a, \quad 
-{1\over 2} + {\tau_1\over \tau_2} {\rm Im}~ u \le {\rm Re}~u< {1\over 2} + 
{\tau_1\over \tau_2} {\rm Im}~ u, \quad u\not\in D_1
\, ,
 \ee 
 we
use the $u$ coordinate system. Finally in the region
\be \label{ezreg}
D_3: a - \tau_2\le {\rm Im} ~z< -b, \quad 
-{1\over 2} + {\tau_1\over \tau_2} {\rm Im}~ z \le {\rm Re}~z< {1\over 2} + 
{\tau_1\over \tau_2} {\rm Im}~ z
\, ,
\ee
we
use the $z$ coordinate system. This has been shown in Fig.~\ref{figtorus}.
The functional relationship between the coordinates
takes the form
\ben \label{eoverlap}
&& \hbox{On $C_1=\p D_1$}: \quad w_1 = u-y\, , \nonumber \\
&& \hbox{On $C_2$}: z = u-\tau\, , \nonumber \\
&& \hbox{On $C_3$}: z = u\, .
\een
Both picture changing operators will be introduced in the $u$ coordinate system
in the region $D_2$.

\subsection{The integration measure}

Let us now follow the prescription of \S\ref{s1gen} and \S\ref{s2gen} to determine the
insertion of $b$-ghost and picture changing operators. First let us ignore the picture
changing operators and determine the $b$-ghost insertions as if we were working
in bosonic string theory. In this case using \refb{eoverlap}
and the analysis of \S\ref{s1gen} we see
that the effect of contraction of $\Omega^{1,2}_p$ with $\p/\p\tau$ is a $b$-ghost
insertion along $C_2$ since the only dependence on $\tau$ arises from the gluing
function along $C_2$. Furthermore since for fixed $y$, $z_{\tau+\delta\tau }
= z_\tau - \delta\tau$, the associated vector field is $v(z)=-1$, showing that
the $b$-ghost insertion corresponding to contraction with $\p/\p\tau$ and
$\p/\p\bar\tau$ is just
\be \label{ebvtau}
(-b_\tau) (-\bar b_\tau), \quad b_\tau\equiv 
\ointop_{C_2} dz b(z), \quad \bar b_\tau\equiv 
\ointop_{C_2} d\bar z \bar b(\bar z) \, .
\ee
In the definition of $b_\tau$ ($\bar b_\tau$)
the integration contour $C_2$ should be oriented 
so that the region $D_3$ lies to its left (right).
On the other hand since the only $y$ dependence of the gluing function is along the
curve $C_1$,   the effect of contraction with $\p/\p y$ is represented by a
contour integral along $C_1$. The associated vector field is $v(w_1)=-1$ and hence
we have the insertion of
\be \label{ebvy}
(-b_y) (-\bar b_y), \quad b_y\equiv 
\int_{C_1} dw_1 b(w_1), \quad \bar b_y\equiv \int_{C_1} d\bar w_1 \bar b(\bar w_1)\, .
\ee
In the definition of $b_y$ ($\bar b_y$)  
the integration contour runs anticlockwise (clockwise) around the puncture
at $y$. 
The $-$ signs in \refb{ebvtau} and \refb{ebvy} reflect the $-$ signs in
the vector fields associated with $\tau$ and $z$ deformation.

Now let us consider the effect of inserting the 
picture changing operators. Both of them are inserted
in the region $D_2$ in the $u$ coordinate system. Of them one location $u_1$ is fixed
while the other one $u_2$ varies with $y$ as $\alpha y$. It follows from the general
prescription of \S\ref{s2gen} that the net insertion of $b$-ghosts and
picture changing operators into the correlation function will be
\be
 b_\tau \bar b_\tau  \VVV(u_1) (\VVV(u_2)  (-b_y) - \p \xi(u_2) (\p u_2/\p y)) (-\bar b_y)
 \ee
Using $u_2=\alpha y$ and the form of the vertex operators given in \refb{evertex} we
can now write the net contribution to the torus 2-point function 
as\footnote{In this expression we have dropped an overall normalization of
$1/\pi^2$ that comes from a combination of two terms. First the  
$(2\pi i)^{-(3g-3+n)}$ factor in \refb{edefomp} gives a factor of $-1/4\pi^2$.
Second the integration measure is really $d\tau\wedge d\bar\tau
\wedge dy\wedge d\bar y$ which translates to $-4 d^2\tau d^2y$. 
Nevertheless we shall be able to check that our procedure agrees with
the one used in \cite{ADS,DIS} including normalization. 
This is described at the end of \S\ref{samplitude}.}
\ben\label{ess3}
 && \int d^2\tau \int d^2 y  \,\Big\langle b_\tau\bar b_\tau 
\VVV(u_1) \bar c(0) c(0) e^{-\phi(0)} V_1(0) e^{ik.X(0)} \nonumber \\ &&
\qquad \qquad  \qquad
\Big(\VVV(\alpha y) + \alpha\, \p \xi (\alpha y) c(y)\Big) e^{-\phi(y)} V_2(y) e^{-ik.X(y)}
\Big\rangle\, ,
\een
where $\partial$ always denotes derivative with respect to the argument.
The final result should be independent of $\alpha$. This will be verified explicitly 
in \S\ref{salpha}.

\subsection{Evaluation of the amplitude} \label{samplitude}

We shall now evaluate \refb{ess3}. First we write
\be \label{exu1}
\VVV(u_1) = \VVV(0) + (\VVV(u_1) -\VVV(0)) = \VVV(0) + \left\{Q_B, \xi(u_1) 
-\xi(0) \right\}\, ,
\ee
where we have used 
\refb{epicture}.
The contribution from the first term can be analyzed by noting that
\ben \label{e7.15}
\VVV(0) \bar c(0) c(0) e^{-\phi(0)} V_1(0) e^{ik.X(0)} 
&=& \bar c(0) c(0) \{ \wt V_1(0) - {i\over 2} k.\psi(0) V_1(0)\} e^{ik.X(0)} 
\nonumber \\ &&
-{1\over 4} \bar c(0) \eta(0) e^{\phi(0)} V_1(0) e^{ik.X(0)}\, ,
\een
where $\psi^\mu$ denote the world-sheet fermions describing
superpartners of the non-compact target space-time coordinates $X^\mu$,
and $\wt V_1$ is the dimension (1,1) vertex operator of the conformal
field theory associated with the compact Calabi-Yau manifold, related to
$V_1$ via the action of superconformal generator $T_F$:
\be \label{ewtv1}
T_F(z) V_1(0) = -{1\over z} \wt V_1(0) + \hbox{non-singular}\, .
\ee
Using \refb{e7.15}  we see that the contribution to \refb{ess3}
from
the first term on the right hand side of \refb{exu1} is given by
\ben \label{ess11}
&& \int d^2\tau \int d^2 y \Big\langle b_\tau\bar b_\tau 
\bigg[\bar c(0) c(0) \{ \wt V_1(0) - {i\over 2} k.\psi(0) V_1(0)\} e^{ik.X(0)} 
-{1\over 4} \bar c(0) \eta(0) e^{\phi(0)} V_1(0) e^{ik.X(0)}
\bigg]
\nonumber \\ && \qquad \qquad  \qquad
\times \Big(\VVV(\alpha y) + \alpha\, \p \xi (\alpha y) c(y)\Big) e^{-\phi(y)} V_2(y) e^{-ik.X(y)}
\Big\rangle\, .
\een
As already mentioned above, and will be verified in \S\ref{salpha},
the total contribution is expected to be independent of $\alpha$. In order to
make contact with the analysis of \cite{ADS,DIS} we shall now take the $\alpha\to 1$ limit, which
amounts to inserting the second picture changing operator at the location $y$ of the
second puncture. 
In this limit the $c\p\xi$ term from $\VVV(\alpha y)$ 
cancels the $\alpha\, \p \xi (\alpha y) c(y)$ term. 
Thus the only term in the second line
of \refb{ess11} that contributes is the term
\be 
\lim_{\alpha\to 1} e^{\phi(\alpha y)} T_F(\alpha y) e^{-\phi(y)} V_2(y) e^{-ik.X(y)}
= \left\{ \wt V_2(y) + {i\over 2} k.\psi(y) V_2(y)\right\} e^{-ik.X(y)}\, ,
\ee
where $\wt V_2(y)$ is defined in the same way as $\wt V_1$ in \refb{ewtv1}
\be \label{ewtv2}
T_F(z) V_2(0) = -{1\over z} \wt V_2(0) + \hbox{non-singular}\, .
\ee
Furthermore $\phi$ charge conservation now allows us to drop the term proportional
to $e^{\phi(0)}$ from the first line of \refb{ess11}. Thus
\refb{ess11} now takes the form
\be \label{ess12}
\int d^2\tau \int d^2 y \left\langle b_\tau\bar b_\tau 
\bar c(0) c(0) \left\{ \wt V_1(0) - {1\over 2} i k.\psi(0) V_1(0)\right\} e^{ik.X(0)}  
\left\{ \wt V_2(y) + {1\over 2} i k.\psi(y) V_2(y)\right\} e^{-ik.X(y)} \right\rangle\, .
\ee
This is precisely the term that was analyzed in \cite{ADS,DIS}. As emphasized there,
if we work with strictly on-shell momentum $k^2=0$ then the result
vanishes. \cite{ADS,DIS} analyzed this by keeping the momenta slightly off-shell and
at the end taking the $k^2\to 0$ limit. We shall come back to discuss this
approach later, but for now we proceed by keeping $k^2=0$ from the beginning.
In that case this term does not contribute.

This leaves us with the contribution from the second term on the
right hand side of \refb{exu1}, and we
need to  take the $\alpha\to 1$ limit at the end.
This can be 
analyzed by deforming the BRST contour and picking up the contribution from
the residues at the rest of the operators inserted in \refb{ess3}. 
Before doing that however we replace $Q_B$ by its holomorphic part $Q_B^R$
since only this part contributes to $\{Q_B, \p\xi\}$. 
We have
\be \label{ezz1}
[Q_B^R, \bar c(0) c(0) e^{-\phi(0)} V_1(0) e^{ik.X(0)} ] = 0\, ,
\ee
\be 
[Q_B^R, \VVV(\alpha y)] = 0\, ,
\ee
\be
[Q_B^R, \p \xi (\alpha y)] = \p \VVV(\alpha y)
\, ,
\ee
\be 
[Q_B^R, c(y) e^{-\phi(y)} V_2(y) e^{-ik.X(y)}] 
= 0\, ,
\ee
\be 
[Q_B^R, e^{-\phi(y)} V_2(y) e^{-ik.X(y)}] 
= \p_y \left( c(y) e^{-\phi(y)} V_2(y) e^{-ik.X(y)} \right)
\, .
\ee
Finally $\{Q_B^R, \bar b_\tau\}$ vanishes and
$\{Q_B^R, b_\tau\}$  generates
total derivative with respect to $\tau$ and integrates to zero.
Thus the net contribution, after adding all the terms, is given by
\ben\label{ess4}
 && \int d^2\tau \int d^2 y  \,
 \Big\langle b_\tau\bar b_\tau 
(\xi(u_1)-\xi(0)) \bar c(0) c(0) e^{-\phi(0)} V_1(0) e^{ik.X(0)} \nonumber \\ &&
\Big[\VVV(\alpha y) \p_y \left( c(y) e^{-\phi(y)} V_2(y) e^{-ik.X(y)} \right)
+ \alpha\, \p \VVV (\alpha y) c(y) e^{-\phi(y)} V_2(y) e^{-ik.X(y)}
\Big]
\, .
\een
This can be rewritten as
\be\label{ess5}
\int d^2\tau \int d^2 y \,  \, \p_y \left[
 \Big\langle b_\tau\bar b_\tau 
\{\xi(u_1)-\xi(0)\} \bar c(0) c(0) e^{-\phi(0)} V_1(0) e^{ik.X(0)} 
\VVV(\alpha y) c(y) e^{-\phi(y)} V_2(y) e^{-ik.X(y)} 
\Big\rangle\right]
\, .
\ee
Note that the only term of $\VVV(\alpha y)$ that contributes is the one with
$\phi$ charge 2:
\be
-{1\over 4} \p\eta b e^{2\phi} - {1\over 4} \p \left(\eta b e^{2\phi}\right)\, .
\ee
Substituting this into \refb{ess5} and using the operator product expansion to
evaluate the $\alpha\to 1$ limit, we get
 \be\label{ess7}
-{1\over 4}\int d^2\tau \int d^2 y   \,  \, \p_y 
 \Big\langle b_\tau\bar b_\tau 
\{\xi(u_1)-\xi(0)\} \bar c(0) c(0) e^{-\phi(0)} V_1(0) e^{ik.X(0)} 
\eta(y) e^{\phi(y)} V_2(y) e^{-ik.X(y)} 
\Big\rangle
\, .
\ee

This is a total derivative in $y$. Thus it can get a non-zero contribution only from
the boundary near $y=0$  if the integrand has a singularity of the form $1/\bar y$. 
Now the only source of $\bar y$ dependence in the above correlator is from the
matter vertex operators $V_1$ and $V_2$;  the $e^{ik\cdot X}$ factors can be
ignored altogether
since any contraction involving
them will pick up $k^2$ factors which vanish by on-shell condition. 
Thus we can simply replace $V_1(0) V_2(y)$
by the part which has a pole of the form $1/\bar y$ and which has a non-vanishing
expectation value on the torus. This is given by\cite{ADS,DIS}
\be \label{eope}
V_1(0) V_2(y) = {q\over \bar y} V_D(0) \, ,
\ee
where $V_D$ is the dimension (1,1) operator representing the vertex operator
of the auxiliary D-field associated with the anomalous U(1) gauge field
and $q$ is the charge carried by the vertex operator under 
this anomalous U(1).  $V_D$ is given by the product of the R-symmetry current
on the right-moving sector of the world-sheet and the left-moving U(1) current
associated with the anomalous U(1) gauge field.
The other relevant operator products are
\be \label{ephiope}
e^{-\phi(0)} e^{\phi(y)} \simeq -y + \OO(y^2)\, ,
\ee
and 
\be 
\{\xi(u_1)-\xi(0)\} \, \eta(y)\simeq y^{-1} + \OO(y^0)\, .
\ee  
Substituting this into \refb{ess7}
we get
\be 
-{q\over 4}\int d^2 \tau\, \langle \Big\langle b_\tau\bar b_\tau 
 \bar c(0) c(0) V_D(0) \rangle \times \int d^2 y \, \p_y \bar y^{-1}\, ,
\ee
where it is understood that 
the integral over $y$ is to be done by putting a cut-off $|y|\ge \eps$ for some
small positive number $\eps$ -- this corresponds to the infrared regularization
$s<\Lambda$ described in \S\ref{sinfra} with the identification $|y|= e^{-s}$,
$\eps = e^{-\Lambda}$ -- and we are supposed to pick up the boundary contribution
from the $|y|=\eps$ end. 
Writing the integral in terms of
$r=|y|$ and $\theta=Arg(y)$ it is easy to see that 
the integral over $y$ receives a contribution of
$-\pi$ from the boundary at $|y|=\eps$.  Thus the 
the final result for the two point function is given by
\be \label{efinresdterm}
{1\over 4} \, \pi \, q\, \int d^2 \tau  \langle \Big\langle b_\tau\bar b_\tau 
 \bar c(0) c(0) V_D(0) \rangle \, .
\ee
This agrees with the result of \cite{1304.2832} up to normalization. We shall check the
normalization shortly.

Note that the entire extra contribution compared to that in \cite{ADS,DIS} for $k^2=0$
came from the need to move the first picture changing operator from 0 to the
position $u_1$, whose effect is given by the second term in the right hand side of
\refb{exu1}. This is needed to ensure that the picture changing operators follow the
correct arrangement in the degeneration limit.

In order to check that the analysis given above captures the complete result, we shall
now verify that \refb{efinresdterm} gives the correct normalization and sign. We shall
do this by comparing the result with that of \cite{ADS,DIS} where the complete
contribution came from \refb{ess12} by keeping $k$ slightly off-shell.\footnote{Although we
are reproducing the computation of  \cite{ADS,DIS}, we should keep in mind that this
is different from the definition of off-shell amplitude 
we have  given earlier. 
Refs.\cite{ADS,DIS} use a coordinate system in which the local coordinates at the punctures are
taken to be $u$ and $u-y$ instead of scaling them by $1/y$ as 
described in footnote \ref{fo5}. Nevertheless
this computation was justified by showing that this produces correctly the location of the
$s$-channel pole in an on-shell four point scattering amplitude. If instead we follow our approach
then the off-shell computation will be very similar to the on-shell computation performed here.
The $y$ dependent scaling will remove the $-k^2$ from the exponent of $y$ in 
\refb{efootref}
and the final result will still come from boundary contributions.}
In this case
the nonvanishing part of the result comes from keeping the second term inside
each curly bracket in \refb{ess12}:
\be
{1\over 4} \int d^2\tau \int d^2 y \left\langle b_\tau\bar b_\tau 
\bar c(0) c(0) k.\psi(0) V_1(0) \, e^{ik.X(0)}  
 k.\psi(y) V_2(y) \, e^{-ik.X(y)} \right\rangle\, .
\ee
Now by Lorentz invariance the $\psi^\mu$, $\psi^\nu$ correlator is proportional
to $\eta^{\mu\nu}$, producing a factor of $k^2$. Since eventually we take the 
$k^2\to 0$ limit we must pick up the singular contribution proportional to
$1/k^2$ from the rest of the terms. This comes from the $y\to 0$ limit of the
integration. Eq.\refb{eope}, \refb{ematterope} together with
\be
e^{ik.X(0)} e^{-ik.X(y)} = |y|^{-k^2} + \hbox{less singular terms}
\ee
and the fact that $V_1$, $V_2$ anti-commute with $\psi^\mu$
now give
\be \label{efootref}
-q\, {k^2\over 8} \int d^2\tau \int d^2 y  \, |y|^{-2-k^2} \, \left\langle b_\tau\bar b_\tau 
\bar c(0) c(0) V_D(0)
\right\rangle\, .
\ee
In the $k^2\to 0$ limit the $y$ integral produces a factor of $-2\pi / k^2$. Thus the
net contribution is
\be
{1\over 4} \, \pi \, q \int d^2\tau \left\langle b_\tau\bar b_\tau 
\bar c(0) c(0) V_D(0)
\right\rangle\, .
\ee
This is in perfect agreement with \refb{efinresdterm}.

\subsection{$\alpha$ independence} \label{salpha}

Finally let us verify that the expression \refb{ess3} is independent of $\alpha$. For this
we take the $\alpha$ derivative of this expression. Acting on $\VVV(\alpha y)$ 
this generates
a $y\{Q_B^R, \p \xi(\alpha y)\}$ while acting on $\alpha \p\xi(\alpha y)$ it gives
$\p\xi(\alpha y) +\alpha y \p^2\xi(\alpha y)$. The BRST contour in 
$\{Q_B^R, \p \xi(\alpha y)\}$ can now be deformed. $\{Q_B^R, \bar b_\tau\}$ vanishes
and the residue from $\{Q_B^R, b_\tau\}$  
generates total derivatives in $\tau$ which integrate to zero.
On the other hand the  $\bar c(0) c(0) e^{-\phi(0)} V_1(0) e^{ik.X(0)}$
is BRST invariant and $\{Q_B^R, e^{-\phi(y)} V_2(y) e^{-ik.X(y)}\}$ generates 
$\p_y \left( c(y) e^{-\phi(y)} V_2(y) e^{-ik.X(y)} \right)$. Combining this with the rest of the
terms gives the $\alpha$ derivative of \refb{ess3} to be
\be\label{ess3change}
 \int d^2\tau \int d^2 y \,  \p_y \bigg[ y  \,\Big\langle b_\tau\bar b_\tau 
\VVV(u_1) \bar c(0) c(0) e^{-\phi(0)} V_1(0) e^{ik.X(0)} \p\xi(\alpha y)
c(y) e^{-\phi(y)} V_2(y) e^{-ik.X(y)}
\Big\rangle \bigg]\, .
\ee
Since this is the integral of
a total derivative, it is given by boundary contribution.
Possible boundary terms could arise from around $y=0$ if the term inside the square
bracket diverged as $1/\bar y$ in this limit. Using \refb{eope}, \refb{ephiope} we see
however that it goes as $y/\bar y$ in the $y\to 0$ limit. Thus there are no non-zero
boundary contributions, showing that \refb{ess3} is indeed independent of $\alpha$.

\subsection{Two loop dilaton tadpole} 

The Fayet-Iliopoulos term is also expected to generate a dilaton tadpole
at two loop level. Formalism involving picture changing operators can also be used to
compute this. This was done in \cite{AtickS}. Here we shall review the basic steps
of \cite{AtickS} so that the
reader can see the close parallel between the computation of one loop mass 
renormalization described above and the two loop dilaton tadpole.

For two loop one point function we need three insertions of picture changing operators.
In the limit of degeneration to two tori, two of the picture changing operators must lie
on the tori that contains the dilaton vertex operator in the $-1$ picture
while the third picture changing
operators will have to lie on the torus without any external state. This was achieved 
in \cite{AtickS} by taking one of the picture changing operators on top of the dilaton vertex
operator to bring it to a 0-picture vertex operator. This is convenient (and is analogous
to taking the $\alpha\to 1$ limit in the one loop analysis) but  is not necessary. We shall
proceed without taking this limit.

The second step involves expressing the $-1$ picture dilaton vertex operator as
$\{Q_S, W\}$ where $Q_S$ is the supersymmetry generator in the $-1/2$ picture
and $W$ is the dilatino vertex operator in the $-1/2$ picture. This can be expressed as
the contour integral of the supersymmetry current around the dilatino vertex operator.

After summing over spin structures 
the correlation function involving  the supersymmetry current satisfies the correct 
periodicity conditions on the genus two Riemann surface.  
In the third step we deform the contour of integration of the supersymmetry current
away from the dilatino vertex operator and try to shrink it to a point. Naively one would
expect that this should be possible leading to vanishing result, but it was found 
in \cite{Verlinde:1987sd,AtickS} 
that the correlation function has spurious poles. Thus the result does not
vanish, but can be expressed as the result of contour integration around the
spurious poles. Let us denote by $C$ the sum of all such contours.

In the next step we move the location of one of the picture changing operators
leaving fixed the position of the contours .
As a result the spurious poles shift. As long as we can ensure that the locations
of all the spurious poles as a function of the location of the supersymmetry current
move outside $C$, the final result will vanish. But in the process of moving the
picture changing operator we pick up a contribution proportional to
$\VVV(z_1) - \VVV(\wt z_1) = \{Q_B, \xi(z_1) - \xi(\wt z_1)\}$ where $z_1$ and
$\wt z_1$ are the initial and final positions of the picture changing operator that is
being moved. Note that in order that the term involving
$\VVV(\wt z_1)$ vanish, we have to ensure that $\wt z_1$ is at a position in which the
spurious poles are outside the contour $C$ for all values of the moduli. In the
degeneration limit this can be achieved if we ensure that $\wt z_1$ 
is on the `wrong side', \i.e.\ the side opposite to that of $z_1$.
In contrast if $z_1$ is on the same side as $\wt z_1$ in this limit then the spurious poles
on the other side will be insensitive to $\wt z_1$ and continue to remain inside the
contour $C$.

In the final step we deform the BRST contour 
in $\{Q_B, \xi(z_1) - \xi(\wt z_1)\}$
and express the result as a total derivative
in the moduli space. The relevant boundary contribution comes from the degeneration
limit described above. The contribution in the degeneration limit gives the expected
contribution to the dilaton tadpole\cite{AtickS}.

The reader would probably have noticed the close parallel between the one loop 
analysis of mass renormalization and the two loop analysis of the dilaton tadpole.
In both cases we move a picture changing operator to the wrong side and show that
the resulting contribution vanishes. Thus the result is given by the difference
between inserting the picture changing operator on the right side and the wrong
side. This in turn is a total derivative in the moduli space and receives contribution
only from the boundary of the moduli space.

\bigskip

{\bf Acknowledgement:}
We thank Nathan Berkovits, Roji Pius, Arnab Rudra, Edward Witten
and Barton Zwiebach for useful discussions, and Barton Zwiebach for his critical
comments on an earlier version of the manuscript.
This work  was
supported in part by the 
DAE project 12-R\&D-HRI-5.02-0303 and J. C. Bose fellowship of 
the Department of Science and Technology, India.




\begin{thebibliography}{99}

\bibitem{nelson} 
  P.~C.~Nelson,
  ``Covariant Insertion of General Vertex Operators,''
  Phys.\ Rev.\ Lett.\  {\bf 62}, 993 (1989).

\bibitem{Vafa1} 
  C.~Vafa,
  ``Operator Formulation on Riemann Surfaces,''
  Phys.\ Lett.\ B {\bf 190}, 47 (1987).

\bibitem{Vafa2} 
  C.~Vafa,
  ``Conformal Theories and Punctured Surfaces,''
  Phys.\ Lett.\ B {\bf 199}, 195 (1987).

\bibitem{Cohen:1985sm} 
  A.~G.~Cohen, G.~W.~Moore, P.~C.~Nelson and J.~Polchinski,
  ``An Off-Shell Propagator for String Theory,''
  Nucl.\ Phys.\ B {\bf 267}, 143 (1986).

\bibitem{Cohen:1986pv} 
  A.~G.~Cohen, G.~W.~Moore, P.~C.~Nelson and J.~Polchinski,
  ``Semi Off-shell String Amplitudes,''
  Nucl.\ Phys.\ B {\bf 281}, 127 (1987).

\bibitem{AG1} 
L.~Alvarez Gaum\'e,
C.~Gomez,
G.~Moore
and C.~Vafa,
``Strings in the operator formalism",  Nucl.\ Phys.\ B {\bf 303}, 455 (1988).

\bibitem{AG2} 
L.~Alvarez Gaum\'e,
C.~Gomez, P.~Nelson,
G.~Sierra
and C.~Vafa,
``Fermionic strings in the operator formalism",  Nucl.\ Phys.\ B {\bf 311}, 333 (1988).


\bibitem{Polchinski:1988jq} 
  J.~Polchinski,
  ``Factorization of Bosonic String Amplitudes,''
  Nucl.\ Phys.\ B {\bf 307}, 61 (1988).



\bibitem{1311.1257} 
  R.~Pius, A.~Rudra and A.~Sen,
  ``Mass Renormalization in String Theory: Special States,''
  arXiv:1311.1257 [hep-th].

\bibitem{1401.7014} 
  R.~Pius, A.~Rudra and A.~Sen,
  ``Mass Renormalization in String Theory: General States,''
  arXiv:1401.7014 [hep-th].

\bibitem{wittensft} 
  E.~Witten,
  ``Noncommutative Geometry and String Field Theory,''
  Nucl.\ Phys.\ B {\bf 268}, 253 (1986).

\bibitem{9206084} 
  B.~Zwiebach,
  ``Closed string field theory: Quantum action and the B-V master equation,''
  Nucl.\ Phys.\ B {\bf 390}, 33 (1993)
  [hep-th/9206084].

\bibitem{0708.2591} 
  L.~Rastelli and B.~Zwiebach,
  ``The Off-shell Veneziano amplitude in Schnabl gauge,''
  JHEP {\bf 0801}, 018 (2008)
  [arXiv:0708.2591 [hep-th]].


\bibitem{1209.5461} 
  E.~Witten,
  ``Superstring Perturbation Theory Revisited,''
  arXiv:1209.5461 [hep-th].

\bibitem{Belopolsky} 
  A.~Belopolsky,
  ``De Rham cohomology of the supermanifolds and superstring BRST cohomology,''
  Phys.\ Lett.\ B {\bf 403}, 47 (1997)
  [hep-th/9609220];
``New geometrical approach to superstrings,''
  hep-th/9703183;

\bibitem{9706033} 
  A.~Belopolsky,
``Picture changing operators in supergeometry and superstring theory,''
  hep-th/9706033.

\bibitem{dp}
  E.~D'Hoker and D.~H.~Phong,
  ``Two loop superstrings. I. Main formulas,''
  Phys.\ Lett.\ B {\bf 529}, 241 (2002)
  [hep-th/0110247].
``II. The Chiral measure on moduli space,''
  Nucl.\ Phys.\ B {\bf 636}, 3 (2002)
  [hep-th/0110283].
``III. Slice independence and absence of ambiguities,''
  Nucl.\ Phys.\ B {\bf 636}, 61 (2002)
  [hep-th/0111016].
``IV: The Cosmological constant and modular forms,''
  Nucl.\ Phys.\ B {\bf 639}, 129 (2002)
  [hep-th/0111040].
``V. Gauge slice independence of the N-point function,''
  Nucl.\ Phys.\ B {\bf 715}, 91 (2005)
  [hep-th/0501196].
``VI: Non-renormalization theorems and the 4-point function,''
  Nucl.\ Phys.\ B {\bf 715}, 3 (2005)
  [hep-th/0501197].
``VII. Cohomology of Chiral Amplitudes,''
  Nucl.\ Phys.\ B {\bf 804}, 421 (2008)
  [arXiv:0711.4314 [hep-th]].


\bibitem{Witten}
E.~Witten,
  ``Notes On Supermanifolds and Integration,''
  arXiv:1209.2199 [hep-th];
 ``Notes On Super Riemann Surfaces And Their Moduli,''
  arXiv:1209.2459 [hep-th];
``Notes On Holomorphic String And Superstring Theory Measures Of Low Genus,''
  arXiv:1306.3621 [hep-th];

\bibitem{donagi-witten} 
  R.~Donagi and E.~Witten,
  ``Supermoduli Space Is Not Projected,''
  arXiv:1304.7798 [hep-th];
``Super Atiyah classes and obstructions to splitting of supermoduli space,''
  arXiv:1404.6257 [hep-th].

\bibitem{wittenssft} 
  E.~Witten,
  ``Interacting Field Theory of Open Superstrings,''
  Nucl.\ Phys.\ B {\bf 276}, 291 (1986).

\bibitem{9202087} 
  R.~Saroja and A.~Sen,
  ``Picture changing operators in closed fermionic string field theory,''
  Phys.\ Lett.\ B {\bf 286}, 256 (1992)
  [hep-th/9202087].

\bibitem{9503099}
  N.~Berkovits,
  ``SuperPoincare invariant superstring field theory,''
  Nucl.\ Phys.\ B {\bf 450} (1995) 90
   [Erratum-ibid.\ B {\bf 459} (1996) 439]
  [hep-th/9503099].

\bibitem{0109100}
  N.~Berkovits,
  ``The Ramond sector of open superstring field theory,''
  JHEP {\bf 0111} (2001) 047
  [hep-th/0109100].

\bibitem{0406212}
  Y.~Okawa and B.~Zwiebach,
  ``Heterotic string field theory,''
  JHEP {\bf 0407} (2004) 042
  [hep-th/0406212].

\bibitem{0409018}
  N.~Berkovits, Y.~Okawa and B.~Zwiebach,
  ``WZW-like action for heterotic string field theory,''
  JHEP {\bf 0411} (2004) 038
  [hep-th/0409018].

\bibitem{1201.1761}
  M.~Kroyter, Y.~Okawa, M.~Schnabl, S.~Torii and B.~Zwiebach,
  ``Open superstring field theory I: gauge fixing, ghost structure, and propagator,''
  JHEP {\bf 1203} (2012) 030
  [arXiv:1201.1761 [hep-th]].

\bibitem{1303.2323}
  B.~Jurco and K.~Muenster,
  ``Type II Superstring Field Theory: Geometric Approach and Operadic Description,''
  JHEP {\bf 1304} (2013) 126
  [arXiv:1303.2323 [hep-th]].

\bibitem{1312.1677}
  Y.~Iimori, T.~Noumi, Y.~Okawa and S.~Torii,
  ``From the Berkovits formulation to the Witten formulation in open superstring field theory,''
  JHEP {\bf 1403} (2014) 044
  [arXiv:1312.1677 [hep-th]].

\bibitem{1312.2948}
  T.~Erler, S.~Konopka and I.~Sachs,
  ``Resolving Witten`s superstring field theory,''
  JHEP {\bf 1404} (2014) 150
  [arXiv:1312.2948 [hep-th]];
``NS-NS Sector of Closed Superstring Field Theory,''
  arXiv:1403.0940 [hep-th].


\bibitem{1312.7197}
  H.~Kunitomo,
  ``The Ramond Sector of Heterotic String Field Theory,''
  PTEP {\bf 2014} 4,  043B01
  [arXiv:1312.7197 [hep-th]].

\bibitem{1403.0940} 
  T.~Erler, S.~Konopka and I.~Sachs,
  ``NS-NS Sector of Closed Superstring Field Theory,''
  arXiv:1403.0940 [hep-th].

\bibitem{FMS} 
  D.~Friedan, E.~J.~Martinec and S.~H.~Shenker,
  ``Conformal Invariance, Supersymmetry and String Theory,''
  Nucl.\ Phys.\ B {\bf 271}, 93 (1986).

\bibitem{Verlinde:1987sd} 
  E.~P.~Verlinde and H.~L.~Verlinde,
  ``Multiloop Calculations in Covariant Superstring Theory,''
  Phys.\ Lett.\ B {\bf 192}, 95 (1987).


\bibitem{DSW}
 M.~Dine, N.~Seiberg and E.~Witten,
  ``Fayet-Iliopoulos Terms in String Theory,''
  Nucl.\ Phys.\ B {\bf 289}, 589 (1987).

\bibitem{ADS}
J.~J.~Atick, L.~J.~Dixon and A.~Sen,
  ``String Calculation of Fayet-Iliopoulos d Terms in Arbitrary Supersymmetric Compactifications,''
  Nucl.\ Phys.\ B {\bf 292}, 109 (1987).

\bibitem{DIS}
M.~Dine, I.~Ichinose and N.~Seiberg,
  ``F Terms and d Terms in String Theory,''
  Nucl.\ Phys.\ B {\bf 293}, 253 (1987).
  
\bibitem{greenseiberg} 
  M.~B.~Green and N.~Seiberg,
  ``Contact Interactions in Superstring Theory,''
  Nucl.\ Phys.\ B {\bf 299}, 559 (1988).

\bibitem{AtickS}
  J.~J.~Atick and A.~Sen,
  ``Two Loop Dilaton Tadpole Induced by Fayet-iliopoulos $D$ Terms in Compactified Heterotic String Theories,''
  Nucl.\ Phys.\ B {\bf 296}, 157 (1988).

\bibitem{1304.2832}
E.~Witten,
``More On Superstring Perturbation Theory,''
  arXiv:1304.2832 [hep-th].

  \bibitem{1403.5494} 
  E.~D'Hoker and D.~H.~Phong,
  ``Two-loop vacuum energy for Calabi-Yau orbifold models,''
  Nucl.\ Phys.\ B {\bf 877}, 343 (2013)
  [arXiv:1307.1749];
  E.~D'Hoker 
  Topics in Two-Loop Superstring Perturbation Theory
arXiv:1403.5494 [hep-th];
  
 \bibitem{1404.5346}
 N.~Berkovits and E.~Witten,
``Supersymmetry Breaking Effects
using the Pure Spinor Formalism of the Superstring",  
arXiv:1404.5346 [hep-th].

\bibitem{kugo}
T.~Kugo, H~Kunitomo and K.~Suehiro,
``Nonpolynomial Closed String Field Theory,''
  Phys.\ Lett.\ B {\bf 226}, 48 (1989);
T.~Kugo and K.~Suehiro,
``Nonpolynomial Closed String Field Theory: Action and Its Gauge Invariance,''
  Nucl.\ Phys.\ B {\bf 337}, 434 (1990).


\bibitem{saadi} 
  M.~Saadi and B.~Zwiebach,
  Annals Phys.\  {\bf 192}, 213 (1989).

\bibitem{1307.5124}
E.~Witten,
``The Feynman $i \epsilon$ in String Theory,''
  arXiv:1307.5124 [hep-th].

\bibitem{ARS} 
  J.~J.~Atick, J.~M.~Rabin and A.~Sen,
  ``An Ambiguity in Fermionic String Perturbation Theory,''
  Nucl.\ Phys.\ B {\bf 299}, 279 (1988).

\bibitem{global} 
  J.~J.~Atick, G.~W.~Moore and A.~Sen,
  ``Some Global Issues in String Perturbation Theory,''
  Nucl.\ Phys.\ B {\bf 308}, 1 (1988).

\bibitem{catoptric} 
  J.~J.~Atick, G.~W.~Moore and A.~Sen,
  ``Catoptric Tadpoles,''
  Nucl.\ Phys.\ B {\bf 307}, 221 (1988).

\bibitem{lechtenfeld}
O.~Lechtenfeld,
``Superconformal Ghost Correlations On Riemann Surfaces,''
  Phys.\ Lett.\ B {\bf 232}, 193 (1989).


\bibitem{morozov}
  A.~Morozov,
  ``STRAIGHTFORWARD PROOF OF LECHTENFELD'S FORMULA FOR BETA, gamma CORRELATOR,''
  Phys.\ Lett.\ B {\bf 234}, 15 (1990)
  [Yad.\ Fiz.\  {\bf 51}, 301 (1990)]
  [Sov.\ J.\ Nucl.\ Phys.\  {\bf 51}, 190 (1990)].
  
\bibitem{VerlindeChiral} 
  E.~P.~Verlinde and H.~L.~Verlinde,
  ``Chiral Bosonization, Determinants and the String Partition Function,''
  Nucl.\ Phys.\ B {\bf 288}, 357 (1987).

\bibitem{Verlinde:1988tx} 
  E.~P.~Verlinde and H.~L.~Verlinde,
  ``Lectures On String Perturbation Theory,''
  IN *TRIESTE 1988, PROCEEDINGS, SUPERSTRINGS '88* 189-250 AND INST. ADV. STUD. PRINCETON - IASSNS-HEP-88-52 (88,REC.MAR.89) 69 P. (906367) (SEE CONFERENCE INDEX)

\bibitem{1411.7478} 
  A.~Sen,
  ``Gauge Invariant 1PI Effective Action for Superstring Field Theory,''
  arXiv:1411.7478 [hep-th].

\bibitem{appear2}
A.~Sen,
``Gauge Invariant 1PI 
Effective Superstring Field Theory: Inclusion of the Ramond Sector,'', to appear.

\bibitem{1404.6254} 
  R.~Pius, A.~Rudra and A.~Sen,
  ``String Perturbation Theory Around Dynamically Shifted Vacuum,''
  arXiv:1404.6254 [hep-th].

\end{thebibliography}
\end{document}